%% file: mainHigherOrderMeshes.tex
\documentclass[pdflatex,sn-mathphys-num]{sn-jnl}

\usepackage{graphicx}%
\usepackage{multirow}%
\usepackage{amsmath,amssymb,amsfonts}%
\usepackage{amsthm}%
\usepackage{mathrsfs}%
\usepackage[title]{appendix}%
\usepackage{xcolor}%
\usepackage{textcomp}%
\usepackage{manyfoot}%
\usepackage{booktabs}%
\usepackage{algorithm}%
\usepackage{algorithmicx}%
\usepackage{algpseudocode}%
\usepackage{listings}%
\usepackage{todonotes}
\usepackage{comment}

\makeatletter
\renewcommand{\todo}[2][]{\tikzexternaldisable\@todo[#1]{#2}\tikzexternalenable}
\makeatother

\theoremstyle{thmstyleone}%
\theoremstyle{thmstyletwo}%

\theoremstyle{thmstylethree}%

\usepackage[labelformat=simple]{subcaption}   	
\usepackage{subfiles} 
\usepackage{ifthen} 
\newboolean{showSubfileBibliography}
\setboolean{showSubfileBibliography}{false} 
\usepackage{tikz}
\usepackage{tikz-3dplot}
\usepackage{pgfplots}
\usepackage{xargs}
\usepackage{forloop}
\usepackage{arrayjobx}
\usepackage{standalone}
\usepackage[capitalise]{cleveref}
\usepackage{mathtools}
\usepackage{siunitx} 
\usepackage{physics}
\usepackage{xfrac}
\newcommand{\MainPath}{.}
\newcommand{\TikzPath}{\MainPath/tikz}
\newcommand{\dirdata}{\MainPath}
\def\myBlack{black}
\def\tikzOff{}
\input{mycommands.tex}
\input{mytikzcommands.tex}
\graphicspath{{.}}
\tikzexternaldisable
\newcommand{\pt}[1]{\boldsymbol{#1}}
\newcommand{\Bspline}{B} 		
\newcommand{\CP}{\pt{c}}		
\newcommand{\curve}{\pt{C}}
\newcommand{\surface}{\pt{S}}
\newcommand{\uusurf}{u} 		
\newcommand{\vvsurf}{v} 		
\newcommand{\uusurfB}{s} 		
\newcommand{\vvsurfB}{t} 		
\newcommand{\pu}{p} 		    
\newcommand{\pusurf}{p} 		
\newcommand{\pvsurf}{q} 		
\newcommand{\visibledomain}{\mathcal{A}^{\textnormal{v}}} 
\newcommand{\TrimCurve}{\pt{C}^{t}}     
\newcommand{\uutrim}{\xi}               
\newcommand{\TrimCurveSubRegion}{\hat{\gamma}}  
\newcommand{\TrimCurveSubRegionMapped}{{\gamma}}  
\newcommand{\MeshEdge}{\tilde{\gamma}}  
\newcommand{\MeshVertex}{\tilde{\pt{v}}}

\newcommand{\IntersectionCurve}{\tilde{\curve}}
\newcommand{\indexA}{i}

\newcommand{\indexB}{j}

\newcommand{\indexC}{k}

\newcommand{\operate}[1]{\mathopen{}\left(#1\right)\mathclose{}} 
\newcommand{\R}{\mathbb{R}}

\newcommand{\TODO}[1]{\color{red}TODO:#1\color{black}}
\newcommand{\myfbox}[1]{#1}
\newcommand{\mynotes}[1]{}

\newcommand{\matr}[1]{\underline{\mathrm{#1}}}
\newcommand{\matrg}[1]{\underline{#1}} 

\newcommand{\BEMRegion}{\Gamma}
\newcommand{\BEMRegionName}{surface region}
\newcommand{\BEMRegionNum}{A}
\newcommand{\BEMRegionIndex}{a}
\newcommand{\BEMRegionIndexB}{b}
\newcommand{\BEMSurface}{\surface}
\newcommand{\BEMSurfaceIndex}{i}
\newcommand{\BEMFEOrder}{\nu}

\begin{document}

\sisetup{
  round-mode=places,
  round-precision=2,
}

\title[CAD-Integrated Electrostatic Boundary Element Simulations with Non-Conforming Higher-Order Meshes]{CAD-Integrated Electrostatic Boundary Element Simulations with Non-Conforming Higher-Order Meshes}


\author*[1]{\fnm{Benjamin} \sur{Marussig}}\email{marussig@tugraz.at}

\author[2]{\fnm{J\"{u}rgen} \sur{Zechner}}

\author[2]{\fnm{Thomas} \sur{R\"{u}berg}}

\author[2]{\fnm{Lars} \sur{Kielhorn}}

\author[1,3]{\fnm{Domagoj} \sur{Bo\v{s}njak}}

\author[3]{\fnm{Thomas-Peter} \sur{Fries}}

\affil[1]{\orgdiv{Institute of Applied Mechanics, Graz Center of Computational Engineering (GCCE)}, \orgname{Graz University of Technology}, \orgaddress{\street{Technikerstaße 4/II}, \city{Graz}, \postcode{8010}, \country{Austria}}}

\affil[2]{\orgname{TailSiT GmbH}, \orgaddress{\street{Nikolaiplatz 4}, \city{Graz}, \postcode{8020}, \country{Austria}}}

\affil[3]{\orgdiv{Institute of Structural Analysis, Graz Center of Computational Engineering (GCCE)}, \orgname{Graz University of Technology}, \orgaddress{\street{Lessingstraße 25/II}, \city{Graz}, \postcode{8010}, \country{Austria}}}


\abstract{
    We present a design through analysis workflow that enables virtual prototyping of electric devices. 
    A CAD plugin establishes the interaction between design and analysis, allowing the preparation of analysis models and the visualization of its results within the design environment. The simulations utilize a fast boundary element method (BEM) that allows for non-conforming and higher-order meshes. 
    Our numerical experiments investigate the accuracy of the approach and its sensitivity to the initial CAD representation. 
    Overall, the workflow enables a close link between design and analysis, where the non-conforming higher-order BEM approach provides accurate results and significantly simplifies the interaction. 
}


\maketitle

\section{Introduction}

\subfile{sectionIntro}

\section{Design through analysis concept}
\label{sec:concept}

\subfile{sectionConcept}

\section{Boundary element formulation}
\label{sec:BEM}

\subfile{sectionBEM.tex}

\section{CAD model enhancement}
\label{sec:CADenhancement}

\subfile{sectionCAD.tex}

\section{Higher-order mesh generation}
\label{sec:HOmeshing}

\subfile{sectionMesh.tex}

\subfile{sectionMeshHO.tex}

\section{Analysis workflow}
\label{sec:workflow}

\subfile{sectionWorkflow.tex}

\section{Numerical experiments}
\label{sec:results}

The following numerical experiments investigate the proposed BEM formulation using high-order non-conforming meshes on CAD objects.
First, \cref{sec:SphericalElectrodes,sec:FloatingPotentials} demonstrate the performance of the BEM solver.
Next, we assess the influence of the trimming curve on the simulation accuracy in \cref{sec:ParallelPlates}.

The entire analysis workflow has been performed with the presented plugin and the numerical examples are documented as Python scripts\footnote{The Python scripts for reproducing the numerical experiments will be made available once the presented paper has been accepted for publication.}.
If not stated otherwise, an isoparametric discretization will be employed, i.e., the order $\pu$ of elements of the geometric mesh and the order $\BEMFEOrder$ of the shape functions for the finite element discretization of the unknown functions on the surface mesh are the same.

\subsection{Spherical electrodes}
\label{sec:SphericalElectrodes}

\subfile{sectionResultsTwoSpheres.tex}

\subsection{Floating potentials}
\label{sec:FloatingPotentials}

\subfile{sectionResultsFloating.tex}

\subsection{Parallel plates}
\label{sec:ParallelPlates}

\subfile{sectionResultsParallelPlatesTrim.tex}

\subsection{High voltage bushing}

\subfile{sectionResultsBushing.tex}

\section{Conclusion}

The proposed design through analysis workflow leverages the properties of the specific use case, namely electrostatic analysis for virtual prototyping (VP) of electric devices, to mitigate the challenges in the interaction between computer-aided design (CAD) and numerical simulations.
In particular, the use of a boundary element method (BEM) allows the use of boundary representations for design and analysis. Furthermore, the BEM formulation can handle non-conforming meshes for simpler mesh generation, utilizes higher-order elements for improved accuracy, and employs the fast multipole method to achieve nearly linear complexity in terms of storage and computational time.
Nonetheless, the CAD models do not provide all the information required for analysis. Thus, we enrich the design representation to improve connectivity data between spline surfaces (and subsequently for their corresponding meshes) for manifold and non-manifold edges.
In addition, we extend the CAD system's meshing capability by enabling the elevation of the mesh order.
The numerical results show the good performance of the simulation tool for various test cases.
All numerical experiments are executed by Python scripts\footnote{The Python scripts will be made available once the paper is accepted for publication.} that call functions of Rhino 7 (the CAD system used) and the proposed VP plugin, thereby documenting the overall workflow.

The presented design through analysis workflow enables VP within a CAD environment on a conventional PC. Currently, the simulations run in parallel to the CAD system on the same machine. A future step may involve cloud-based BEM simulations to expand the size of the VP problems considered and utilize dedicated hardware, such as HPC clusters.
Another open topic is data handling and file management since tracking changes in CAD models and analysis data is not directly addressed by the plugin. However, this missing feature does not limit the VP plugin's capabilities, as users can establish it via well-established Git tools.

\backmatter

\bmhead{Acknowledgements}

This work was supported by the Österreichische Forschungsförderungsgesellschaft (FFG), Austria under the grand number: 883886.

\bibliography{VEGA}

\end{document}

%% file: sectionIntro.tex
Virtual prototyping utilizes computational engineering analysis to assess the behavior of a designed product before it is constructed, thereby eliminating the need for cumbersome and costly physical mock-ups. Usually, computer-aided design (CAD) tools drive the product development process. The resulting CAD models provide the most exact geometry representation available; however, they are generally not analysis-suitable. Hence, virtual prototyping requires design through analysis workflows to effectively integrate CAD and analysis.

\subsection{Challenges regarding design}
\label{sec:ChallengesDesign}
There are several challenges to be addressed to obtain a robust integration.
First of all, 
CAD systems are not flawless,
and as highlighted by \citet{Piegl2005}: ``\emph{By far the most challenging issue in CAD is robustness}.''
Today, most commercial CAD modelers represent a solid by its boundary representation (b-rep), 
which consists of intersecting non-uniform rational B-spline (NURBS) surfaces.
Hence, the canonical solid modeling operation is to compute the intersection between two parametric surfaces \cite{Hoffmann1989,Patrikalakis1993}. 
The algebraic complexity of such a surface-to-surface intersection (SSI) increases rapidly with the degree of the surfaces involved \cite{Sederberg1983,Sederberg1984,Katz1988}, which makes analytic approaches to this problem impractical.
Thus, different approximate SSI approaches have been proposed \cite{Mortenson1997,Patrikalakis2009}. They result in various independent approximations of the actual intersection rather than an unambiguous solution.
The intersection's representation in the parameter space of a surface allows for  specifying a visualized and hidden surface area.
\begin{figure}[b!]
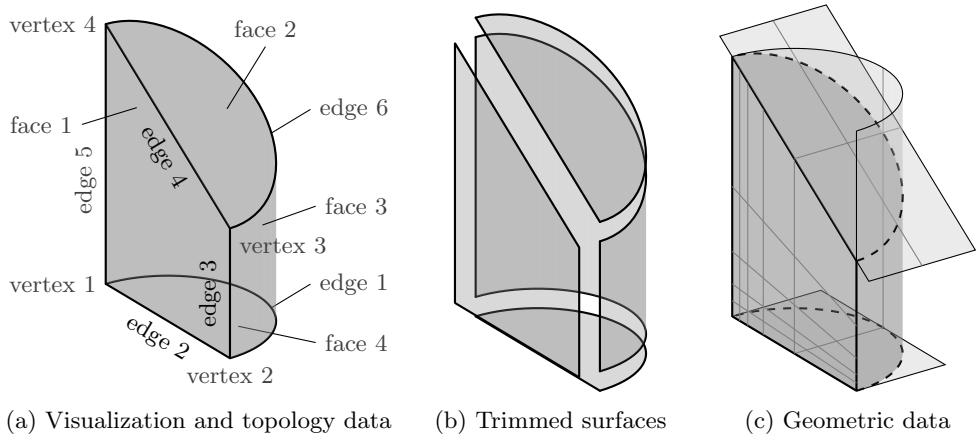

  \centering
  \begin{subfigure}[b]{0.4\textwidth}
	
    \def\tkzscale{0.95}
    \centering
    \tikzsetnextfilename{BRepModelVisualization}
    \input{\TikzPath/BRepModelVisualization}
    \subcaption{Visualization and topology data}
    \label{fig:challenges_visualization}

  \end{subfigure}
  \begin{subfigure}[b]{0.29\textwidth}
	
    \def\tkzscale{0.95}
    \centering
    \tikzsetnextfilename{BRepModelPart}
    \input{\TikzPath/BRepModelPart}
    \subcaption{Trimmed surfaces}
    \label{fig:challenges_parts}

  \end{subfigure}
  \begin{subfigure}[b]{0.29\textwidth}
	
    \def\tkzscale{0.95}
    \centering
    \tikzsetnextfilename{BRepModelSketck}
    \input{\TikzPath/BRepModelSketck}
    \subcaption{Geometric data}
    \label{fig:challenges_parameter}

  \end{subfigure}
  \caption{Different views of a CAD solid model: (a) the visible part of the object and its points, curves, and surfaces associated with its topological entities, i.e., vertices, edges, and faces;
    (b) the individual trimmed surfaces; and (c) their underlying geometric data where dashed lines mark the approximated intersection curves and gray lines indicate the underlying tensor product basis. 
  }
  \label{fig:challenges}
\end{figure}
The resulting CAD objects are also referred to as trimmed surfaces, and various related challenges are detailed in \cite{Marussig2018a}. 
To obtain a solid model, the connectivity between intersecting surfaces is established by assigning the approximate intersection curves (which do not coincide) to a single topological entity. 
\Cref{fig:challenges} illustrates these different aspects of a CAD solid model.
The crucial discrepancy, which still exists, is 
that solid modeling is concerned with the use of \emph{unambiguous} representations, but SSI 
approximations 
do not provide a unique definition of an intersection.
In other words, all CAD modeling approaches have to deal with imprecise data and, thus, fail to guarantee exact topological consistency \cite{Farouki2004}.

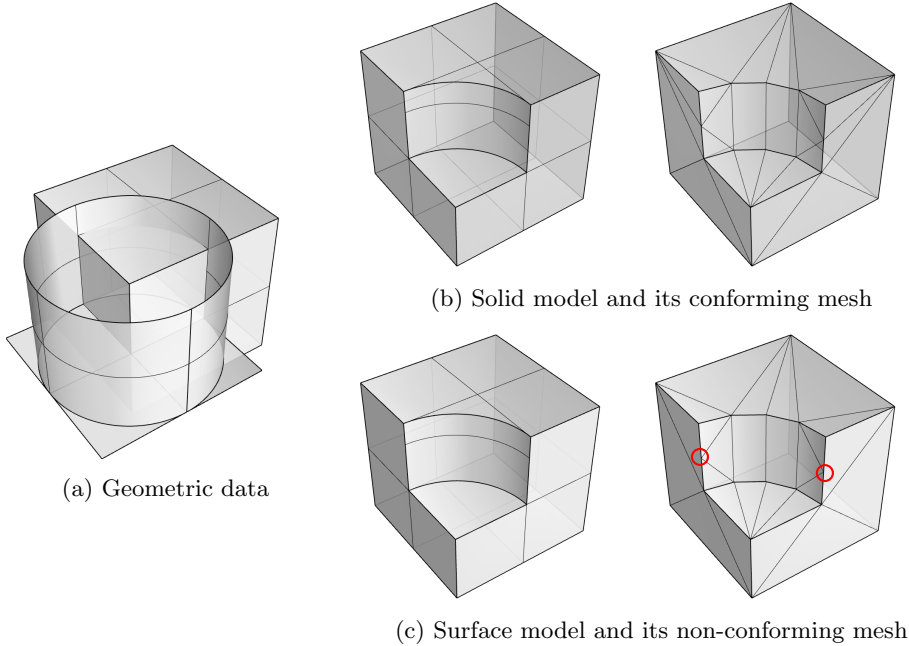
\begin{figure}
    \centering
    \begin{minipage}{0.32\textwidth} 
        \begin{subfigure}[b]{1.0\textwidth}
        \myfbox{\includegraphics[trim={2.2cm 1.4cm 2.0cm 3.8cm},clip,width=0.92\textwidth]{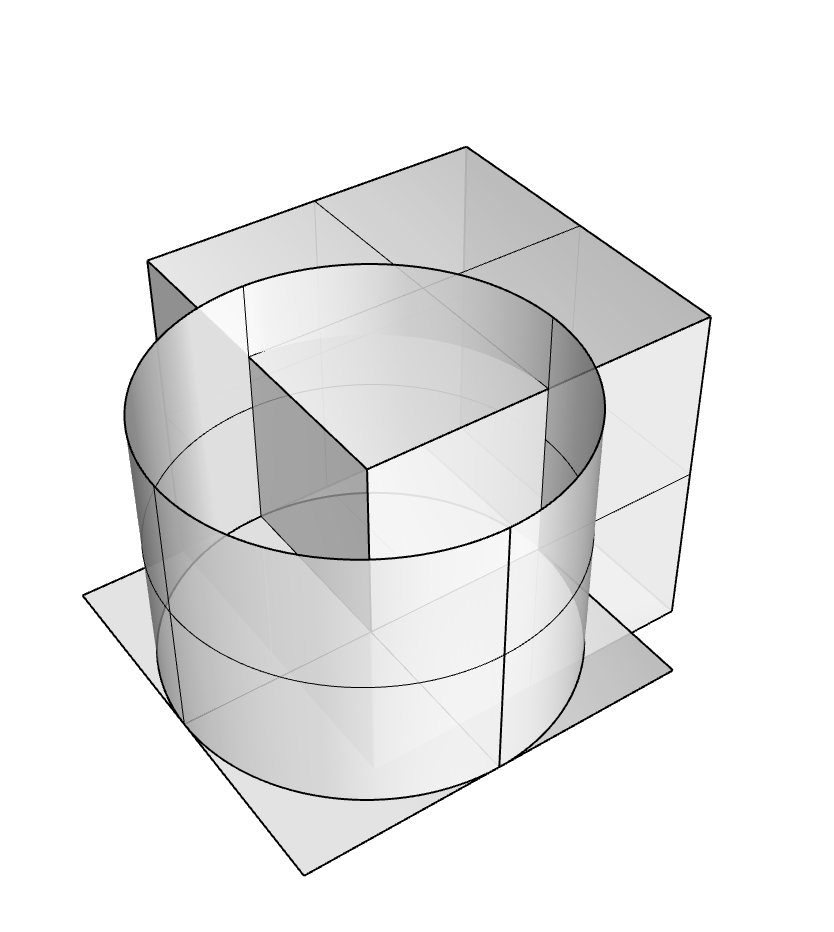}}
        \subcaption{Geometric data}
    \end{subfigure}
    \end{minipage}
    \begin{minipage}{0.64\textwidth} 
    \begin{subfigure}[b]{\textwidth}    
        \myfbox{\includegraphics[trim={2.2cm 4.0cm 2.0cm 3.8cm},clip,width=0.45\textwidth]{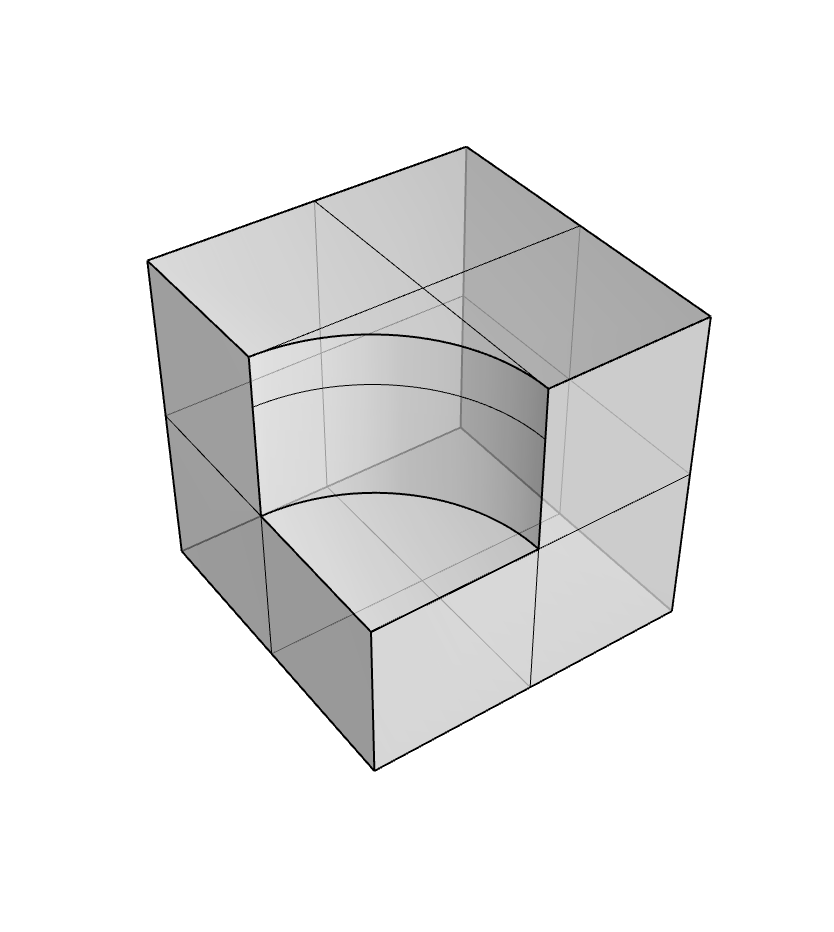}}  
      \myfbox{\includegraphics[trim={2.2cm 4.0cm 2.0cm 3.8cm},clip,width=0.45\textwidth]{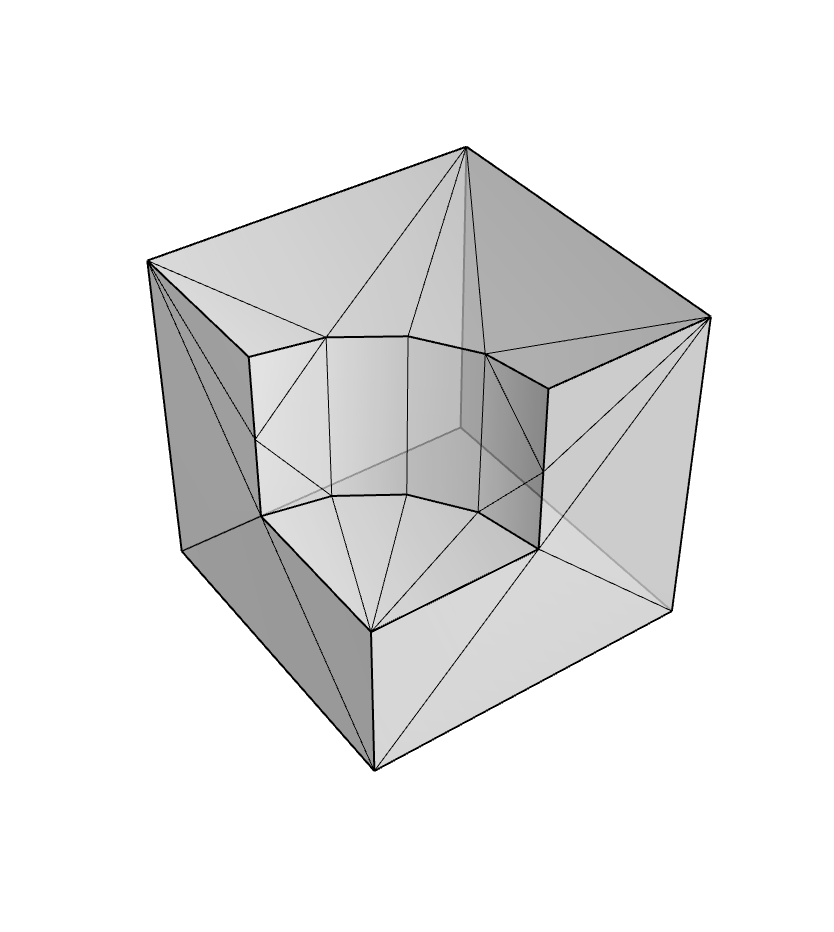}}
      \subcaption{Solid model and its conforming mesh}    
      \label{fig:BoxCylinderBooleanModelMesh}
    \end{subfigure}    
    \begin{subfigure}[b]{\textwidth} 
      \myfbox{\includegraphics[trim={2.2cm 4.0cm 2.0cm 3.8cm},clip,width=0.45\textwidth]{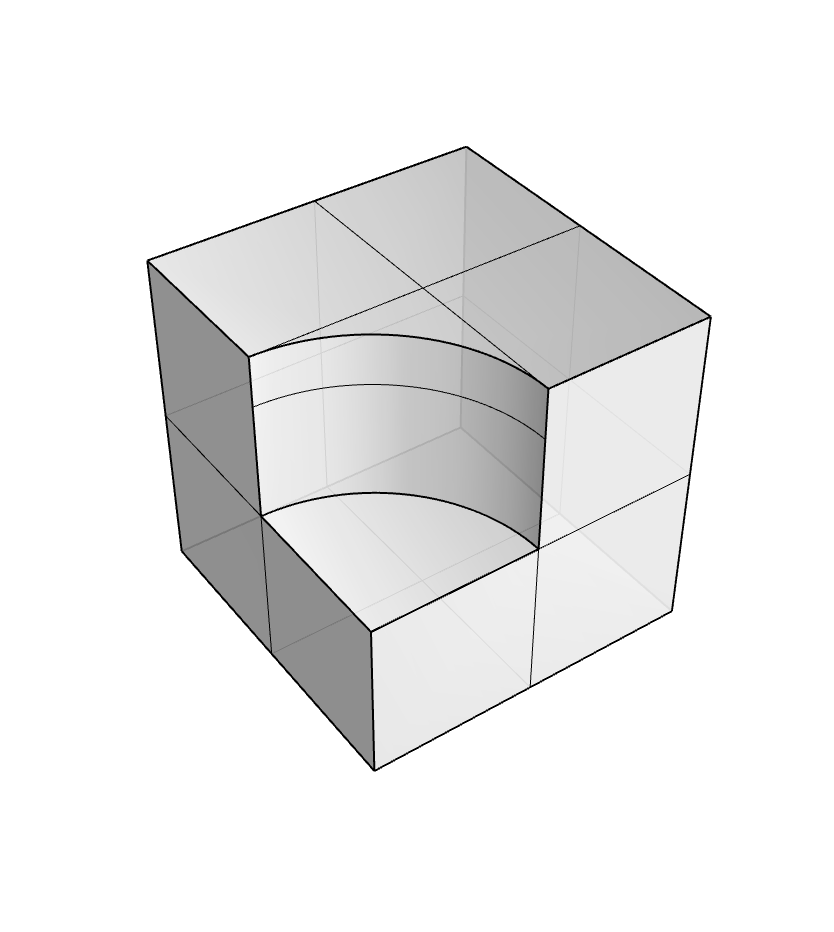}}
      \begin{tikzpicture}[scale=1]
        \myaddgraphic{BoxCylinderBoolean_surface_model_mesh}{trim={2.2cm 4.0cm 2.0cm 3.8cm},clip,width=0.45\textwidth}{0}
        {
          \draw[thick,red] (0.25,0.54)  circle (3pt); 
          \draw[thick,red] (0.685,0.48) circle (3pt);
        }
        {}
      \end{tikzpicture}
      \subcaption{Surface model and its non-conforming mesh}    
      \label{fig:BoxCylinderBooleanModelMeshSurface}
    \end{subfigure}
    \end{minipage}
    \caption{
      The same (a) geometry and its models with different topological data: (b) a solid model or (c) a surface model. 
      The missing topology
       of (c) is reflected in its mesh which is non-conforming; the red circles highlight the hanging mesh nodes.
    }
    \label{fig:BoxCylinderBooleanSpline}
\end{figure}

Especially for cross-disciplinary collaboration, an aggravating factor is that CAD model flaws are not apparent to the user. This statement is particularly true for topological information. As illustrated in \cref{fig:BoxCylinderBooleanSpline}, a b-rep's underlying geometric data (a) usually differs from the model's visualization (b,c); furthermore, the difference between a solid model (i.e., one with topology/connectivity information between intersecting surfaces) and a surface model (i.e., one where each surface has only naked edges with any connection to another surfaces) is not visible. Thus, a user may destroy essential information when editing a model without noticing it.

\begin{figure}
    \centering
    \begin{subfigure}[b]{0.32\textwidth}    
        \myfbox{\includegraphics[trim={1.5cm 1.0cm 1.5cm 1.0cm},clip,width=0.9\textwidth]{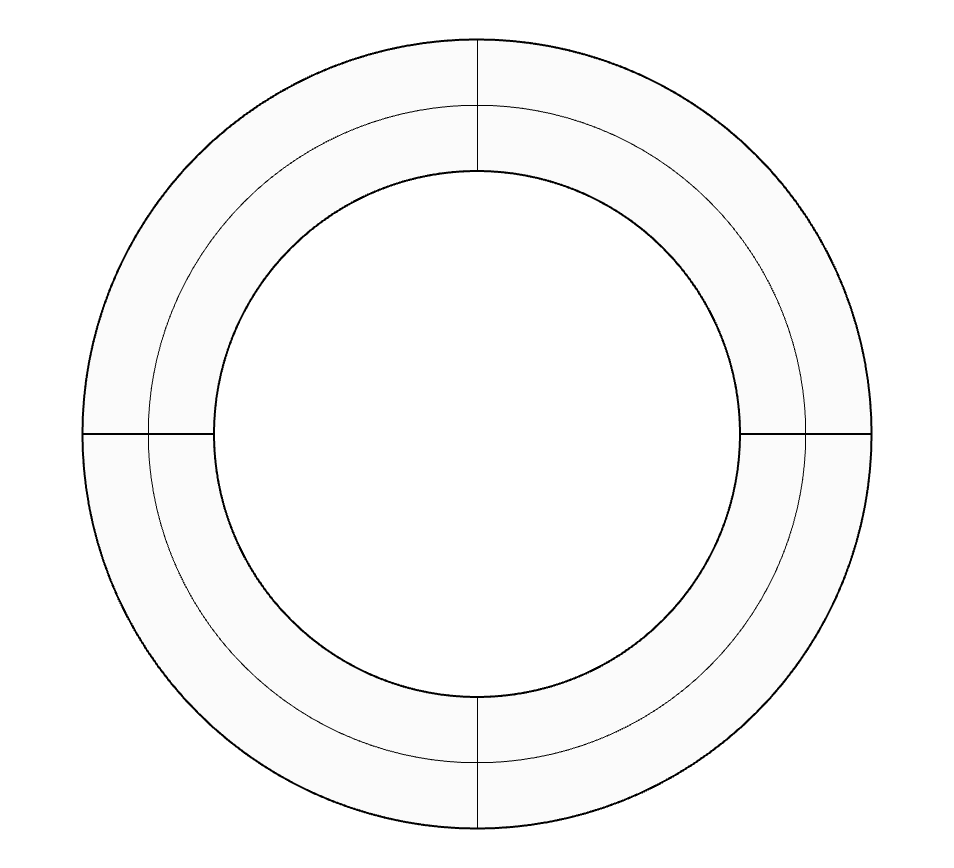}}
    \subcaption{Initial CAD model}    
    \end{subfigure}
    \begin{subfigure}[b]{0.32\textwidth}  
        \myfbox{\includegraphics[trim={1.5cm 1.0cm 1.5cm 1.0cm},clip,width=0.9\textwidth]{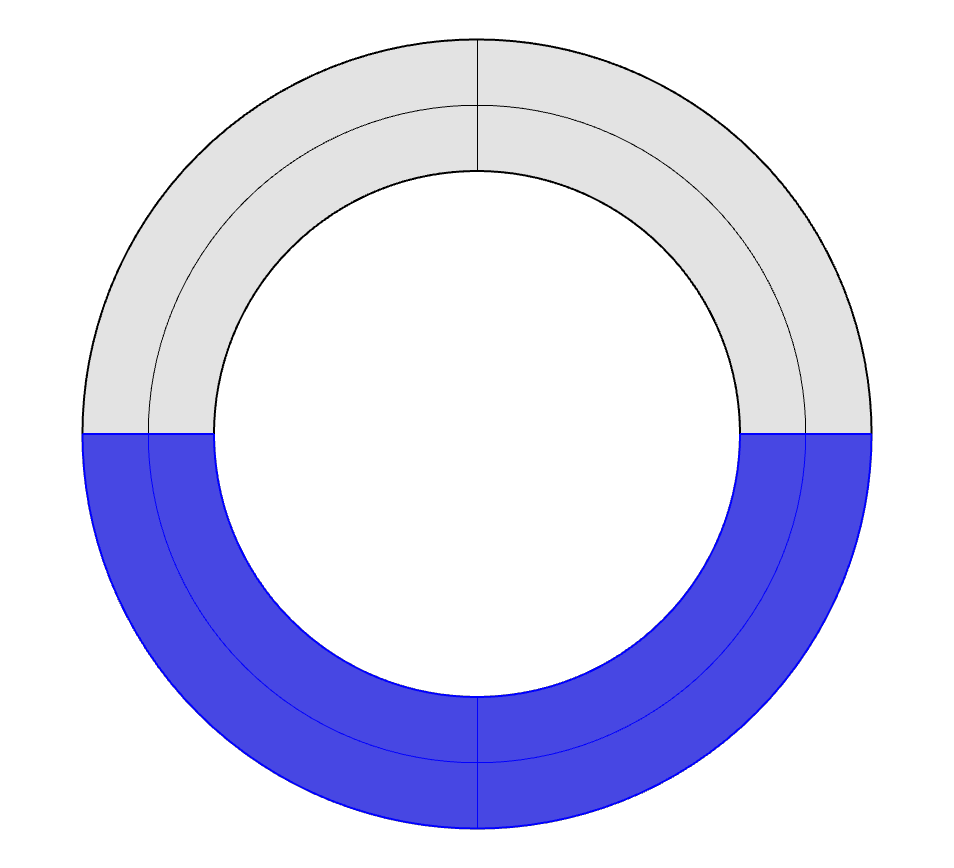}}
        \label{fig:castingIGES}
        \subcaption{IGES export/import}
    \end{subfigure}
    \begin{subfigure}[b]{0.32\textwidth}  
        \myfbox{\includegraphics[trim={1.5cm 1.0cm 1.5cm 1.0cm},clip,width=0.9\textwidth]{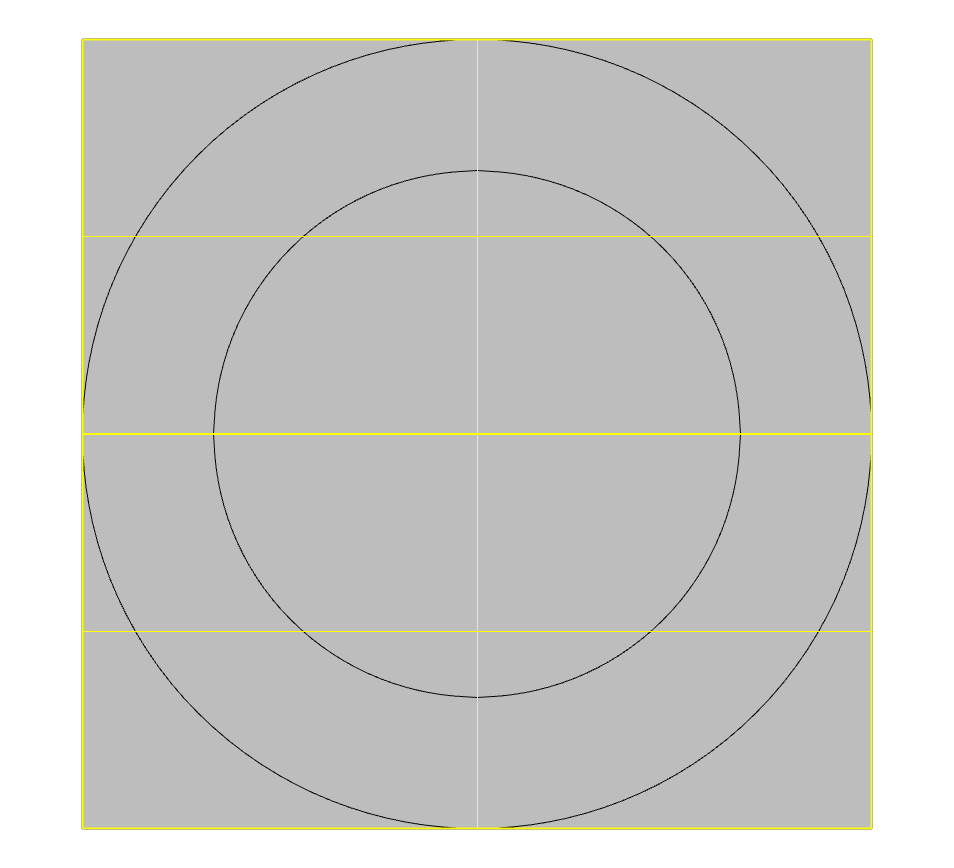}}
        \label{fig:castingSTEP}
        \subcaption{STEP export/import}
    \end{subfigure}
    \caption{Unexpected behaviour during the translation of a CAD model to a downstream application: (a) the original CAD model of a planar ring represented by two regular splines, (b) the exported model data using IGES, or (c) STEP; both reopened in the same CAD system.}
    \label{fig:casting}
\end{figure}

Robustness issues often become apparent when considering  \emph{interoperability} and \emph{data transfer} of CAD models.
For instance, the missing topology data from the b-rep shown in \cref{fig:BoxCylinderBooleanModelMeshSurface} manifests when used as the basis for mesh generation, where the mesh for the surface model is non-conforming. 
While various strategies for data exchange have been proposed \cite{Bianconi2006} and model quality 
testing tools have been provided \cite{GonzalezLluch2017}, the underlying problem usually remains: 
Since there is no canonical representation of trimmed solid models, different systems likely employ independent data structures and robustness checks~\cite{Hoschek1992,Corney2001}. Consequently, data exchange involves a translation process, which opens room for misinterpretation or even loss of model information. 
\cref{fig:casting} illustrates a simple example: 
a planar ring specified by two NURBS surfaces. 
%
The model was designed using the CAD system Rhino 7 and exported using the default settings to the neutral data formats IGES\footnote{Initial Graphics Exchange Specification} and STEP\footnote{Standard for the Exchange of Product Model Data}.
Subsequently, these files have been imported again into Rhino 7 leading to the models shown in \cref{fig:castingIGES,fig:castingSTEP}.
The IGES output neglects the topology information and provides two independent surfaces (indicated by the different colors). The STEP output, on the other hand, maintains the topology information but casts the ring to a single trimmed surface, i.e., the gray plane with the yellow parametrization and the black trim curves.
This transfer challenge makes the treatment of trimmed models an essential aspect of design and analysis interoperability,
particularly when the transfer involves another software tool,  
which requires a closed, complete model representation.

\subsection{Meshing challenges} 
Generating an analysis-suitable representation from a CAD model requires consideration of simulation aspects, model interpretation, and conversion challenges.
This translation process
has become, by far, the most time-consuming task of a numerical simulation, as highlighted by several articles and surveys \cite{Sederberg2008,Hardwick2005,Jackson2013,KubotekUSA2006}.
A significant amount of effort (both semi-automated and/or manual) is required to simplify CAD models \cite{Thakur2009} and make them “watertight,” enabling their use in current simulation software.
Geometry repair and corrections are hereby key tasks,
as concluded by \citet{Botsch2010}: ``\emph{it is tempting to think that model repair is a necessity that will disappear [\dots]. However, [\dots] model repair will remain one of the most enduring problems and it will be critical for further streamlining the geometry processing pipeline}.''

Polygonal meshes are the most prominent geometric models in computational engineering analysis.
Generating a mesh of trimmed b-rep models is prone to producing undesired artifacts 
such as gaps and small overlaps, holes and isles, singular vertices, inconsistent orientation, complex edges, intersections, and so on \cite{Botsch2010}. 
Detecting and resolving such unwanted features is cumbersome, independent of the element type used for the mesh. 
The acceleration of these repair steps is prohibited as long as user intervention is required, as is usually the case. 
The interested reader is referred to \cite{Attene2013} for a structured overview of mesh repairing techniques, a classification of mesh defects, and a collection of related software tools\footnote{\myhref{https://www.meshrepair.org/}{https://www.meshrepair.org/}, accessed online 14.4.2025}. 
The consideration of subdomains (e.g., to assign different material properties) yields non-manifold assemblies, making the meshing even more difficult \cite{Qian2012,Qian2010,Zhang2010,Zhang2006}.
Hexahedral meshes are usually considered superior to tetrahedral meshes in terms of performance and accuracy, but hex meshing is more demanding \cite{Pietroni2022}.
Yet, even in tet meshing, there are complex tasks such as resolving topology ambiguity \cite{Zhang2012,Zhang2012a}.
While powerful meshing algorithms and prominent software tools, such as Gmsh and Cubit, exist, there are still theoretical, algorithmic, and practical challenges, as detailed in \cite{Pietroni2022}.

Besides polygonal meshes, analysis-suitable splines are an active research area aiming to utilize the representative power and smoothness of splines for numerical simulations, i.e., isogeometric analysis (IGA).
The corresponding reparametrization pipelines may convert a trimmed CAD model to a set of non-trimmed NURBS surfaces \cite{Harbrecht2010,Urick2019,Shepherd2022} or other spline representations \cite{Lai2016,Lai2017,Yu2022,Yu2022a,Marsala2024}.  
Often, hex meshing techniques are a crucial intermediate step of these reconstructions. 

While standard finite element methods (FEM) with volumetric conformal meshes are perhaps the most general simulation approach, the discrepancy to CAD model representations complicates 
the realization of a robust design through analysis workflow. 
Alternative simulation schemes can help overcome this potential obstacle in virtual prototyping.
For instance, non-conforming coupling techniques or discontinuous Galerkin methods may simplify the meshing process by not requiring  conformal meshes..
Immersed boundary methods, on the other hand, significantly reduce the effort of generating volumetric meshes by intersecting the CAD's b-rep with a regular background mesh \cite{Wassermann2019,Messmer2022}.
Here, the analysis has to deal with the resulting cut elements.  
FEM shell analysis allows reducing the geometric representation to an object's (mid) surface. 
In other words, the analysis is performed on a (non-closed) b-rep, which allows including CAD models directly into a design through analysis workflow \cite{Teschemacher2018,Teschemacher2022,Hao2023}.
Boundary element methods (BEM) also enable a reduction of the spatial dimension of the analysis domain, i.e.,  they employ b-reps for simulations. 
In contrast to shell models, BEM usually requires a closed b-rep that encloses the domain of interest.
This includes the treatment of infinite domains by flipping the direction of the outward normal vector.
Due to the close relation between CAD and BEM representations, their combination was a candidate for bridging the gap between solid modeling and engineering analysis early on \cite{Casale1989,Casale1992,Yu1994,stout1999cfd} and rejuvenated with the rise of IGA, considering watertight \cite{Marussig2019,Cervera2005,Sun2018,Takahashi2018,Marussig2015,Doelz2020,Torchio2023} as well as trimmed geometries \cite{Marussig2016b,Marussig2016a,Beer2019}.
However, a prerequisite for BEM is the knowledge of the fundamental solutions of the problem's underlying partial different equations.
In short, while FEM with volumetric conformal meshes is the most general simulation approach, the discrepancy with CAD model representations complicates the parametric linking between design and analysis. 
Alternative analysis schemes can simplify this interaction at the cost of complicating the numerical simulation method.

\subsection{Motivation} 
Despite the great interest from the industry, current virtual prototyping, based on computer-aided drafting in CAD and product testing using computational engineering analysis, is far from perfect.
Often, it involves a broad spectrum of available tools and concepts for this cross-disciplinary collaboration, making the implementation particularly challenging for small and medium-sized enterprises, which frequently lack the necessary prerequisites \cite{Danjou2007}.
Even if these preconditions are met, the exchange of CAD and simulation data of state-of-the-art industry solutions is fragile and error-prone. 
An excellent example that supports this claim is the interoperability disaster of Airbus: The development of the A380 airplane was delayed by almost two years, and ultimately, it was about 6.1 billion dollars over budget.  
The root of the problem was traced back to a single decision: using two different \emph{versions} of a CAD system, which resulted in design inconsistencies, mismatched calculations, and configuration management failures \cite{A380CAD2006}.

In this paper, we present a design through analysis workflow  that enables the virtual prototyping of electric devices. CAD models are subjected to electrostatic simulations to obtain the surrounding electric fields. 
While some preliminary results on the required simulation setup have already been presented in our conference contribution \cite{Marussig2024}, this work provides an extension to higher-order discretizations, a BEM suited for non-conforming meshes which is accelerated by the fast multipole method (FMM),
the treatment of non-manifold models, and a more thorough discussion on the concept for design and analysis interoperability. 
In particular, \cref{sec:concept} gives a concise overview of the different components of the proposed concept.
\Cref{sec:BEM} details the physical problem and the employed BEM formulation.
Then, we outline the enhancements required on the CAD model and mesh generation side in \cref{sec:CADenhancement,sec:HOmeshing}, respectively.
\Cref{sec:workflow} summarizes the final analysis workflow employed for the numerical experiments presented in \cref{sec:results}.

\ifthenelse{\boolean{showSubfileBibliography}}{
    \bibliography{VEGA}%
}{}

%% file: sectionConcept.tex
The proposed design through analysis workflow targets virtual prototyping for electrostatic problems.
The underlying strategy is to leverage the properties of this specific use case to bring the requirements of design and analysis models closer together, thereby minimizing the impediment to interoperability. 
In particular, the workflow unites the following components:
\begin{itemize}
    \item \textbf{Data exchange via CAD plugin:} The interface to the simulation is set up as a plugin of the CAD software used for design. Hence, the model preparations for analysis, the assignment of metadata required for the simulation, and the visualization of analysis results rely on the same data structures, which significantly mitigates translation problems. 
    \item \textbf{B-reps for analysis:} We employ a BEM formulation to compute the electric fields, which facilitates the use of b-rep models for design and analysis. This close relation allows a simpler parametric linking between the CAD model and its analysis-suitable counterpart. Furthermore, computing the fields in the exterior domain does not require additional discretization.
    \item \textbf{Fast BEM solver:} We apply the fast multipole method (FMM) to obtain almost linear complexity concerning the BEM simulations' storage and computational time. In addition, the implementation is parallelized. 
    These performance aspects are essential for performing virtual prototyping of practical problems on standard hardware. 
    \item \textbf{Enhanced CAD connectivity data:} Other ingredients are enhancements of the CAD model which (i) establish connectivity data first between the CAD surfaces and then between their corresponding analysis meshes, even if the CAD model lacks the topology information, and (ii) allow  generation of non-manifold meshes.
    \item \textbf{Non-conforming higher-order meshes:} The simulations rely on a non-conforming surface mesh generation for each trimmed surface independently, thereby minimizing meshing complexity. 
    Furthermore, we may elevate these meshes to higher-order ones to reduce the geometric error with fewer degrees of freedom.
    \item \textbf{Automation and workflow scripting:}  The CAD software used allows the execution of commands via Python scripts. We utilize this feature to parametrize and document the steps taken in the virtual prototyping process.
\end{itemize}
The resulting virtual prototyping tool utilizes Rhino~7 as CAD software, and the presented plugin will be made available at the platform Food4Rhino\footnote{\myhref{https://www.food4rhino.com/}{https://www.food4rhino.com/}, accessed online 27.6.2025}~under the name VEGA (Virtual prototyping by Electrostatic Geometry-aware Analysis).


%% file: sectionBEM.tex
\begin{figure}
  \centering
  \includegraphics[width=1.0\textwidth]{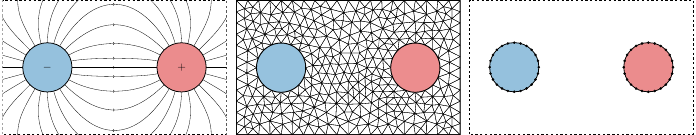}
  
  \caption{Electric field between two electrodes (left), its truncated
    discretization for the analysis with the FEM (middle) and the
    surface discretization required for the simulation with the BEM.}
  \label{fig:BEMvsFEM}
\end{figure}

In this section, we present the electrostatic transmission problem and
derive an indirect boundary integral formulation. A finite element
discretization then leads to the boundary element method (BEM). The
motivation for this choice of method is rooted in the surface-only
discretization, which perfectly fits into the CAD-integration proposed
in this work, and the ease of tackling unbounded problems in
comparison to volume-based discretization methods. \Cref{fig:BEMvsFEM}
illustrates the advantage of electric field computations with BEM. For
related works,
confer~\cite{munger2021dielectric,Blaszczyk2010,Amann2014}.

\subsection{Electrostatic transmission problem} \label{sec:electrostatics}

The electrostatic transmission problem is obtained from Maxwell's
equations by assuming time-independent conditions and the absence of
electric currents. The remaining field equations
are~\cite{jackson2012classical}
\begin{equation}
  \label{eq:electrostatics}
  \div \mathbf{D} = \rho\,, \quad
  \curl \mathbf{E} = \mathbf{0}
  \quad \text{and} \quad
  \mathbf{D} = \varepsilon \mathbf{E}
\end{equation}
with a given (time-independent) volume charge density
$\rho$. $\mathbf{D}$ is the displacement field and $\mathbf{E}$ the
electric field. Since the latter is irrotational, we postulate a
potential $u$ such that $\mathbf{E} = - \grad u$. The electric
permittivity can be written as $\varepsilon = \varepsilon_r
\varepsilon_0$ with the vacuum permittivity $\varepsilon_0$ and the
dimensionless relative permittivity $\varepsilon_r$. We obtain the
electrostatic field equation
\begin{equation}
  \label{eq:electrostatics2}
  -\div \varepsilon_r \grad u = \frac{\rho}{\varepsilon_0}\,.
\end{equation}
It is further assumed that $\varepsilon_r > 0$ is domain-wise
constant. Moreover, we only deal with surface charges but do not face
volume charges, hence $\rho = 0$. The above equations are valid in the
entire three-dimensional space $\mathbb{R}^3$.

\begin{figure}
  \centering 
  \centering
  \def\tkzscale{0.675}
  \tikzsetnextfilename{domainDecomposition}
  \input{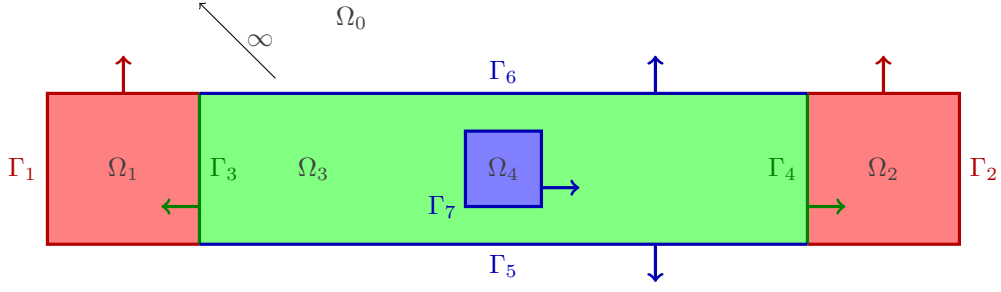}
  \caption{%
    Example constellation: two electrodes $\Omega_1$ and $\Omega_2$, a
    dielectric $\Omega_3$ and a floating potential $\Omega_4$, all
    surrounded by an unbounded air region $\Omega_0$; the thick lines
    show the skeleton $\Gamma$ with its individual {\BEMRegionName}s
    $\BEMRegion_\BEMRegionIndex$, $1 \leq \BEMRegionIndex \leq 7$ and the arrows indicate the surface
    normal vectors; some examples for the index sets and functions are
    $\mathbb{\BEMRegionNum}(3) = \{3,4,5,6,7\}$, $\mathrm{sign}_3(5) = +$,
    $\mathrm{sign}_3(7) = -$, $\mathrm{dom}(6) = \{0,3\}$,
    $\mathrm{opp}_3(3) = 1$, $\mathrm{opp}_3(7) =4$.}
  \label{fig:dd}

\end{figure}

This space is now decomposed into $N_\Omega$ domains $\Omega_n$,
$0 \leq n < N_\Omega$, each with its constant relative permittivity
$\varepsilon_{r,n}$ and we use the convention that $\Omega_0$ denotes
the all-surrounding and unbounded air region with
$\varepsilon_{r,0} = 1$. Let $\partial \Omega_n$ denote the boundary
of the domain $\Omega_n$ and $\Gamma$ the skeleton of the
decomposition, that is the union of all $\partial \Omega_n$. 
This two-dimensional entity $\Gamma$ is now naturally decomposed into
$\BEMRegionNum$ {\BEMRegionName}s $\BEMRegion_\BEMRegionIndex$ which
consist of one or several surface patches
$\BEMSurface_\BEMSurfaceIndex$ as later introduced in~\cref{sec:CADenhancement}. 
Each such {\BEMRegionName} forms the interface between two domains
$\Omega_{n}$ and $\Omega_{m}$.  By convention, the surface normal
$\mathbf{n}_\BEMRegionIndex$ of {\BEMRegionName} $\BEMRegion_\BEMRegionIndex$ points from $\Omega_{n}$ to
$\Omega_{m}$ if $n > m$. Moreover, we can create the index set
$\mathbb{\BEMRegionNum}(n)$ containing all $1 \leq \BEMRegionIndex \leq \BEMRegionNum$ such that
$\BEMRegion_\BEMRegionIndex \cap \partial \Omega_n \neq \emptyset$.  Moreover,
$\mathrm{sign}_n(\BEMRegionIndex) \in \{+,-\}$ provides the sign of the orientation
of $\BEMRegion_\BEMRegionIndex$ with respect to $\Omega_n$. The function
$\{n^+,n^-\} = \mathrm{dom}(\BEMRegionIndex)$ gives the domain indices on the exterior
and interior sides of the {\BEMRegionName} $\BEMRegion_\BEMRegionIndex$.  Finally, one can
establish the function $m = \mathrm{opp}_n(\BEMRegionIndex)$, $0 \leq m < N_\Omega$ which
provides the domain index $m$ of the domain that is on the opposite
side of $\BEMRegion_\BEMRegionIndex$ when viewed from $\Omega_n$.  An example of such a
constellation is given in \cref{fig:dd}.

With the above equations and the domain decomposition at hand, we can
now formulate the electrostatic transmission problem. In each domain,
the potential $u_n(\mathbf{x})$ fulfils Laplace's equation
\begin{equation}
  \label{eq:laplace}
  - \Delta u_n = 0 \qquad \mathbf{x} \,\in\, \Omega_n \\
\end{equation}
and at each {\BEMRegionName} 
we have the transmission conditions
\begin{equation}
  \label{eq:transmission}
  \mathbf{E}_{n_1} \times \mathbf{n}_\BEMRegionIndex = \mathbf{E}_{n_2} \times
  \mathbf{n}_\BEMRegionIndex
  \quad \text{and} \quad
  \mathbf{D}_{n_1} \cdot \mathbf{n}_\BEMRegionIndex = \mathbf{D}_{n_2} \cdot
  \mathbf{n}_\BEMRegionIndex
  \qquad \mathbf{x} \, \in \, \BEMRegion_\BEMRegionIndex\,.
\end{equation}
The uniquely defined normal vector $\mathbf{n}_\BEMRegionIndex$ on $\BEMRegion_\BEMRegionIndex$ allows
for the two function limits
\begin{equation}
  \label{eq:limits}
  f^{\pm}(\mathbf{x}) = \lim_{\delta \to 0} f(\mathbf{x} \pm \delta
  \mathbf{n}_\BEMRegionIndex) \qquad \mathbf{x} \, \in \, \BEMRegion_\BEMRegionIndex\,,
\end{equation}
referred to as exterior ($+$) and interior ($-$) limits,
respectively. With this notion, we can define the Dirichlet and
Neumann traces on {\BEMRegionName} $\BEMRegion_\BEMRegionIndex$ of the function $u$,
\begin{equation}
  \label{eq:traces}
  \gamma_{D,\BEMRegionIndex}^\pm u(\mathbf{x}) = u^\pm(\mathbf{x}) \qquad \text{and} \qquad
  \gamma_{N,\BEMRegionIndex}^\pm u(\mathbf{x}) = \left[ \grad u \right]^\pm(\mathbf{x}) \cdot
  \mathbf{n}_\BEMRegionIndex(\mathbf{x})
  \qquad \mathbf{x} \, \in \, \BEMRegion_\BEMRegionIndex\,.
\end{equation}
It can be shown, that the above transmission
conditions~\labelcref{eq:transmission} are equivalent to the conditions
\begin{equation}
  \label{eq:transmission2}
  \gamma_{D,\BEMRegionIndex}^+ u - \gamma_{D,\BEMRegionIndex}^- u = 0 \qquad \text{and} \qquad
  \varepsilon_{r,n^+} \gamma_{N,\BEMRegionIndex}^+ u - \varepsilon_{r,n^-} \gamma_{N,\BEMRegionIndex}^- u = 0\,,
\end{equation}
where $\{n^+,n^-\} = \mathrm{dom}(\BEMRegionIndex)$. 

A special case are conducting domains ($\varepsilon \to \infty$) on
whose {\BEMRegionName} we obtain a constant potential. Such domains are either
electrodes with prescribed potential $g_n$ or floating potentials with
a constant, but yet unknown potential $\alpha_n$. For the latter, we
have the additional equation that the total charge is prescribed
\begin{equation}
  \label{eq:totalCharge}
  Q_n = \int_{\partial \Omega_n} \mathbf{D} \cdot \mathbf{n} \dd s
  = - \varepsilon_0 \sum_{\BEMRegionIndex \in \mathbb{\BEMRegionNum}(n)} \mathrm{sign}_n(\BEMRegionIndex) \varepsilon_{r,\mathrm{opp}_n(\BEMRegionIndex)} \int_{\BEMRegion_{\BEMRegionIndex}}
  \gamma^{\mathrm{sign}_n(\BEMRegionIndex)}_{N,\BEMRegionIndex} u \dd s\,.
\end{equation}

\subsection{Boundary integral operators} \label{sec:bio}

In order to solve the domain-wise Laplace equations~\labelcref{eq:laplace}
we introduce a global surface density function $\sigma(x)$ on the
skeleton $\Gamma$ and express the solution in each domain by means of
the single layer potential~\cite{sauter2010}
\begin{equation}
  \label{eq:slp}
  u(x) = 
  \left( \mathcal{S} \sigma \right)(\mathbf{x}) =
  \int_\Gamma U(\mathbf{x},\mathbf{y}) \sigma(\mathbf{y}) \dd s_y
\end{equation}
with the Laplace equation's fundamental solution $U(\mathbf{x},\mathbf{y}) = \tfrac{1}{4\pi|\mathbf{x}-\mathbf{y}|}$. Using the
restriction $\sigma_\BEMRegionIndexB(\mathbf{x}) = \sigma(\mathbf{x})$ if
$\mathbf{x} \in \BEMRegion_\BEMRegionIndexB$, we can decompose the single layer potential
as
\begin{equation}
  \label{eq:slpdecomp}
  u(x) =
  \sum_{\BEMRegionIndexB=1}^\BEMRegionNum \left( \mathcal{S}_\BEMRegionIndexB \sigma_\BEMRegionIndexB \right)(\mathbf{x}) =
  \sum_{\BEMRegionIndexB=1}^\BEMRegionNum \int_{\BEMRegion_\BEMRegionIndexB} U(\mathbf{x},\mathbf{y})
  \sigma_\BEMRegionIndexB(\mathbf{y}) \dd s_y\,.
\end{equation}
In order to comply with the transmission conditions, the
traces~\labelcref{eq:traces} are applied to this expression and yield
\begin{equation}
  \label{eq:bio}
  \begin{aligned}
    \gamma_{D,\BEMRegionIndex}^\pm u(\mathbf{x}) &= \gamma_{D,\BEMRegionIndex}^\pm \left(\sum_{\BEMRegionIndexB=1}^\BEMRegionNum\mathcal{S}_\BEMRegionIndexB \sigma_\BEMRegionIndexB\right)(\mathbf{x})
                                     = \sum_{\BEMRegionIndexB=1}^\BEMRegionNum \left( \mathcal{V}_{{\BEMRegionIndex}{\BEMRegionIndexB}}
                                     \sigma_\BEMRegionIndexB \right)(\mathbf{x})\\
    \gamma_{N,\BEMRegionIndex}^\pm u(\mathbf{x}) &= \gamma_{N,\BEMRegionIndex}^\pm \left( \sum_{\BEMRegionIndexB=1}^\BEMRegionNum
                                     \mathcal{S}_\BEMRegionIndexB \sigma_\BEMRegionIndexB \right)(\mathbf{x})
                                     = \sum_{\BEMRegionIndexB=1}^\BEMRegionNum \left[ \left( \mp
                                     \frac{\delta_{{\BEMRegionIndex}{\BEMRegionIndexB}}}{2}  \mathcal{I} +
                                     \mathcal{K}^\prime_{{\BEMRegionIndex}{\BEMRegionIndexB}} \right) \sigma_\BEMRegionIndexB \right](\mathbf{x})
  \end{aligned}
\end{equation}
where the evaluation point $\mathbf{x}$ lies inside $\BEMRegion_\BEMRegionIndex$ and the
latter is assumed to be smooth in the neighbourhood of
$\mathbf{x}$. Here, we have introduced the Kronecker delta notation,
$\delta_{{\BEMRegionIndex}{\BEMRegionIndexB}} = 1$ if and only if $\BEMRegionIndex = \BEMRegionIndexB$ and zero otherwise. The above
traces give rise to the definition of the following boundary integral
operators: single layer operator $\mathcal{V}_{{\BEMRegionIndex}{\BEMRegionIndexB}}$, adjoint double
layer operator $\mathcal{K}^\prime_{{\BEMRegionIndex}{\BEMRegionIndexB}}$ and the identity
$\mathcal{I}$. Moreover, the operators have the subscripts ${\BEMRegionIndex}{\BEMRegionIndexB}$ which
indicate that the boundary integral was taken over the {\BEMRegionName}
$\BEMRegion_\BEMRegionIndexB$ and the evaluation point $\mathbf{x}$ lies on $\BEMRegion_\BEMRegionIndex$.

\subsection{Transmission and boundary conditions} \label{sec:tbc}

Given the above introduced domain decomposition, we can classify each
{\BEMRegionName} $\BEMRegion_\BEMRegionIndex$ to be either bounding an electrode ($E$),
bounding a floating potential ($F$), or be an interface between two
dielectric domains ($D$), where we consider the surrounding
air in $\Omega_0$ as dielectric, too. In the first two cases,
we have a Dirichlet boundary condition of the form $\gamma_{D,\BEMRegionIndex}
u(x) = g_\BEMRegionIndex$ with prescribed electric potential $g_\BEMRegionIndex$ on that {\BEMRegionName} or of the form $\gamma_{D,\BEMRegionIndex} u(x) = \alpha_\BEMRegionIndex$ with the yet
unknown floating potential $\alpha_\BEMRegionIndex$. Note that in case of the
Dirichlet trace $\gamma_D$, the side of the trace is not
important and, hence, the superscript $+$ or $-$ is omitted here.
The Dirichlet conditions are expressed as
\begin{equation}
  \label{eq:dirichlet}
  \gamma_{D,\BEMRegionIndex} u(\mathbf{x}) = \sum_{\BEMRegionIndexB=1}^\BEMRegionNum \left(\mathcal{V}_{{\BEMRegionIndex}{\BEMRegionIndexB}}
    \sigma_\BEMRegionIndexB\right) (\mathbf{x}) =
  \left\{
    \begin{aligned}
      &g_n &&\BEMRegionIndex \in E \\
      &\alpha_n &&\BEMRegionIndex \in F
    \end{aligned}
  \right. \qquad \forall \BEMRegionIndex \in \mathbb{\BEMRegionNum}(n) \,.
\end{equation}

In case of the transmission boundary between two dielectric domains
$\Omega_n$ and $\Omega_m$, we basically have to insert the integral
operators~\labelcref{eq:bio} into the transmission
conditions~\labelcref{eq:transmission2},
\begin{equation}
  \label{eq:transmission3}
  \begin{aligned}
    \sum_{\BEMRegionIndexB=1}^\BEMRegionNum \left(\mathcal{V}_{{\BEMRegionIndex}{\BEMRegionIndexB}} \sigma_\BEMRegionIndexB\right) (\mathrm{x})
    -
    \sum_{\BEMRegionIndexB=1}^\BEMRegionNum \left(\mathcal{V}_{{\BEMRegionIndex}{\BEMRegionIndexB}} \sigma_\BEMRegionIndexB\right) (\mathrm{x})
    &= 0 \\
    \sum_{\BEMRegionIndexB=1}^\BEMRegionNum \varepsilon_{r,n} \left( - \frac{\delta_{{\BEMRegionIndex}{\BEMRegionIndexB}}}{2} \mathcal{I}
    \sigma_\BEMRegionIndexB +
    \mathcal{K}^\prime_{{\BEMRegionIndex}{\BEMRegionIndexB}} \sigma_\BEMRegionIndexB \right)(\mathbf{x}) -
    \sum_{\BEMRegionIndexB=1}^\BEMRegionNum \varepsilon_{r,m} \left( \frac{\delta_{{\BEMRegionIndex}{\BEMRegionIndexB}}}{2} \mathcal{I}
    \sigma_\BEMRegionIndexB +
    \mathcal{K}^\prime_{{\BEMRegionIndex}{\BEMRegionIndexB}} \sigma_\BEMRegionIndexB \right)(\mathbf{x}) &= 0
  \end{aligned}
\end{equation}
with $\{n,m\} = \mathrm{dom}(\BEMRegionIndex)$ in the second equation.  The first of
these conditions is trivially satisfied and the second can be
re-formulated as
\begin{equation}
  \label{eq:transmission4}
  - \lambda_{\BEMRegionIndex} (\mathcal{I} \sigma_\BEMRegionIndex)(\mathbf{x}) + \sum_{\BEMRegionIndexB=1}^\BEMRegionNum
  \left( \mathcal{K}^\prime_{{\BEMRegionIndex}{\BEMRegionIndexB}} \sigma_\BEMRegionIndexB \right)(\mathbf{x}) = 0 \,,
\end{equation}
where we have introduced the dimensionless parameter
\begin{equation}
  \label{eq:lambda}
  \lambda_{\BEMRegionIndex} = \frac{\varepsilon_{r,n} + \varepsilon_{r,m}}{2(
    \varepsilon_{r,n} - \varepsilon_{r,m})}
  \quad \text{with} \quad
  \{n,m\} = \mathrm{dom(\BEMRegionIndex)}\,.
\end{equation}
It is assumed here, that always
$\varepsilon_{r,m} \neq \varepsilon_{r,n}$ when $\Omega_m$ and
$\Omega_n$ have a common interface.

In the case of a floating potential, the Neumann trace as in \cref{eq:bio} has to be inserted into the total charge
calculation according to~\labelcref{eq:totalCharge}
\begin{equation}
  \label{eq:totalCharge2}
  \sum_{\BEMRegionIndex \in \mathbb{\BEMRegionNum}(n)} \varepsilon_{r,\mathrm{opp}_n(\BEMRegionIndex)}
  \int_{\BEMRegion_{\BEMRegionIndex}}
  \sum_{\BEMRegionIndexB=1}^\BEMRegionNum \left[ - \mathcal{I} \frac{\delta_{{\BEMRegionIndex}{\BEMRegionIndexB}}}{2}  + \mathcal{K}^\prime_{{\BEMRegionIndex}{\BEMRegionIndexB}} \right](\sigma_\BEMRegionIndexB) \dd s = Q_n\,.
\end{equation}
Note that \cref{eq:dirichlet,eq:totalCharge2} 
remain valid in the case that electrodes and floating potentials are
just conducting sheets~\cite{munger2021dielectric} which are also
referred to as screens~\cite{sauter2010}.

\subsection{Galerkin boundary element method} \label{sec:gbem}

The aim is now to translate the 
\cref{eq:dirichlet,eq:transmission4,eq:totalCharge2}
into a linear
system of equations. To this end, we use a finite element
discretization of the unknown function $\sigma(x)$ on the surface
meshes as generated in \cref{sec:HOmeshing}. 
Due to the specific choice of a single layer potential,
the discretization has to be only $L_2$-conforming~\cite{Amann2014}
which allows for inter-element discontinuities. We use $\BEMFEOrder$-th order
shape functions, $\BEMFEOrder \geq 0$, which are defined on each element. For
$\BEMFEOrder = 0$ this would be element-wise constant functions, for $\BEMFEOrder = 1$
(bi-)linear functions and so forth. Therefore each element has either
$(\BEMFEOrder+1)^2$ (quadrilateral) or $(\BEMFEOrder+1)(\BEMFEOrder+2)/2$ (triangle) degrees
of freedom~\cite{steinbach2007numerical}. In total, the mesh has
$N_\BEMFEOrder$ degrees of freedom and we can write
\begin{equation}
  \label{eq:femapprox}
  \sigma(\mathbf{x}) \approx \sum_{i=1}^{N_\BEMFEOrder} \sigma_i \varphi^\BEMFEOrder_i(\mathbf{x})\,.
\end{equation}
It is also assumed that the used shape functions form a partition of
unity, that is they all sum up to one. This assumption is not
mandatory, but simplifies the charge
condition~\labelcref{eq:totalCharge2}.  In~\cite{Heuer2013,Heuer2016}, it
is shown how to discretise the hypersingular BEM operator in a
similar, non-conforming context which is a more subtle task, because
that operator requires continuous shape functions.

In order to obtain linear equations, the residuals
of~\labelcref{eq:dirichlet} and~\labelcref{eq:transmission4} are weighted with
$\varphi^\BEMFEOrder_j(x)$ and integrated over the surface. As the order of
the shape functions is globally fixed, we will omit the superscript
$\BEMFEOrder$ in the following. Let $\mathbb{I}(\BEMRegionIndex)$ be the set of all indices
$i$ such that $\mathrm{supp}(\varphi_i) \in \BEMRegion_\BEMRegionIndex$, that is the
shape functions whose support is in $\BEMRegion_\BEMRegionIndex$.  The
discretization~\labelcref{eq:femapprox} gives rise to the following system
matrices
\begin{equation}
  \label{eq:sysmats}
  \begin{aligned}
    \matr{V}_{{\BEMRegionIndex}{\BEMRegionIndexB}}[i,j] &= \int_{\BEMRegion_\BEMRegionIndex} \varphi_i(\mathbf{x}) \left(\mathcal{V}_{{\BEMRegionIndex}{\BEMRegionIndexB}}  \varphi_j \right)(\mathbf{x}) \dd s
    = \int_{\BEMRegion_\BEMRegionIndex} \int_{\BEMRegion_\BEMRegionIndexB} \varphi_i(\mathbf{x}) U(\mathbf{x},\mathbf{y}) \varphi_j(\mathbf{y}) \dd s_y \dd s_x\\
    \matr{K}_{{\BEMRegionIndex}{\BEMRegionIndexB}}[i,j] &= \int_{\BEMRegion_\BEMRegionIndex} \varphi_i(\mathbf{x}) \left(\mathcal{K}^\prime_{{\BEMRegionIndex}{\BEMRegionIndexB}}  \varphi_j \right)(\mathbf{x}) \dd s
                       = \int_{\BEMRegion_\BEMRegionIndex} \int_{\BEMRegion_\BEMRegionIndexB} \varphi_i(\mathbf{x}) \grad_x [U(\mathbf{x},\mathbf{y})\cdot \mathbf{n}_\BEMRegionIndex(\mathbf{x})] \varphi_j(\mathbf{y)} \dd s_y \dd s_x
    \\
    \matr{M}_\BEMRegionIndex[i,j] &= \int_{\BEMRegion_\BEMRegionIndex} \varphi_i(\mathbf{x}) \varphi_j(\mathbf{x}) \dd s
  \end{aligned}
\end{equation}
In addition, we introduce the abbreviation
$\tilde{\matr{K}}_{{\BEMRegionIndex}{\BEMRegionIndexB}} = -\delta_{{\BEMRegionIndex}{\BEMRegionIndexB}} \lambda_\BEMRegionIndex \matr{M}_{{\BEMRegionIndex}{\BEMRegionIndexB}} + \matr{K}_{{\BEMRegionIndex}{\BEMRegionIndexB}}$ with
the definition of $\lambda_\BEMRegionIndex$ as in~\labelcref{eq:lambda}.  Furthermore,
we use the following vectors
\begin{equation}
  \label{eq:sysvecs}
  \begin{aligned}
    \matr{h}_n[i] &= \sum_{\BEMRegionIndex\in \mathbb{\BEMRegionNum}(n)} \int_{\BEMRegion_\BEMRegionIndex} \varphi_i(\mathbf{x}) \dd s
                    \,, \quad
                    \matr{g}_n = g_n \matr{h}_n\\
    \matr{f}_n[i] &= \sum_{\BEMRegionIndex\in \mathbb{\BEMRegionNum}(n)} \sum_{j\in \mathbb{I}(\BEMRegionIndex)} \varepsilon_{r,\mathrm{opp}_n(\BEMRegionIndex)} \sum_{\BEMRegionIndexB=1}^\BEMRegionNum \left[ - \frac{\delta_{{\BEMRegionIndex}{\BEMRegionIndexB}}}{2} \matr{M}_\BEMRegionIndex[j,i] + \matr{K}_{{\BEMRegionIndex}{\BEMRegionIndexB}}[j,i] \right]\,,
  \end{aligned}
\end{equation}
where we need $\matr{h}_n$ for each electrode ($n \in E$) and floating
potential ($n \in F$), and $\matr{f}_n$ for each floating potential
($n \in F$). Grouping all block matrices and vectors according to the
three classes $E$, $F$ and $D$ yields the system of equations
\begin{equation}
  \label{eq:syseq}
  \begin{bmatrix}
    \tilde{\matr{K}}_{DD} &\matr{K}_{DE} &\matr{K}_{DF} &\matr{0}\\
    \matr{V}_{ED} &\matr{V}_{EE} &\matr{V}_{EF} &\matr{0} \\
    \matr{V}_{FD} &\matr{V}_{FE} &\matr{V}_{FF} &-\matr{H}_F \\
    \matr{F}_{D} &\matr{F}_{E} &\matr{F}_F &\matr{0}
  \end{bmatrix}
  \begin{bmatrix}
    \matrg{\sigma}_D \\ \matrg{\sigma}_E \\ \matrg{\sigma}_F \\ \matrg{\alpha}
  \end{bmatrix} =
  \begin{bmatrix}
    \matr{0} \\ \matr{g}_E \\ \matr{0} \\ \matr{0}
  \end{bmatrix}\,.
\end{equation}
This system has the dimension of $N = N_\BEMFEOrder+N_f$,
with $N_f$ denoting the number of floating potentials, and  is fully populated.

\subsection{Quadrature} \label{sec:quad}

\begin{figure}
  \centering 
  \centering
  \def\tkzscale{0.675}
  \tikzsetnextfilename{quadrature}
  \input{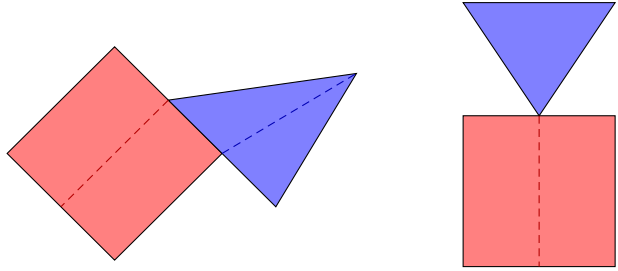}
  \caption{%
    Two additional cases due to non-conforming meshes: overlapping
    edges (left) and vertex-to-edge (right).}
  \label{fig:quad}

\end{figure}

Let $\matr{A}[i,j]$ denote a generic matrix coefficient from the above
system of equations~\labelcref{eq:syseq} representing one of the two
involved integral operators. The support of the shape functions
$\varphi_i$ and $\varphi_j$ are the elements $\tau_{i}$ and
$\tau_{j}$, respectively. For simplicity, we assume here that $\BEMFEOrder=0$
which implies that these elements are only equal for $i=j$. This is
not necessarily the case of $\BEMFEOrder >0$, but does not change the
following observations. These two elements are in a
specific geometric relation to each other that can be classified by
considering their intersection
\begin{equation}
  \label{eq:isec}
  \eta_{ij} = \overline{\tau}_i \cap \overline{\tau}_j\,.
\end{equation}
We either have that
$\eta_{ij} = \overline{\tau}_i = \overline{\tau}_j$, (identical
elements) that $\eta_{ij}$ is a line (overlapping edges), that
$\eta_{ij}$ is a point (vertex-to-vertex or vertex-to-edge), or that
$\eta_{ij}$ is empty (regular case). The first three cases (non-empty
$\eta_{ij}$) are the so-called singular cases in which the integral
operators' kernel function can be become arbitrarily large as the
points $\mathbf{x}$ and $\mathbf{y}$ get close to each other. In order
to tackle this problem, in~\cite{sauter2010} quadrature rules have
been devised that address these situations and allow to compute the
above integrals with high accuracy. In the context of this work, at
non-conforming interfaces between patches the adjacent elements are
subdivided such that those quadrature rules can be re-used.  As an
example, consider the situations depicted in \cref{fig:quad}. In the left
image, edges overlap and both elements are subdivided such that we have
now $2\times 2$ quadrature interactions with one case of
edge-adjacency, two cases of vertex-adjacency and one regular
case. Similarly, the situation in the right image can be converted to
two cases of vertex-adjacency by proper subdivision of the red
element. Note that this subdivision is only for quadrature purpose and
it is thus robust with respect very small sub-elements.

In addition to the intricacies of the singular integrals, the regular
case also requires some care. Here, it can happen that two elements
can become arbitrarily close to each other without ever falling into
the on of singular cases. In this situation, we subdivided each
quadrature element such that the ratio of relative distance to element
size becomes larger than some prescribed tolerance.

\subsection{Fast multipole method} \label{sec:fmm}

As pointed out above, the system matrix in~\eqref{eq:syseq} is fully
populated. Denoting by $N$ its dimension, it thus has a setup and
storage complexity of $\mathcal{O}(N^2)$ and a direct solution of that
system would require $\mathcal{O}(N^3)$ operations. This poses a
severe limit on the applicability of the method to problems relevant
in engineering practice. An established remedy to this problem is the
use of the fast multipole method (FMM) introduced
by~\cite{greengard1997new}.

\begin{figure}
  \centering
  \includegraphics[width=0.6\textwidth]{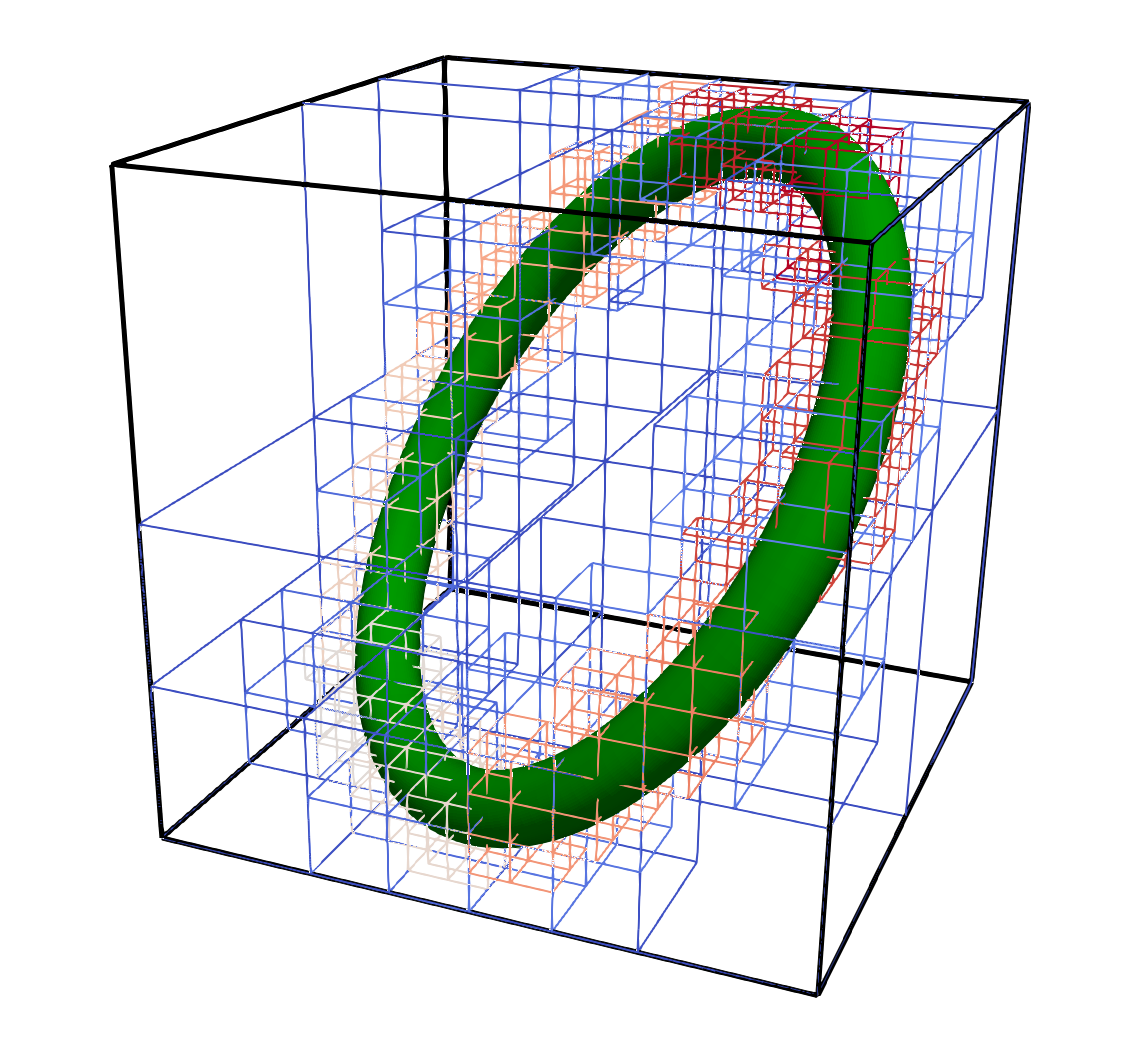}
  \caption{Octree cluster for a torus.}
  \label{fig:torus}
\end{figure}

The method begins by creating an adaptive octree for the geometry of
the surface mesh in which each octand is subdivided until it contains
less than a prescribed number of elements. Such a tree is depicted in
\cref{fig:torus}. The octants that are not subdivided are
called \emph{leaves} of the tree. The logic of the method is that each
leaf has a nearfield which contains the leaf itself and all directly
adjacent ones. The matrix entries for this nearfield are computed
directly and stored in a sparse matrix. In a matrix vector product,
the matrix blocks that correspond to interactions between leaves that
are not part of the nearfield are approximated. In order to do so, a
series expansion of the kernel function
$\tfrac{1}{|\mathbf{x}-\mathbf{y}|}$ is used which is based on
spherical harmonics. In addition to this series expansion, multi-level
transfer operators are designed such that the data transfer from in-
to output vector in a matrix-vector product can be carried out with
almost linear complexity.

Obviously, the use of this matrix approximation does not allow for a
direct solution of the system and therefore an iterative solver is
used. Due to the lack of symmetry of system~\labelcref{eq:syseq}, we use
the GMRES method. Since, the matrices stemming from integral operators
are typically well-conditioned it was sufficient for this work to use
a Jacobi preconditioner for accelerating the convergence of the
solution process.

\subsection{Post-processing} \label{sec:postproc}

Once system~\labelcref{eq:syseq} has been solved, the unknown density
function $\sigma$ has been determined on $\Gamma$, but it typically
does not have a physical character. In order to compute meaningful
physical data, one has to employ further computational steps. The
potential $u(\mathbf{x})$ or the electric field
$\mathbf{E}(\mathbf{x}) = - \grad u(\mathbf{x})$ can be evaluated at
any point $\mathbf{x} \notin \Gamma$ by evaluating the single layer
potential~\labelcref{eq:slp} or its gradient. Note that if the evaluation
point $\mathbf{x}$ gets close to $\Gamma$ special quadrature
techniques are needed in order to obtain good accuracy of the result.

Often, one is interested in the surface data, the potential
$\gamma_D u$ or the surface charge density
$\mathbf{D}^\pm \cdot \mathbf{n} = \varepsilon \gamma^\pm_N u$. In order to
compute these data, we approximate the needed traces in the same space
as $\sigma$, see~\eqref{eq:femapprox}, and express the weighted
residuals of the traces~\labelcref{eq:bio}. This leads to the systems of
equations
\begin{equation}
  \label{eq:syseqpost}
  \begin{aligned}
    \matr{M}_\BEMRegionIndex \matr{u}_\BEMRegionIndex &= \sum_{\BEMRegionIndexB=1}^\BEMRegionNum \matr{V}_{{\BEMRegionIndex}{\BEMRegionIndexB}}
                            \matr{\sigma}_\BEMRegionIndexB\\
    \matr{M}_\BEMRegionIndex \matr{q}^\pm_\BEMRegionIndex &= \sum_{\BEMRegionIndexB=1}^\BEMRegionNum \left[ \pm \frac{\delta_{{\BEMRegionIndex}{\BEMRegionIndexB}}}{2}
                                \matr{M}_\BEMRegionIndex + \matr{K}_{{\BEMRegionIndex}{\BEMRegionIndexB}}\right] \matr{\sigma}_\BEMRegionIndexB
  \end{aligned}
\end{equation}
which can be solved element-wise. The matrices $\matr{M}_\BEMRegionIndex$ are
particularly simple to invert as they are block-diagonal where each
block has the size of degrees of freedom per element. The resulting
vectors $\matr{u}_\BEMRegionIndex$ and $\matr{q}_\BEMRegionIndex^\pm$ are the coefficients of the
approximate traces $\gamma_D u$ and $\gamma_N^\pm u$,
respectively. Note that the latter has different interior and exterior
traces. Once the Neumann trace is known,
\cref{eq:totalCharge} allows to compute the total charge of
a domain.

\ifthenelse{\boolean{showSubfileBibliography}}{
    \bibliography{VEGA}%
}{}

%% file: sectionCAD.tex
The proposed CAD enhancements provides the necessary connectivity data for analysis.
Hence, we recap the related difficulty when dealing with trimmed surfaces before discussing the implemented measures to facilitate an analysis-suitable representation.

\subsection{Trimmed CAD models}

Let us recall some preliminaries of trimmed CAD models.
In general, they are b-reps consisting of a collection of surface patches.
The predominant surface representation is a tensor product B-spline patch given by
\begin{align}
    \label{eq:BsplinePatch}
    \surface\operate{\uusurf,\vvsurf} =
    \sum_{\indexA} \sum_{\indexB}
    \Bspline_{\indexA,\pusurf} \operate{\uusurf}  \Bspline_{\indexB,\pvsurf}\operate{\vvsurf}    \: \CP_{\indexA,\indexB}
\end{align}
where $\Bspline_{\indexA,\pusurf}$ and $\Bspline_{\indexB,\pvsurf}$ are univariate B-splines of degrees $\pusurf$ and $\pvsurf$, specified by separate knot vectors for the parametric directions $\uusurf$ and $\vvsurf$, respectively.
The tensor product of these univariate B-splines forms the basis of the patch, and the final geometry is determined by the associated control points $\CP_{\indexA,\indexB}$ in model space $\R^3$.
By introducing weighted control points $\CP^w_{\indexA,\indexB} \in \R^4$, B-splines can be generalized to NURBS (non-uniform rational B-splines), which allows the exact definition of conic sections and quadric surfaces.
In either case, representation~\labelcref{eq:BsplinePatch} provides surfaces with an intrinsic four-sided topology due to its tensor product structure.

Trimmed patches circumvent this restriction by defining a visible area~$\visibledomain$ over a surface $\surface\operate{\uusurf,\vvsurf}$ of arbitrary shape.
Therefore, so-called \emph{trimming curves} $\TrimCurve\operate{\uutrim}$ are introduced in the parametric space of $\surface\operate{\uusurf,\vvsurf}$.
They often result from surface-to-surface intersections (SSI).
\begin{figure}
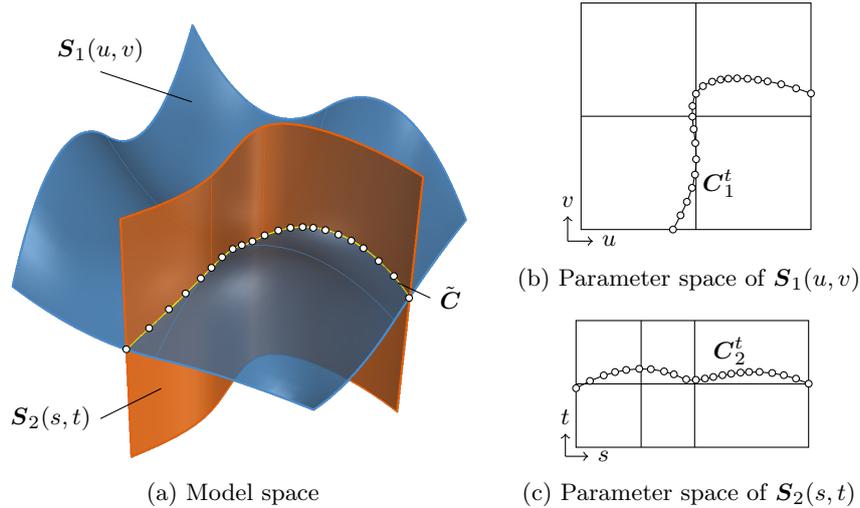

  \centering
  \begin{minipage}[b]{0.45\textwidth}
	
    \def\tkzscale{0.20}
    \centering
    \tikzsetnextfilename{intersectionCurveExample}
    \input{\TikzPath/intersectionCurveExample}
    \subcaption{
            Model space
        }
    \label{fig:intersectionCurveExampleModelSpace}

  \end{minipage}
  \begin{minipage}[b]{0.45\textwidth}
	
    \def\tkzscale{0.11}
    \centering
    \tikzsetnextfilename{intersectionCurveExampleParameter}
    \input{\TikzPath/intersectionCurveExampleParameter}
    \subcaption{
            Parameter space of $\surface_1\operate{\uusurf,\vvsurf}$
        }
    \label{fig:intersectionCurveExampleParameterA}

    \def\tkzscale{0.1}
    \centering
    \tikzsetnextfilename{intersectionCurveExampleParameterB}
    \input{\TikzPath/intersectionCurveExampleParameterB}
    \subcaption{
            Parameter space of $\surface_2\operate{\uusurfB,\vvsurfB}$
        }
    \label{fig:intersectionCurveExampleParameterB}

  \end{minipage}
  \caption{Curve interpolation of an ordered point set (indicated by white circles) to obtain the independent approximations $\IntersectionCurve$, $\TrimCurve_1$, and $\TrimCurve_2$ of the intersection between the two surface patches $\surface_1\operate{\uusurf,\vvsurf}$ and $\surface_2\operate{\uusurfB,\vvsurfB}$.}
  \label{fig:intersectionCurveExample}
\end{figure}
As indicated in \cref{fig:intersectionCurveExample}, the SSI algorithm provides an ordered point set along the intersection of two patches $\surface_1\operate{\uusurf,\vvsurf}$ and $\surface_2\operate{\uusurfB,\vvsurfB}$, which is projected into the surfaces' parameter spaces.
The subsequent curve interpolations of these points is performed in (a) the model space and the parameter space of (b) $\surface_1\operate{\uusurf,\vvsurf}$ and (c) $\surface_2\operate{\uusurfB,\vvsurfB}$ leading to the curves $\IntersectionCurve$, $\TrimCurve_1$, and $\TrimCurve_2$, respectively.
Usually, these curves are given as B-spline curves and the point data is usually discarded once they are constructed.
Note that there is no mathematical relation between the curves and they do \emph{not} define the same object \cite{Farin2002,Corney2001}.
In case of solid models, they are associated to the same topological entity, i.e., the models edge.
However, this topology data is not apparent to non-expert users as discussed in \cref{sec:ChallengesDesign} (cf.~\cref{fig:BoxCylinderBooleanSpline}).
Hence, we aim to enhance CAD models by establishing an connection between trimming curves of intersecting surfaces that allows applying our BEM formulation introduced in \cref{sec:BEM} to them.

\subsection{Trimmed CAD model enhancement}

We assume that the CAD object providing the geometric information for the engineering analysis consists of one or several surface models.
Thus, we must
(i)
detect the connection between surface edges (trimming curves and patch boundaries),
(ii)
capture the `engineering user intent',
and (iii)
enable non-manifold edges
to obtain an analysis-suitable model representation.

\paragraph{Connection between surface edges}
If a parametric value, $\vvsurf^{iso}$, is fixed, the surface patch \labelcref{eq:BsplinePatch} results in an \emph{isocurve} in the $\uusurf$-direction, and vice versa.
By setting up isocurves for each knot value of a patch, we obtain a grid that reflects where continuity changes can occur.
When establishing the connectivity between surface edges, we transfer these essential geometric features by
computing vertices, $\pt{v}_i$, that mark the closest points of a surface's isocurves 
to the other surface's edge. 
To be precise, we store these $\pt{v}_i$ only if the distance to the projection to another edge is below a user-defined \emph{join tolerance}, which is greater or equal to the model tolerance of the SSI operations used for setting up the original CAD object.
\begin{figure}
    
  \centering
  \def\tkzscale{1.0}
  \tikzsetnextfilename{VEGASurfaceEnhancementTrimPointData.tex}
  \input{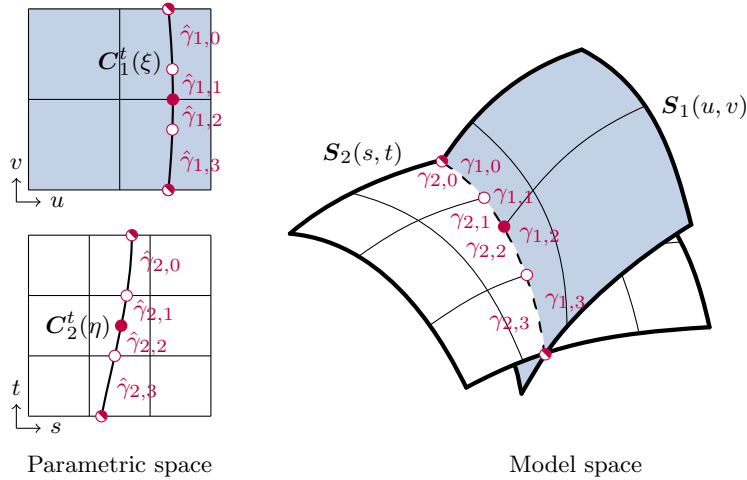}
  \caption{
        Enhanced edge data (displayed in purple) for conventional trimming curves $\TrimCurve_i$ of two intersecting surfaces $\surface_i$.
        The dots along the intersection illustrate the corresponding point data in the model and parametric spaces; purple and white color indicate that they originate from $\surface_1$ and $\surface_2$, respectively.
        They introduce segments $\TrimCurveSubRegion_{i,k}$ that project relevant features of each surface.
     }
  \label{fig:ConnectingCurvesEnhanced}

\end{figure}
\cref{fig:ConnectingCurvesEnhanced} considers the case where two trimming curves $\TrimCurve_i$ 
define the surfaces' shared edge.
Note that the other surface edges, i.e., the patch boundaries, are neglected since at least one distance of the image of successive $\pt{v}_i$ and their counterparts projected onto the other surface's boundary do not meet the join tolerance.
The accumulation of these dots divides each $\TrimCurve_i$ into four segments $\TrimCurveSubRegion_{i,k}$, $k=1,\dots,4$. Adjacent segments still define different curves, and their images in model space $\TrimCurveSubRegionMapped_{i,k}$ and $\TrimCurveSubRegionMapped_{j,k}$ do not coincide; however, their ends mark critical points on one of the surfaces involved.
We denote these $\TrimCurveSubRegion_{i,k}$ as \emph{enhanced edges} since they enhance each surface edge (trimming curves and patch boundaries) with information about their neighbors $\TrimCurveSubRegion_{j,k}$ and provide a connection between each other.
To be precise, 
each $\TrimCurveSubRegion_{i,k}$ contains the surface edge it is part of, the related parametric domain of this edge, and the adjacent surface.

Robustness is of utmost importance for the usability of the enhancement process.
Conceptually, the enhanced edges are generated within the CAD tool (Rhino 7) using its inherent routines, such as the provided closest point projection.
Thus, the CAD model data are enhanced by utilizing the validation checks of the design tool and do not require any translation.
Nevertheless, correctly identifying neighboring edges, including the correct start and end points of enhanced edge pairs, can become challenging, particularly if an edge's curvature is high.
We mitigate the risk of incorrect detection by projecting points  $\pt{v}_i$ back to their parent surface, which may start an iterative process if the re-projected points do not coincide with the original ones.
In addition, an edge $\TrimCurveSubRegion$ may be bisected adaptively to resolve its complexity.

Finally, we note that setting up these enhanced edges $\TrimCurveSubRegion$ can also be an initial step in deriving watertight spline models as detailed in \cite{Urick2019,Marussig2019}.
However, using analysis-suitable splines is not the direction of this paper, as we aim to utilize the high-performance BEM solver tailored to polygonal meshes.

\paragraph{Capturing user intent}

Since enhanced edges not only enrich the geometric model data but also provide information on which surfaces and edges are relevant for the simulation, they represent the first step towards cross-disciplinary collaboration between design and analysis. Hence, it is vital to capture the user's intent during this stage of virtual prototyping, which we support by the following measures.

\begin{figure}
    \centering
    \begin{subfigure}[b]{0.48\textwidth}
        \myfbox{\includegraphics[trim={7.5cm 0.0cm 9cm 3cm},clip,width=0.95\textwidth]{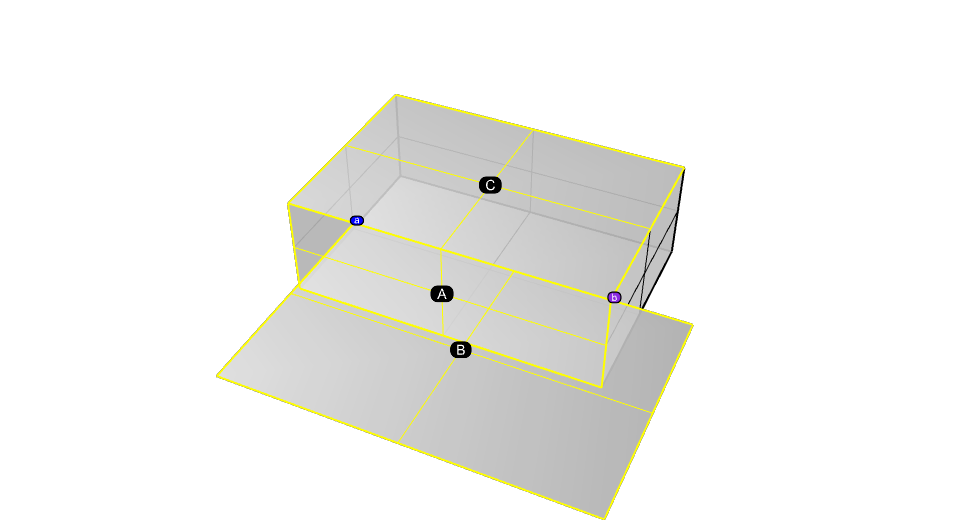}}
        \subcaption{Non-manifold edge}
        \label{fig:EnhancedCADUINonManifold}
    \end{subfigure}
    \begin{subfigure}[b]{0.48\textwidth}
        \myfbox{\includegraphics[trim={7.5cm 0.0cm 9cm 3cm},clip,width=0.95\textwidth]{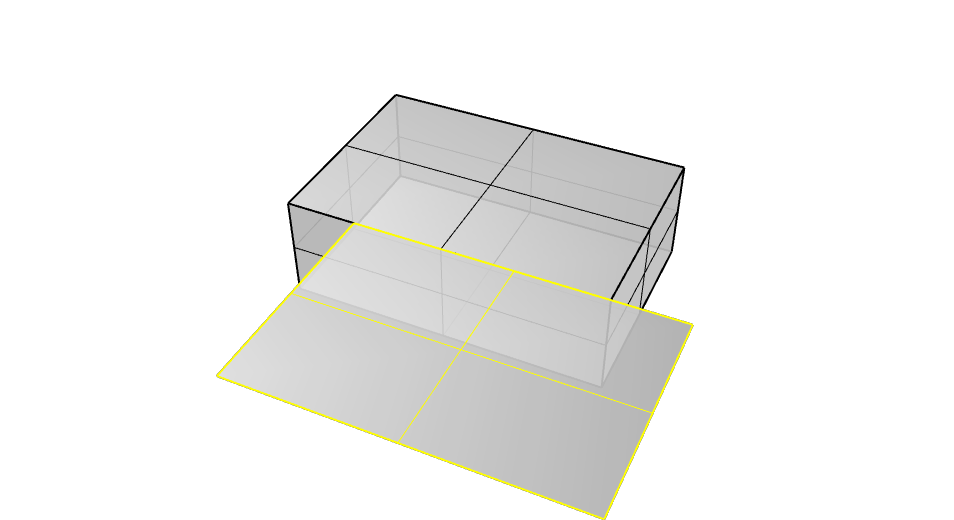}}
        \subcaption{Isolated surface}
        \label{fig:EnhancedCADUIIsolatedSurfaces}
    \end{subfigure}
    \caption{User interaction to clarify ambiguous cases: (a) detection of a non-manifold edge $a$-$b$ potentially connecting the surfaces $A$, $B$, and $C$; (b) if the user decides that $A$ and $C$ are relevant for $a$-$b$, $B$ has no connection to another surface, and thus, it is suggested to be removed from the analysis group.}
    \label{fig:EnhancedCADUI}
\end{figure}

First, we introduce \emph{analysis geometry groups}. They are snapshots of the CAD model parts considered connected during analysis. After establishing the corresponding enhanced edges, the related surfaces are grouped into a single object. Similar to solid b-reps in CAD, these groups are uneditable to prevent geometry manipulations, which would destroy the connectivity data. Moreover, a group may be stored as a copy of the original surfaces when a user wants to continue editing the CAD geometry. In this case, the group serves as a snapshot of the design stage.
Second, when setting up an analysis group, the user classifies it as either \emph{open} or \emph{closed}. For the latter, it is straightforward to validate whether the obtained enhanced edges reflect the user intent since each edge must connect exactly two surfaces.
Consequently, a closed group encloses a bounded domain. The normal vectors of each surface are labeled to indicate whether they point outside or inside this domain.
Open groups, on the other hand, are more general since they contain naked edges, i.e., ones not connected to another surface. Thus, capturing the user intent is more challenging, and ambiguous cases may arise. These cases are highlighted to the user to provide a decision on how to proceed.
For instance, the user must ensure that a surface is part of a group if it has any connection to other surfaces.
Furthermore, we introduce a \emph{search tolerance} that is larger than the actual join tolerance to obtain a larger set of potential surface connections.
The search tolerance can be set based on the surface edge type so that it is larger for trimming curves than for patch boundaries to account for the higher geometric accuracy in the latter case.
In any case, if surface edges are within this search tolerance but are not within the join tolerance, the situation is considered ambiguous and displayed to the user.
Note that non-manifold edges, i.e., edges associated with more than two surfaces, are not permitted in open and closed groups. Such edges will also be highlighted to the user, as illustrated in \cref{fig:EnhancedCADUINonManifold}.
Suppose the edge $a$-$b$  in this example connects the surfaces labeled $A$ and $C$; surface $B$ is isolated from the others, which triggers the subsequent query for the user shown in \cref{fig:EnhancedCADUIIsolatedSurfaces}.

\paragraph{Non-manifold edges}

Considering any point on a manifold surface, there exists a neighborhood that is homeomorphic to the plane \cite{Hoffmann1989}.
In other words, we can locally map the surface into a plane without introducing holes or singular points.
The topology of a manifold behaves like the usual 2D plane, and the implementation of Boolean operations is well-understood for manifold polyhedral objects \cite{Maentylae1988}. In the case of freeform surfaces such as B-spline patches, the trimming challenges must be addressed, as discussed above.
Many CAD systems restrict themselves to manifold models due to their sound theoretical foundations, and they are usually sufficient for representing b-reps.
This is also the case for the software used for the proposed plugin.
For analysis purposes, however, a b-rep may need to be split into subdomains to assign different material parameters, for instance, which inevitably introduces non-manifold features in a geometric model.

\begin{figure}[h]
    \centering
    \begin{minipage}[b]{0.48\textwidth}
    	
    \def\tkzscale{0.55}
    \centering
    \tikzsetnextfilename{VEGANonManifoldEdgeSelection}
    \input{\TikzPath/VEGANonManifoldEdgeSelection}
    \subcaption{
            Edge selection
        }
    \label{fig:NonManifoldModelSelection}

    \end{minipage}
    \begin{minipage}[b]{0.48\textwidth}
        \myfbox{\includegraphics[trim={18.8cm 6.5cm 19.5cm 6.75cm},clip,scale=0.5]{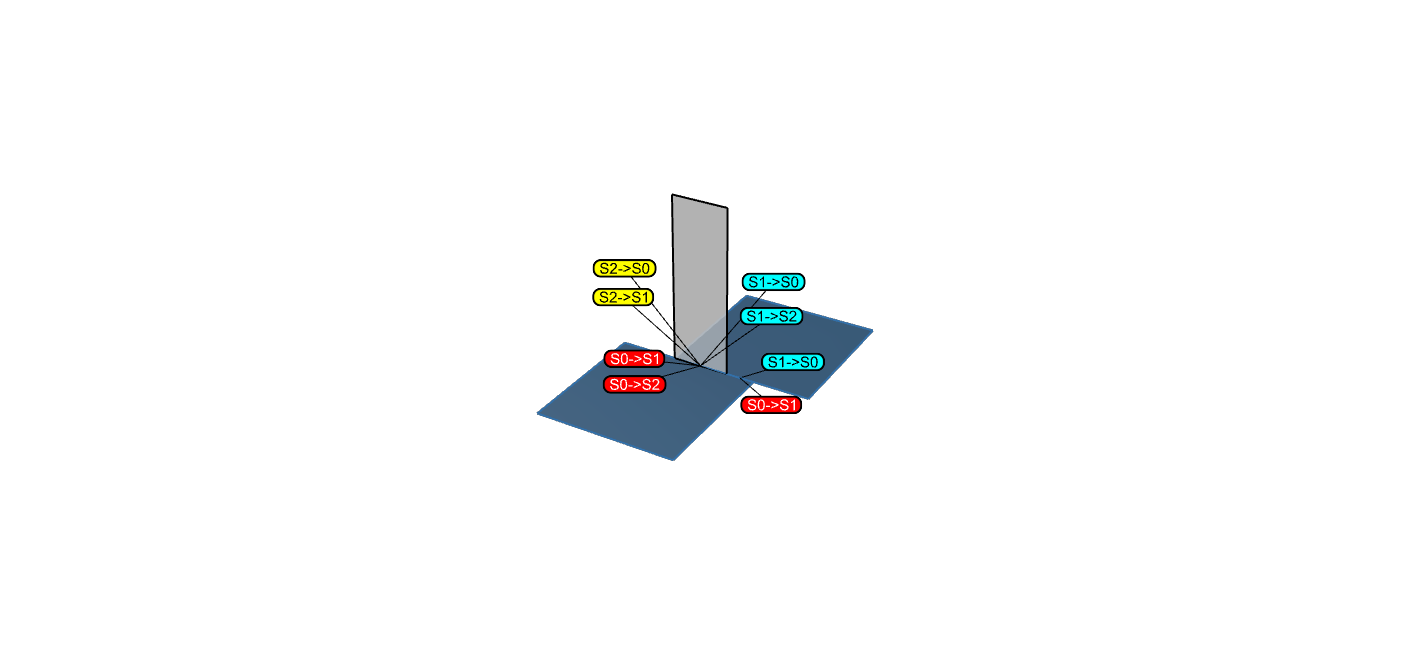}}
        \subcaption{Non-manifold enhanced edge data}
        \label{fig:NonManifoldModelEdges}
    \end{minipage}
    \caption{Establishing a non-manifold edge between three surfaces by joining two open manifold analysis groups indicated in blue and gray, respectively: (a) the user selects the non-manifold edge $D$ from $\surface_2$ to be connected to the group of $\surface_0$ and $\surface_1$; which leads to (b) the updated edge data. The data point labels denote the surfaces $i$ and $j$ of the associated edge, S$i{\rightarrow}Sj$; red, cyan, and yellow colors indicate that the data belongs to $\surface_0$, $\surface_1$, and $\surface_2$, respectively.} 
    \label{fig:NonManifoldModel}
\end{figure}

We address this problem by considering non-manifold objects as an interconnection of several manifold objects given as open or closed analysis geometry groups.
For this interconnection, we allow intersections of groups to sets of dimension 1 or 0.
As a result, groups can only join along surface edges.
The interconnection is reflected in updating the enhanced edge data by adding another adjacent surface.
To prevent any misinterpretation, we request the user to specify the non-manifold edges for the interconnection, as shown in \cref{fig:NonManifoldModelSelection}.
In addition, \cref{fig:NonManifoldModelEdges} highlights the resulting connectivity data between the three surfaces involved.
Note that updating enhanced edge data may require splitting a $\TrimCurveSubRegion$.
For the example shown, the blue analysis group consisted of two surface patches $\surface_i$ with a single $\TrimCurveSubRegion_{i,1}$ on each patch, $i=0,1$.
These $\TrimCurveSubRegion_{i,1}$ are divided into two due to the interconnection with the gray analysis group, as indicated by the data labels between $\surface_0$ and $\surface_1$ in \cref{fig:NonManifoldModelEdges}. Note that number of data labels increases for the non-manifold edge since all three surfaces are involved. 

\ifthenelse{\boolean{showSubfileBibliography}}{
    \bibliography{VEGA}%
}{}

%% file: sectionMesh.tex
The CAD model enhancement described in \cref{sec:CADenhancement} provides the connectivity data mandatory for correctly capturing the object's behavior in the simulation.
However, the geometric representation is not analysis-suitable in general.
Hence, we apply a meshing step to obtain the final geometric analysis model for the electrostatic simulations.
By leveraging the discontinuous BEM formulation introduced in \cref{sec:BEM}, 
the meshing effort is reduced to a minimum.
The meshes can be non-conforming since they inherit the connectivity data required for the BEM analysis from the enhanced CAD model data.
Furthermore, the order $\pu$ of the meshes can be increased from linear to higher orders to improve the accuracy of the geometric analysis model.
This improvement also reduces the gaps and overlaps between surfaces, yielding a more robust quadrature procedure along non-conforming edges.
Note that the order $\pu$ of the geometric discretization can differ from the order $\BEMFEOrder$ of the shape functions for the discretization of the BEM fields.

\subsection{Non-conforming surface meshes}

Since BEM relies on surface meshes and the utilized formulation allows for non-conforming edges between elements, the task boils down to generating a mesh for a single (trimmed) surface.
In the first step, we employ the CAD tool's internal mesh generators.
The default one generates linear triangles, which may be combined into quadrilaterals. The version used for the plugin (Rhino 7) also introduced the possibility of linear quad meshes with optimized topology.
In any case, the outcome of the first step is a collection of meshes that are conforming within their corresponding surface patch but non-conforming along the edges between adjacent surfaces.

The second step transforms the connectivity information from the enhanced CAD model edges $\TrimCurveSubRegion$ to the non-conforming mesh edges $\MeshEdge$.
Therefore, the vertex $\MeshVertex$ of a $\MeshEdge_\indexA$ is projected to the image of the closest enhanced patch edge $\TrimCurveSubRegionMapped_\indexB$, which yields the projected point  $\MeshVertex^p$ and its intrinsic coordinate within $\TrimCurveSubRegion_\indexB$. 
From $\TrimCurveSubRegion_\indexB$, we extract the connected  $\TrimCurveSubRegion_\indexC$ of all adjacent surfaces and project $\MeshVertex^p$ to the associated naked mesh edges $\MeshEdge_\indexC$.
This final projection provides the location of the vertex $\MeshVertex$ on the adjacent non-conforming edges $\TrimCurveSubRegion_\indexC$ which is required for the simulation.
\Cref{fig:EnhancedTrimmedCylinder} shows an example for the mesh generation:
\begin{figure}
    \centering
    \begin{subfigure}[b]{0.42\textwidth}    
        \myfbox{\includegraphics[trim={1.9cm 7.8cm 1.5cm 6.0cm},clip,scale=0.4]{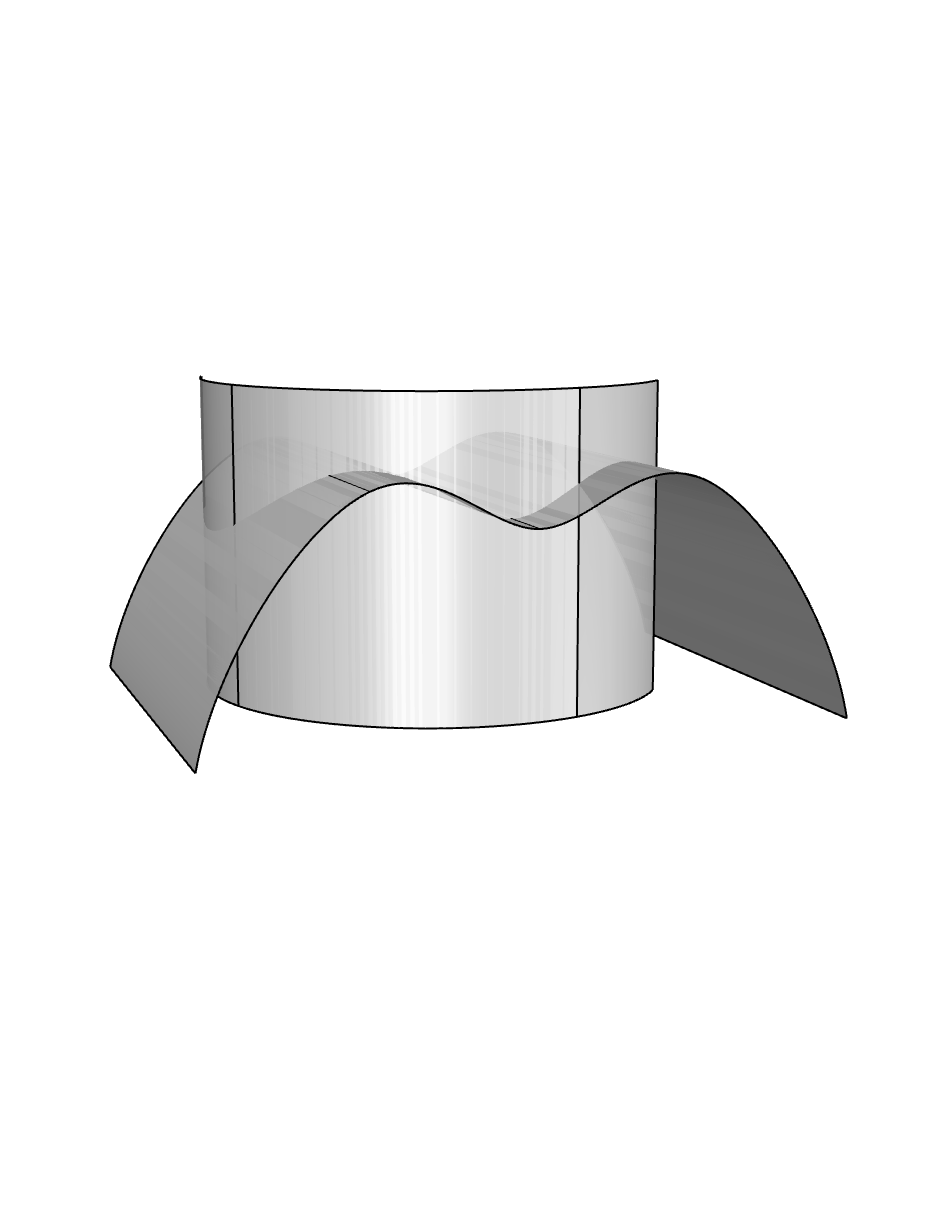}}
    \subcaption{Geometric data}    
    \end{subfigure}
    \begin{subfigure}[b]{0.28\textwidth}  
        \myfbox{\includegraphics[trim={3.0cm 7.8cm 5.0cm 6.0cm},clip,scale=0.4]{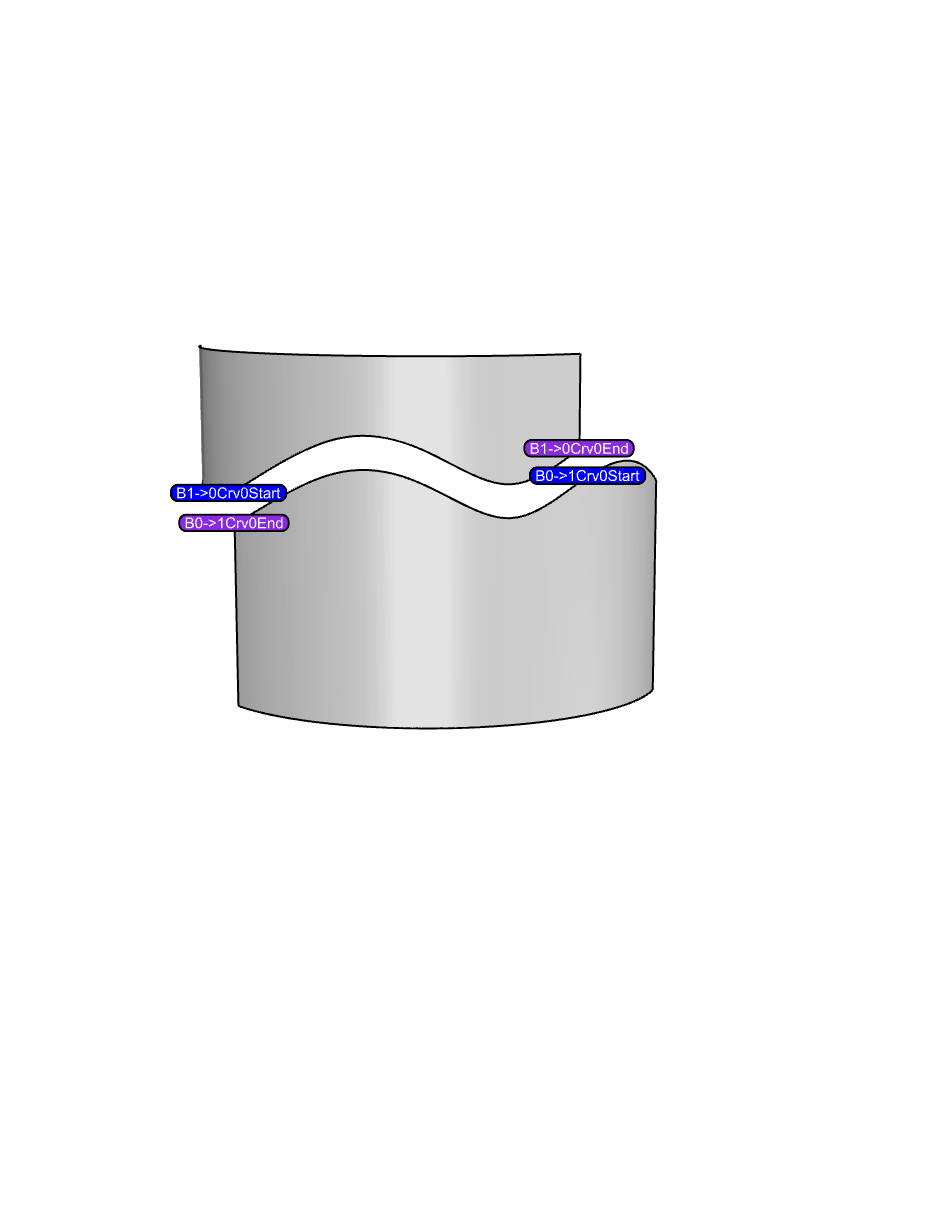}}
        \subcaption{Enhanced CAD model}
        \label{fig:EnhancedTrimmedCylinderEnhancedCADmodel}
    \end{subfigure}
    \begin{subfigure}[b]{0.28\textwidth}
        \myfbox{\includegraphics[trim={3.5cm 7.8cm 5.0cm 6.0cm},clip,scale=0.4]{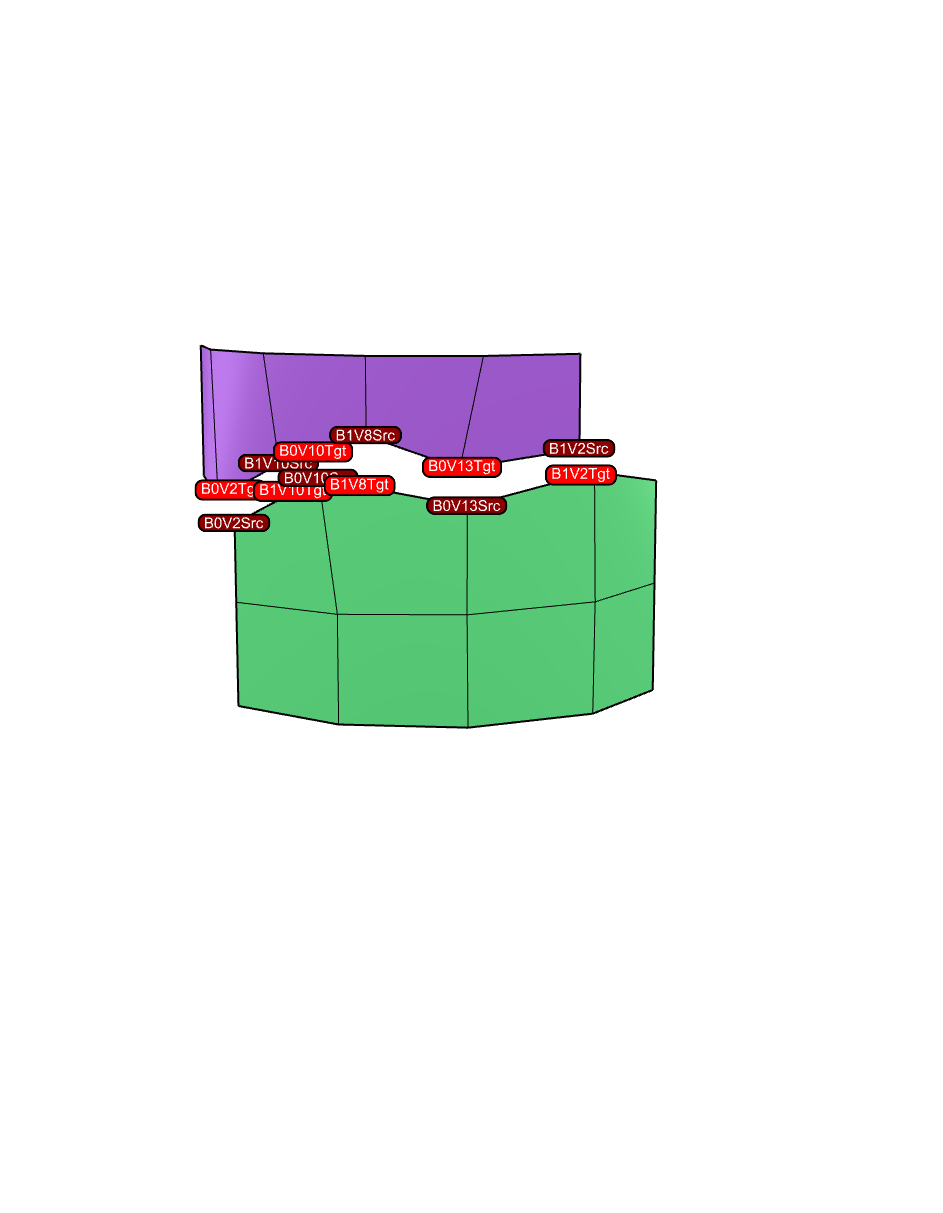}}
        \subcaption{Enhanced mesh}
    \end{subfigure}
    \caption{Enhanced meshes of an open analysis geometry group: (a) the underlying geometry of the CAD model, (b) the trimmed CAD model with the enhanced edge data, and (c) the related non-conforming mesh where naked edge vertices (dark red) have been projected to the adjacent edges (light red). In (b) and (c), the gap between the trimmed patches has been increased for better visibility. }
    \label{fig:EnhancedTrimmedCylinder}
\end{figure}
(a) The CAD model consists of two overlapping cylindrical surfaces trimmed by a third one; (b) the valid domains of the cylindrical surfaces are connected via their individual enhanced edge; (c) the information is inherited to the mesh edges.

\ifthenelse{\boolean{showSubfileBibliography}}{
    \bibliography{VEGA}%
}{}

%% file: sectionMeshHO.tex
\subsection{Order elevation of the mesh}

\begin{figure}
    \centering
    \begin{subfigure}[b]{0.3\textwidth}    
        \myfbox{\includegraphics[width=1.0\textwidth]{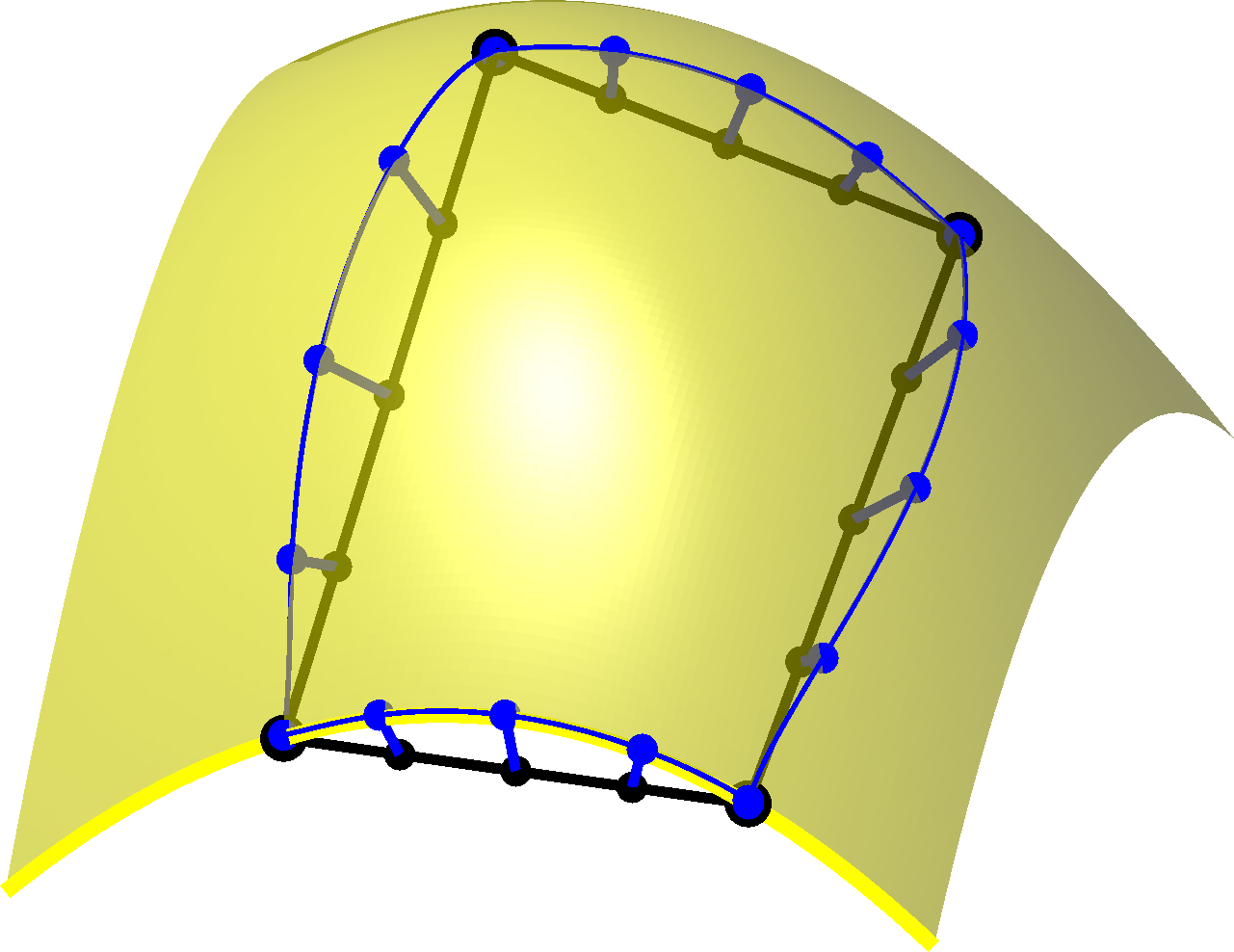}}        
        \subcaption{Proj.~of edge nodes}
        \label{fig:VisOrderElevation1} 
    \end{subfigure}
    \begin{subfigure}[b]{0.3\textwidth}  
        \myfbox{\includegraphics[width=1.0\textwidth]{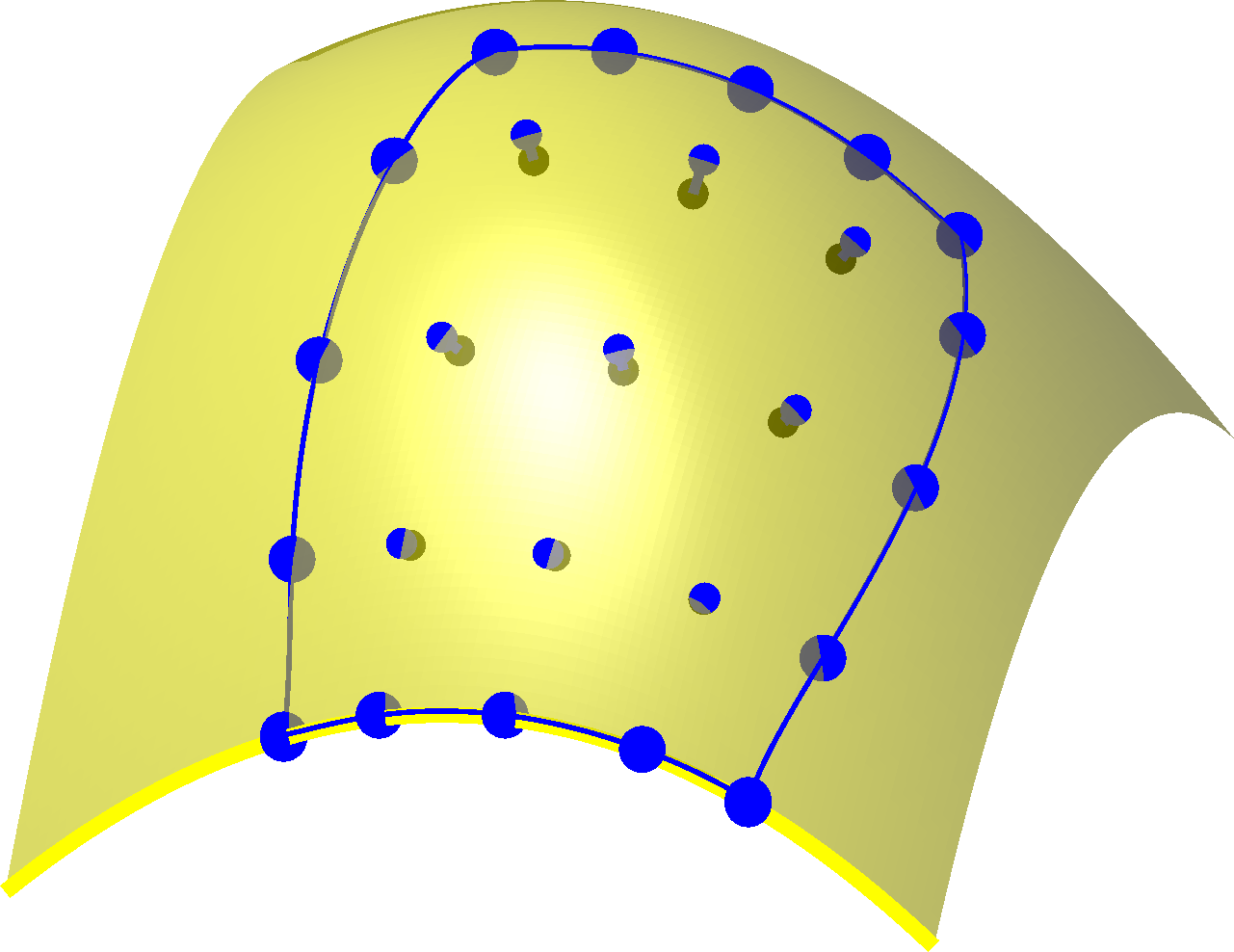}}
        \subcaption{Proj.~of inner nodes}
        \label{fig:VisOrderElevation2}
    \end{subfigure}
    \begin{subfigure}[b]{0.3\textwidth}  
        \myfbox{\includegraphics[width=1.0\textwidth]{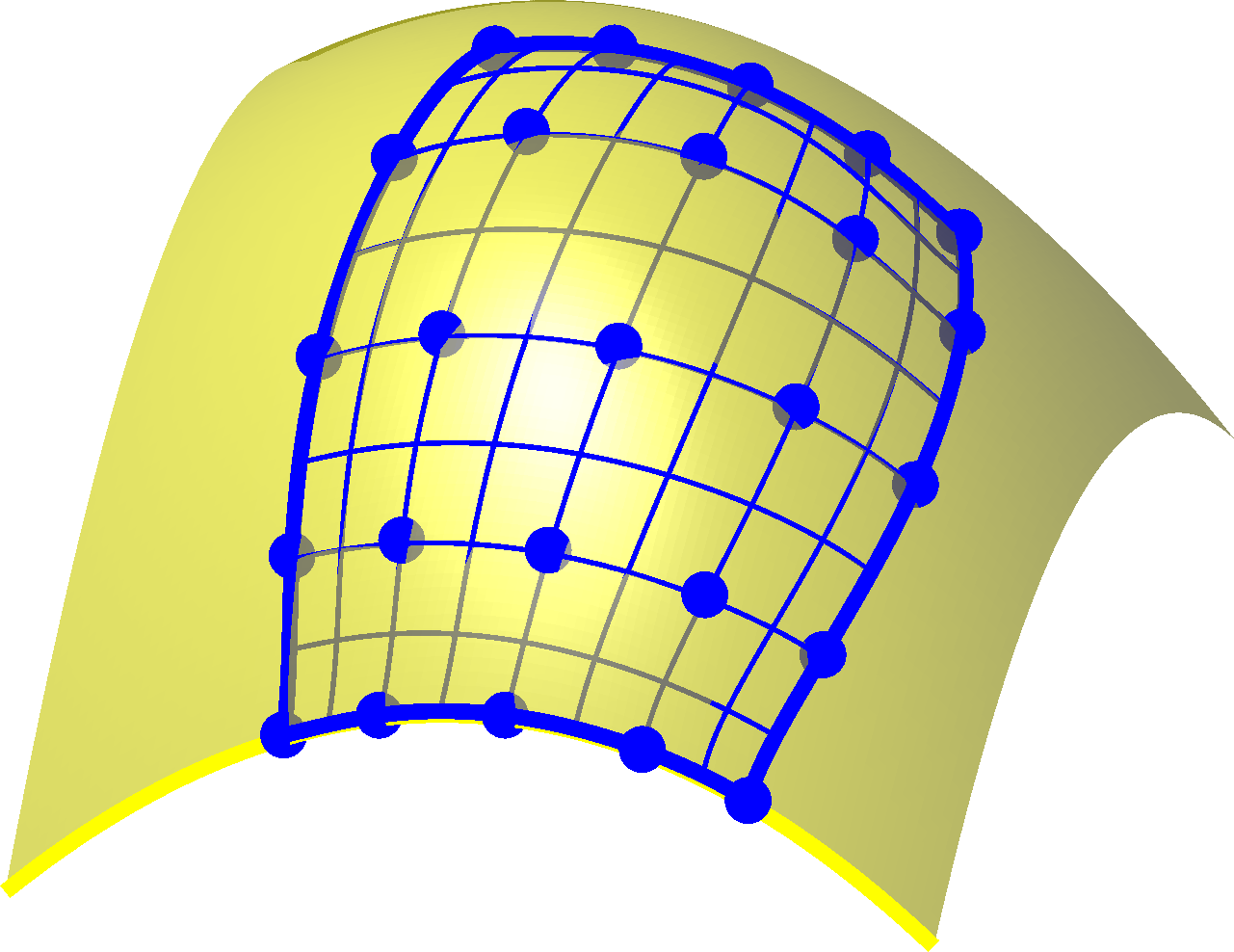}}
        \subcaption{Higher-order element}
        \label{fig:VisOrderElevation3}
    \end{subfigure}
    \caption{Detail of a patch shown as a yellow surface with boundary (model edge) as a yellow line. A linear quad element is shown via black contour lines in (a), one of the quad edges belongs to the model edge. The interpolated $\pu+1$ nodes on the edges are seen as small black dots in (a), they are projected onto the patch, resulting in the blue dots. Based on the edge nodes, start guesses for the inner nodes are generated based on transfinite maps and shown as black dots in (b). They result in the blue dots after the projection onto the surface. (c) shows the resulting higher-order element implied by the nodes seen as blue dots.}
    \label{fig:VisOrderElevation}
\end{figure}

The CAD system's mesh generators provide (bi-)linear meshes, which
are now improved by elevating the order. Higher-order meshes enable substantially more accurate geometry representations and achieve superior results when approximating boundary value problems. They enable the BEM employed herein to achieve higher-order convergence rates as shown later in \cref{sec:results}. For the order elevation, we follow previous works in \cite{Fries2015,Omerovic_2016a,Fries_2016b} and perform the following steps element by element:

Suppose we are starting from a linear quad element, $\pu+1$ nodes on the quad edges are placed by linear interpolation, when $\pu$ is the desired order of the quad element. For quad edges that correspond to model edges (i.e., boundary sections of the underlying patch), a closest-point projection of the nodes to the respective \emph{model edge} is made and the enhanced CAD data of the model edges is recovered as outlined above. For quad edges that are inside the patch, a closest-point projection of the edge nodes to the patch \emph{surface} is performed. The outcome of this step are nodes on the patch that represent the outer contour of the higher-order element to be generated next, see \cref{fig:VisOrderElevation1}.

\begin{figure}[h]
    \centering
    \begin{subfigure}[b]{0.48\textwidth} 
        \centering   
        \myfbox{\includegraphics[trim={8cm 7cm 10cm 8cm},clip,width=0.7\textwidth]{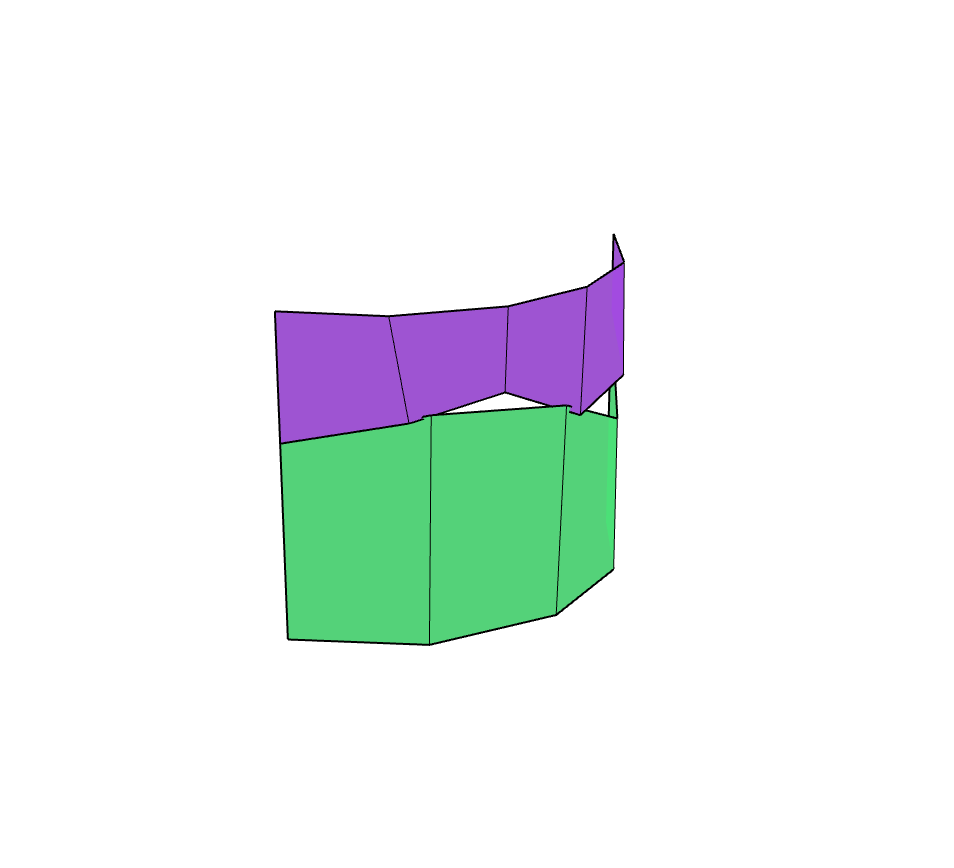}}        
        \subcaption{Linear non-conforming mesh}
    \end{subfigure}
    \begin{subfigure}[b]{0.48\textwidth}  
        \centering
        \myfbox{\includegraphics[trim={8cm 7cm 10cm 8cm},clip,width=0.7\textwidth]{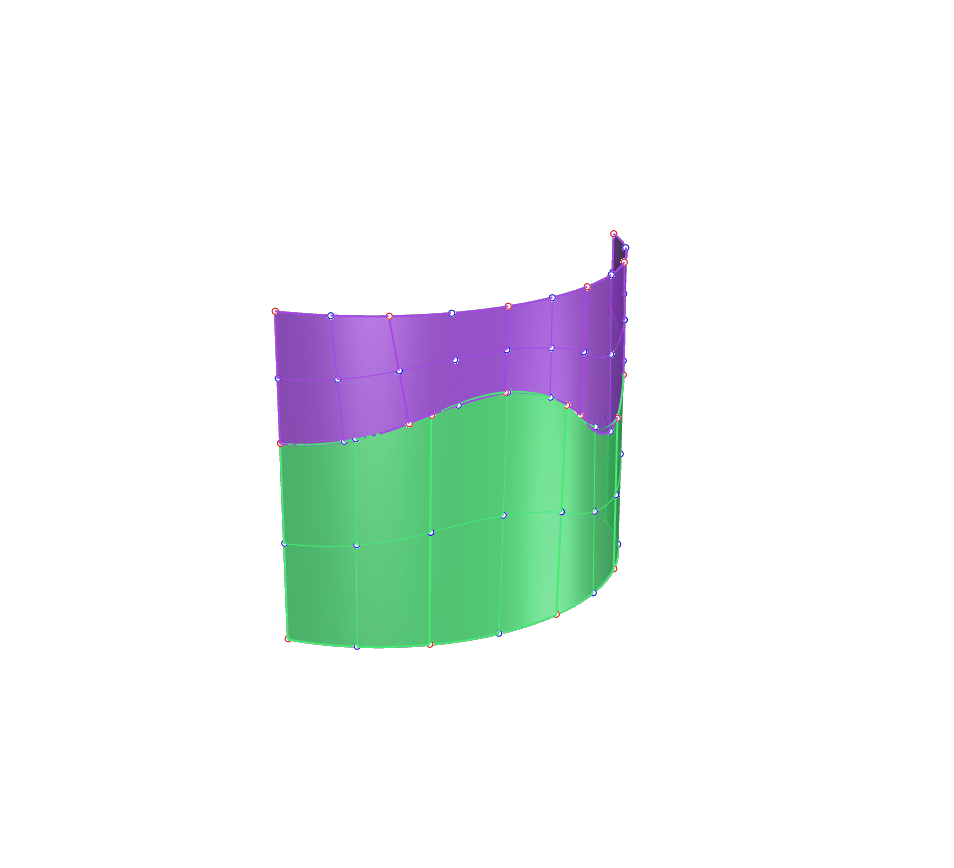}}
        \subcaption{Quadratic non-conforming mesh}
    \end{subfigure}
    \caption{A linear non-conforming mesh (a) and its order-elevated counterpart (b).}
    \label{fig:HOMesh}
\end{figure}

Based on the projected $p+1$ nodes on every quad edge, start values are generated for the inner nodes of the higher-order element based on transfinite maps \cite{Gordon_1973a,Gordon_1973b}. These inner nodes are projected by closest-point projections to the patch surface, see \cref{fig:VisOrderElevation2}. Note that although all nodes of the generated higher-order element are exactly on the patch (up to rounding-off accuracy, $\leq10^{-12}$) as seen in \cref{fig:VisOrderElevation3}, the element is still only an approximation of the patch, yet with higher-order accuracy. Although the procedure is outlined for elevating to any element order $p$, we shall later restrict ourselves to quadratic orders of the quads, $p=2$, because already this moderate order elevation shows substantial improvements in the geometry representation, see \cref{fig:HOMesh}, and approximation properties, see \cref{sec:results}.

\ifthenelse{\boolean{showSubfileBibliography}}{
    \bibliography{VEGA}%
}{}

%% file: sectionWorkflow.tex
The mesh generation described in \cref{sec:HOmeshing} yields analysis-suitable representations, taking  into account the connectivity data of the CAD model enhancement from \cref{sec:CADenhancement}.
It remains to incorporate the analysis data  of the problem (i.e., boundary conditions, material parameters) into the resulting higher-order non-conforming meshes. 
Therefore, our CAD plugin introduces \emph{analysis domains} of types electrode, dielectric, or floating. They establish a data structure containing the related material parameters and known boundary conditions.
Geometry groups resulting from the CAD model enhancement can be assigned to set up the analysis domains, which provides the association of the analysis data with the geometric information and the mesh.
Besides the analysis domains, there are also specific commands to specify \emph{solver settings}, such as FMM leave size and FMM degree, as well as the number of CPUs that shall be used, among others.

The CAD plugin manages the exchange with the BEM solver. It is
realized by exporting the geometric, analysis, and solver data into \emph{YAML files} used as input for the solver. These files also serve to document the simulations performed. The simulation itself runs as an independent task, so it does not block working with the CAD model. Once finished, the user is notified, and the numerical results can be displayed within the CAD software. In addition, there is a VTK export for more advanced visualizations with a postprocessor such as Paraview.

The user can execute all commands directly or from a Python script. 
The latter utilizes the Python scripting feature of Rhino 7. 
This option allows scripting the entire workflow, including CAD model generation, enhancement, meshing, analysis, and result visualization. 
Introducing variables enables an automated variation of the model or the analysis. 
Especially in combination with a version control system like Git, workflow scripting is a powerful tool for systematically manipulating and documenting the virtual prototyping process.


%% file: sectionResultsTwoSpheres.tex
The performance of the proposed non-conformal higher-order BEM solver is in the focus of this example.
We consider two spherical electrodes with radius $r=0.5$ in a vacuum with a voltage of $\pm 1$, respectively. 
The distance between the spheres' centers is $d=1.25$.
To assess the quality of the results, we use the capacitance $C$
of two spheres with equal radius $r$ 
\begin{align}
C= 2 \pi \varepsilon r \sum_{n=1}^{\infty }\frac{\sinh \left( \ln \left( D+\sqrt{D^2-1}\right) \right) }{\sinh \left( n\ln \left( D+\sqrt{ D^2-1}\right) \right)}  
\end{align}
with $D=\frac{d}{2r}$ and  the relative permittivity $\varepsilon$~\cite{RAWLINS1985}. In the implementation, the sum is computed until its increments fall below a threshold of $10^{-8}$. 

\begin{figure}
    \centering
    \myfbox{\includegraphics[trim={19cm 7cm 17cm 5cm},clip,width=0.78
    \textwidth]{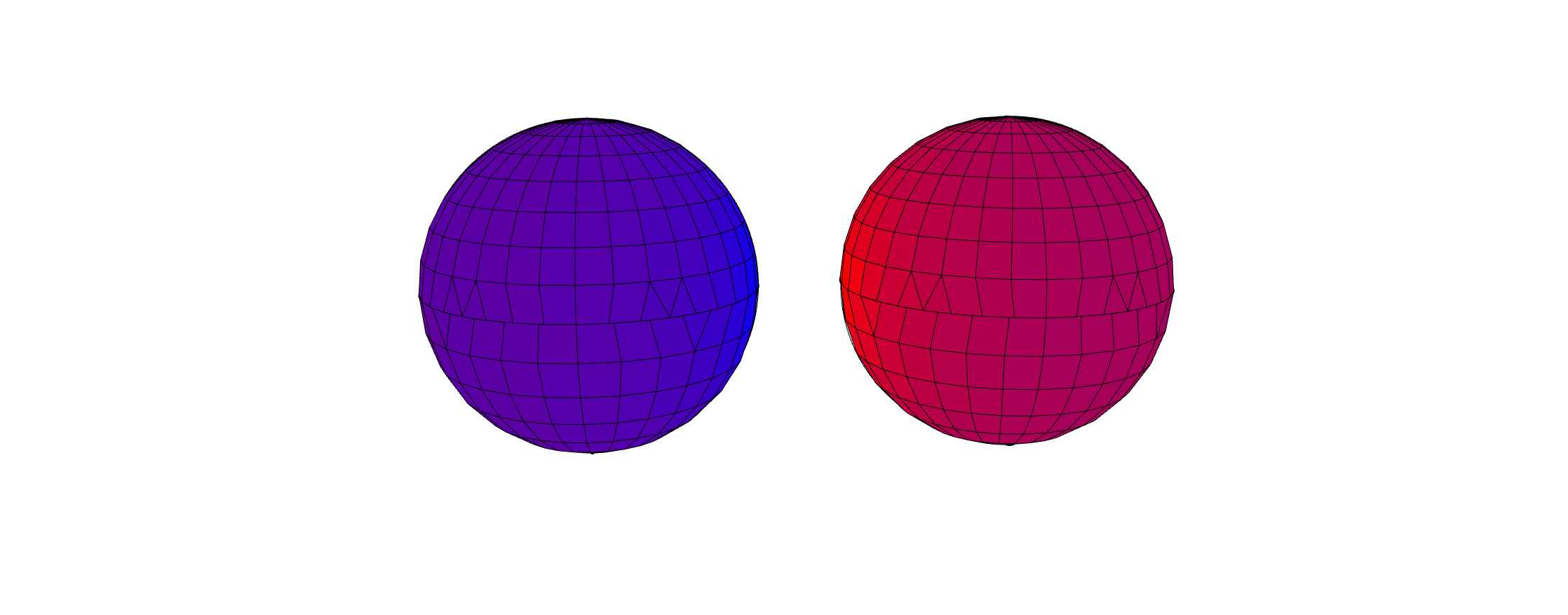}}
    \myfbox{\includegraphics[width=0.12\textwidth]{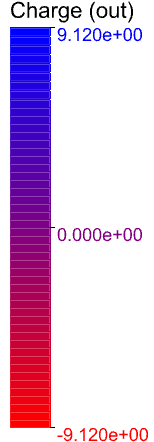}}
    
    \caption{
        Capacitance of two spheres: computed surface charge using a linear ($p=1$) non-conforming mesh with 932 nodes resulting in 932 triangles, 779 quadrilateral, and 3596 degrees of freedom (DoF).
        }   
    \label{fig:TwoSpheresChargeOut}  
\end{figure}

Each sphere is exactly represented by two hemispheres modeled defined by trimmed NURBS surfaces. At the poles the surface edges degenerate to a point and the trimmed boundaries specify naked edges coinciding along the spheres equator.
We employ the Rhino mesh generator that joins triangular elements to quadrilateral where possible. 
Subjecting the resulting linear meshes ($p=1$) to order elevation yields their quadratic counterparts ($p=2$).
When using the same mesh settings for all hemispheres, we obtain conforming meshes. For non-conforming meshes, on the other hand, the target element size $h$ is reduced for one of the hemispheres.
\Cref{fig:TwoSpheresChargeOut} illustrates an example of the resulting discretization using linear meshes. 

\begin{figure}
    \begin{tikzpicture}[every node/.style={black},]
        \begin{axis}[
            xmode=log,
            ymode=log, 
            xlabel={$h^{-1}$}, ylabel={$\epsilon_{rel}$}, %
            legend style={draw=none, /tikz/every even column/.append style={column sep=8pt}},
            scale=1.0,
            title={Capacitance of two spheres}, 
            legend style={at={(1.05,0.5)},anchor=west},
        ]

        \addplot table [ x expr = 1/\thisrowno{0}, y index = 2 ]{\dirdata/TwoSpheresConvergenceStudy_CombinedQuads_Linear_p1_v1.0_2025-06-03_183441.log};
        \addplot table [ x expr = 1/\thisrowno{0}, y index = 2 ]{\dirdata/TwoSpheresConvergenceStudy_CombinedQuads_Linear_p1_v0.6_2025-06-03_183441.log};
        \upperSlopeTriangleFlip{2}{0.15}{1}{0.15}

        \addplot table [ x expr = 1/\thisrowno{0}, y index = 2 ]{\dirdata/TwoSpheresConvergenceStudy_CombinedQuads_Quadratic_p2_v1.0_2025-06-03_183441.log};
        \addplot table [ x expr = 1/\thisrowno{0}, y index = 2 ]{\dirdata/TwoSpheresConvergenceStudy_CombinedQuads_Quadratic_p2_v0.6_2025-06-03_183441.log};
        \lowerSlopeTriangleFlip{3}{0.00002}{1}{0.15}

        \legend{
            $p=1$ conform,
            $p=1$ non-conf.,
            $p=2$ conform,
            $p=2$ non-conf.,                
            };
    
    \end{axis}
    \end{tikzpicture}
    \caption{Capacitance of two spheres:
        relative error with respect to the target element size $h$ for conformal (conform) and non-conformal (non-conf.) meshes with different polynomial oder $p$.   }
    \label{fig:TwoSpheresConvergenceStudyIsoparamertric}
\end{figure}
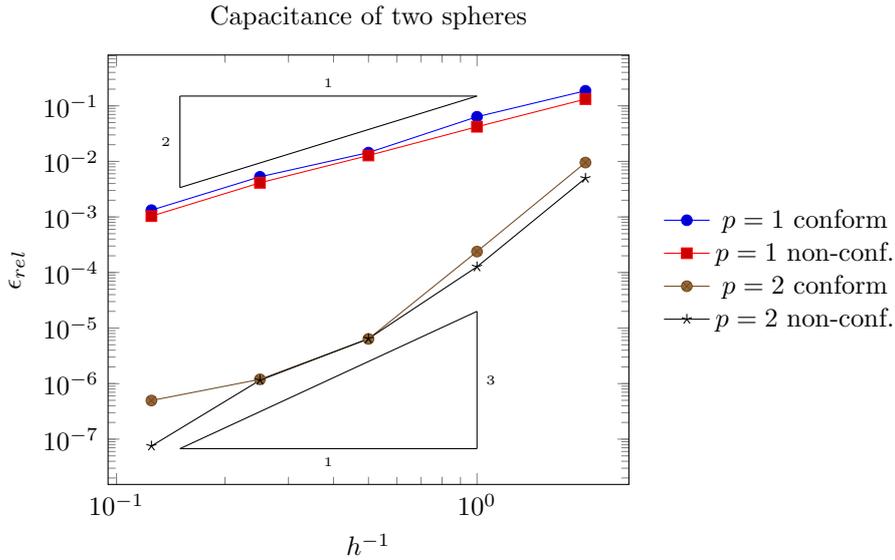

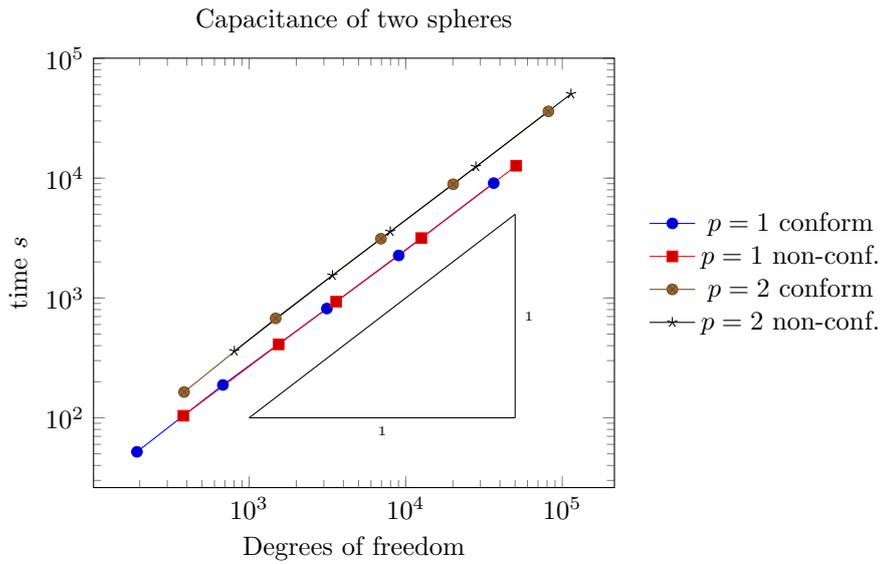
\begin{figure}
    \begin{tikzpicture}[every node/.style={black},]

        \begin{axis}[
            xmode=log,
            ymode=log, 
            xlabel={Degrees of freedom}, ylabel={time $s$}, %
            legend style={draw=none, /tikz/every even column/.append style={column sep=8pt}},
            scale=1.0,
            title={Capacitance of two spheres}, 
            legend style={at={(1.05,0.5)},anchor=west},
        ]

        \addplot table [ x index = 9, y index = 6 ]{\dirdata/TwoSpheresConvergenceStudy_CombinedQuads_Linear_p1_v1.0_2025-06-03_183441.log};
        \addplot table [ x index = 9, y index = 6 ]{\dirdata/TwoSpheresConvergenceStudy_CombinedQuads_Linear_p1_v0.6_2025-06-03_183441.log};

        \addplot table [ x index = 9, y index = 6 ]{\dirdata/TwoSpheresConvergenceStudy_CombinedQuads_Quadratic_p2_v1.0_2025-06-03_183441.log};
        \addplot table [ x index = 9, y index = 6 ]{\dirdata/TwoSpheresConvergenceStudy_CombinedQuads_Quadratic_p2_v0.6_2025-06-03_183441.log};
    
        \lowerSlopeTriangleFlip{1}{0.1}{50000}{1000}

        \legend{
            $p=1$ conform,
            $p=1$ non-conf., 
            $p=2$ conform,
            $p=2$ non-conf.,                    
            };
    
    \end{axis}
    \end{tikzpicture}
    \caption{Capacitance of two spheres:
    scaling of the computational time related to the degrees of freedom of the various conformal (conform) and non-conformal (non-conf.) meshes with different polynomial oder $p$.
    }
    \label{fig:TwoSpheresTimingsScalingIsoparamertric}
\end{figure}

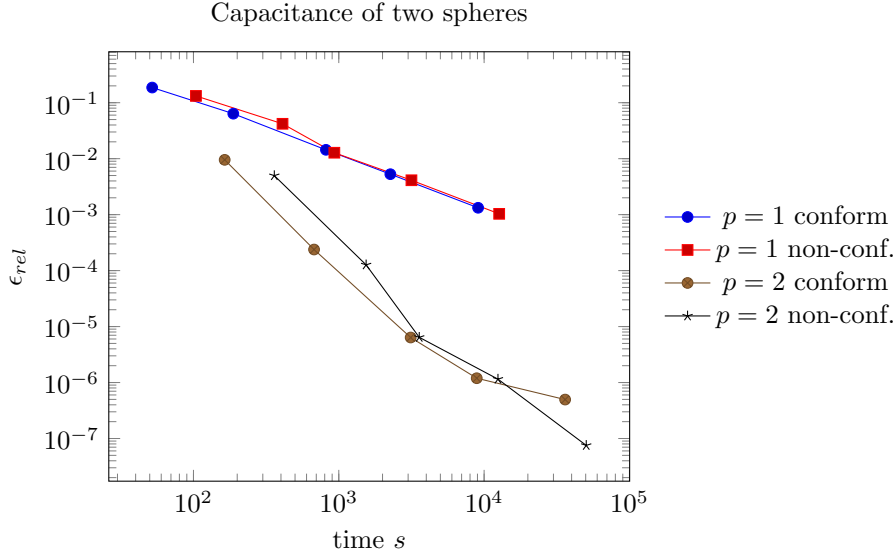
\begin{figure}
    \begin{tikzpicture}[every node/.style={black},]

        \begin{axis}[
            xmode=log,
            ymode=log, 
            xlabel={time $s$}, ylabel={$\epsilon_{rel}$}, %
            legend style={draw=none, /tikz/every even column/.append style={column sep=8pt}},
            scale=1.0,
            title={Capacitance of two spheres}, 
            legend style={at={(1.05,0.5)},anchor=west},
        ]

        \addplot table [ x index = 6, y index = 2 ]{\dirdata/TwoSpheresConvergenceStudy_CombinedQuads_Linear_p1_v1.0_2025-06-03_183441.log};
        \addplot table [ x index = 6, y index = 2 ]{\dirdata/TwoSpheresConvergenceStudy_CombinedQuads_Linear_p1_v0.6_2025-06-03_183441.log};

        \addplot table [ x index = 6, y index = 2 ]{\dirdata/TwoSpheresConvergenceStudy_CombinedQuads_Quadratic_p2_v1.0_2025-06-03_183441.log};
        \addplot table [ x index = 6, y index = 2 ]{\dirdata/TwoSpheresConvergenceStudy_CombinedQuads_Quadratic_p2_v0.6_2025-06-03_183441.log};

        \legend{
            $p=1$ conform,
            $p=1$ non-conf.,
            $p=2$ conform,
            $p=2$ non-conf.,                                
            };
    
    \end{axis}
    \end{tikzpicture}
    \caption{
    Capacitance of two spheres:
    relative error related to the computational time of the employed conformal (conform) and non-conformal (non-conf.) meshes with different polynomial oder $p$.    
    }
    \label{fig:TwoSpheresTimingsDoFsIsoparamertric}
\end{figure}

Furthermore, the color coding illustrates the surface charge obtained by the BEM simulation.
\Cref{fig:TwoSpheresConvergenceStudyIsoparamertric} summarizes the overall convergence behavior for conforming and non-conforming meshes of order $p=1,2$.
The related computational timings are reported in \cref{fig:TwoSpheresTimingsScalingIsoparamertric} and \cref{fig:TwoSpheresTimingsDoFsIsoparamertric}
%

These numerical results demonstrated the computational efficiency of the higher-order representation.
For $p=2$, the relative error for the finest meshes is around $10^{-7}$, while it is several orders of magnitude higher in the linear case.
Furthermore, there is no loss in accuracy when using non-conformal meshes. 
The corresponding higher computational time can traced back to the larger number of degrees of freedom due to our discretization setup.
In general, the linear scaling of the fast BEM formulation is evident. 
The FMM introduces approximation errors, which become apparent for the chosen settings at $10^{-7}$, as indicated by the graph for the conformal higher-order mesh.

\ifthenelse{\boolean{showSubfileBibliography}}{
    \bibliography{VEGA}%
}{}

%% file: sectionResultsFloating.tex
The previous example verified the convergence behavior of the presented BEM formulation but did not contain floating potentials. Thus, the examples in this section focus on them by comparing our results with ones reported in \cite{Amann2014,Blaszczyk2010}.
In particular, we are looking at three test cases:
\begin{itemize}
    \item Two spheres far: Two spheres with the same radius $r=\SI{0.1}{\meter}$, and the closest distance between them is $d=r$.
    \item Two spheres near: Two spheres with the same radius $r=\SI{0.1}{\meter}$, and the closest distance between them is $d=r/4$.
    \item Spheres and bicone: A sphere with radius $r=\SI{0.1}{\meter}$ and a bicone where the height and radius of the base is $r$. One spike of the bicone points towards the sphere, and its distance to it is $d=r/4$.
\end{itemize}
In all cases, one sphere is an electrode with a given electric potential $g_E=\SI{100}{\kilo\volt}$, and the surrounding air has a relative permittivity $\varepsilon_0=1$. The second object is modeled as a floating potential with a constant potential $\alpha$ to be calculated. 
As in \cite{Amann2014,Blaszczyk2010}, the approximate solutions of a 2D axial symmetric charge simulation solver \cite{Andjelic1992} are used for comparison.
The reference solutions reported are:
\begin{itemize}
    \item $\alpha_{ref}=\SI{33.9}{\kilo\volt}$ for the `two spheres far' test case
    \item $\alpha_{ref}=\SI{48.7}{\kilo\volt}$ for the `two spheres near' test case 
    \item $\alpha_{ref}=\SI{45.7}{\kilo\volt}$ for the `spheres and bicone' test case
\end{itemize}

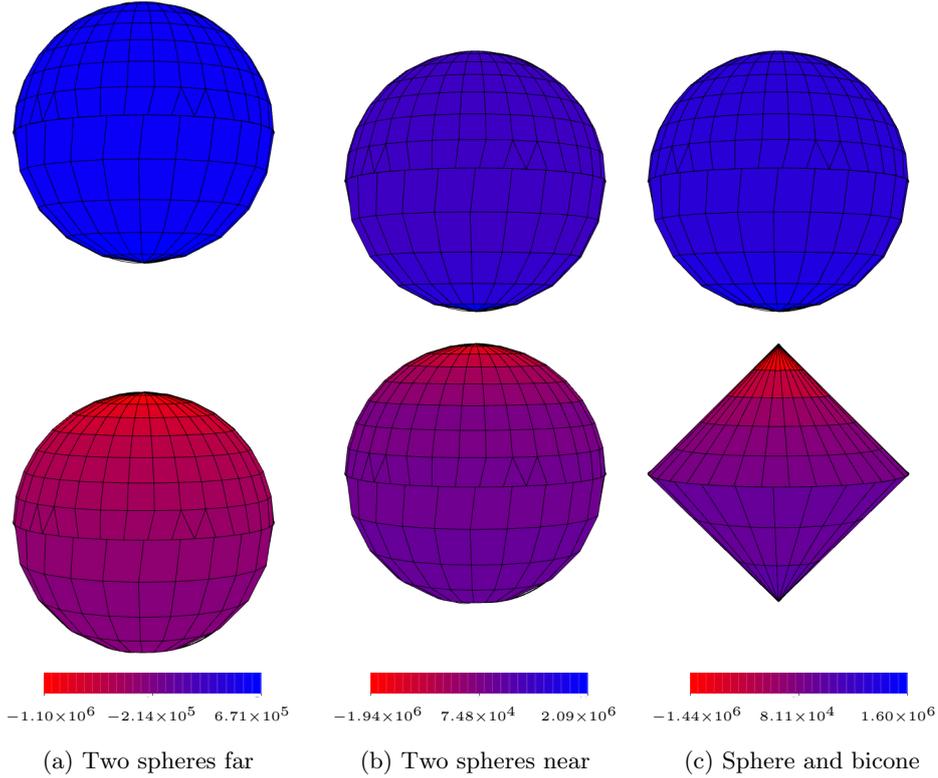
\begin{figure}
    \centering
    \begin{subfigure}[b]{0.32\textwidth}    
        \centering
        \myfbox{\includegraphics[trim={24cm 0cm 24cm 0cm},clip,width=0.85\textwidth]{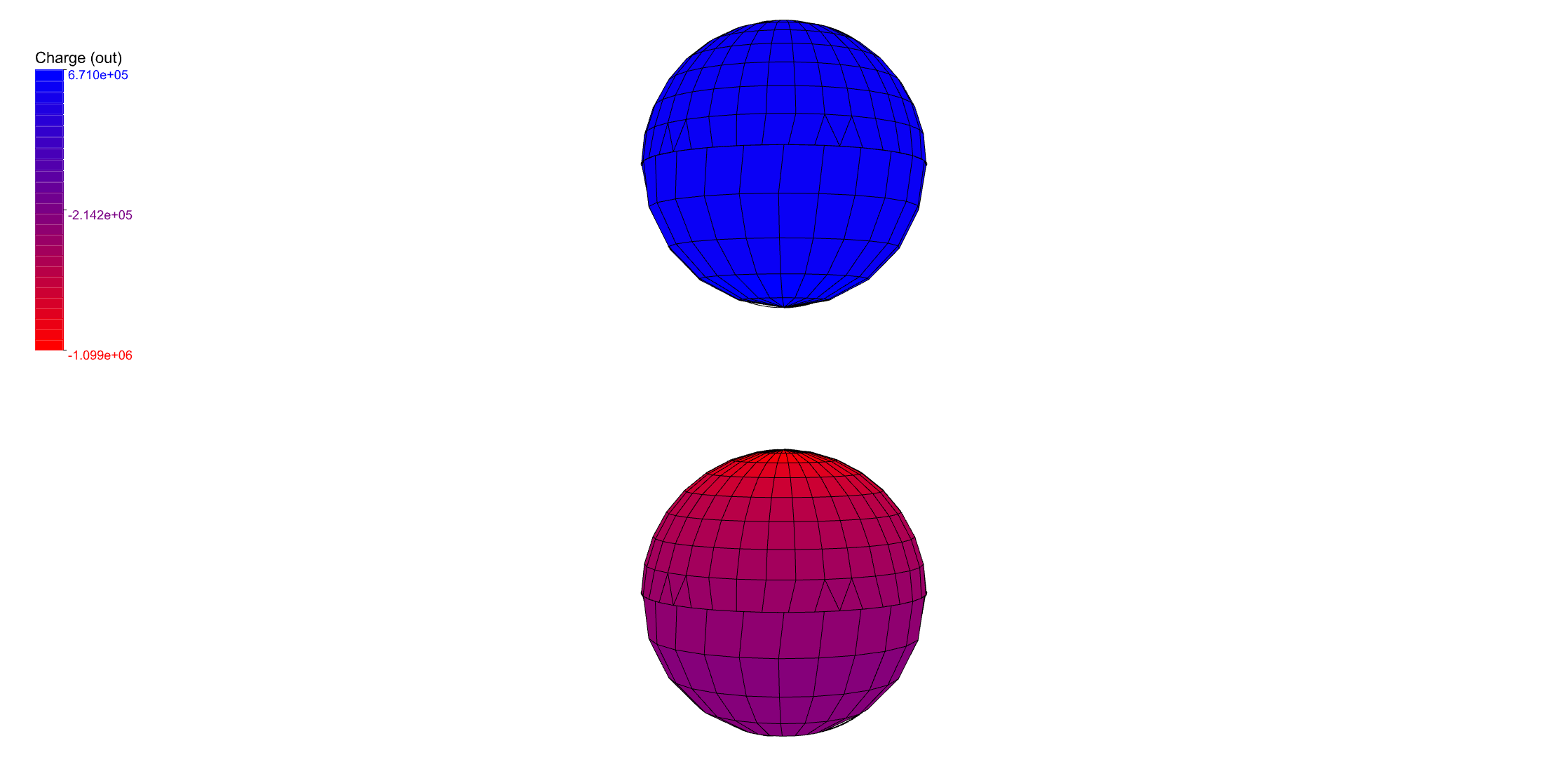}
        }  

        \begin{tikzpicture}[every node/.style={\myBlack}]
            \myaddgraphic{FloatingStudy_SphereFar_Quadratic_p2_n6_v0-6_2025-06-24_173043}{trim={1.4cm 15.2cm 56.6cm 2.5cm},clip,height=0.7\textwidth,angle=-90}{0}
            {
                \node[below] at (0.05,0){\tiny\num[tight-spacing = true]{-1.099e6}};
                \node[below] at (0.5,0){\tiny\num[tight-spacing = true]{-2.142e5}};
                \node[below] at (0.95,0){\tiny\num[tight-spacing = true]{6.710e5}};
            }{}
        \end{tikzpicture}
        \subcaption{Two spheres far}
        \label{fig:FloatingStudySphereFar} 
    \end{subfigure}
    \begin{subfigure}[b]{0.32\textwidth}  
        \centering
        \myfbox{\includegraphics[trim={24cm 0cm 24cm 0cm},clip,width=0.85\textwidth]{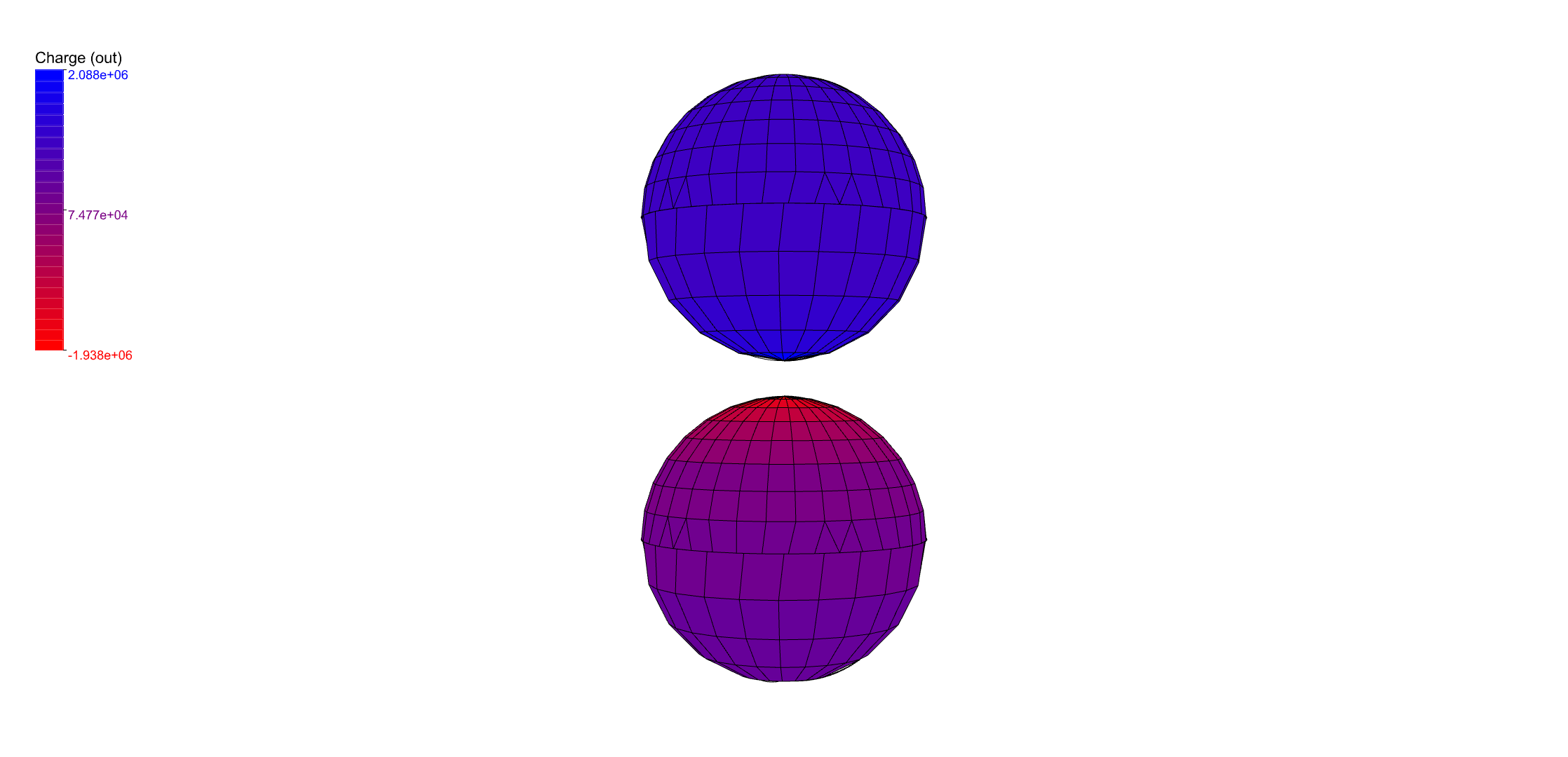}}
            
        \begin{tikzpicture}[every node/.style={\myBlack}]
            \myaddgraphic{FloatingStudy_SphereNear_Quadratic_p2_n6_v0-6_2025-06-24_173204}{trim={1.4cm 15.2cm 56.6cm 2.5cm},clip,height=0.7\textwidth,angle=-90}{0}
            {
                \node[below] at (0.05,0){\tiny\num[tight-spacing = true]{-1.938e6}};
                \node[below] at (0.5,0){\tiny\num[tight-spacing = true]{7.477e4}};
                \node[below] at (0.95,0){\tiny\num[tight-spacing = true]{2.088e6}};
            }{}
        \end{tikzpicture}
        \subcaption{Two spheres near}
        \label{fig:FloatingStudySphereNear}
    \end{subfigure}
    \begin{subfigure}[b]{0.32\textwidth}  
        \myfbox{\includegraphics[trim={24cm 0cm 24cm 0cm},clip,width=0.85\textwidth]{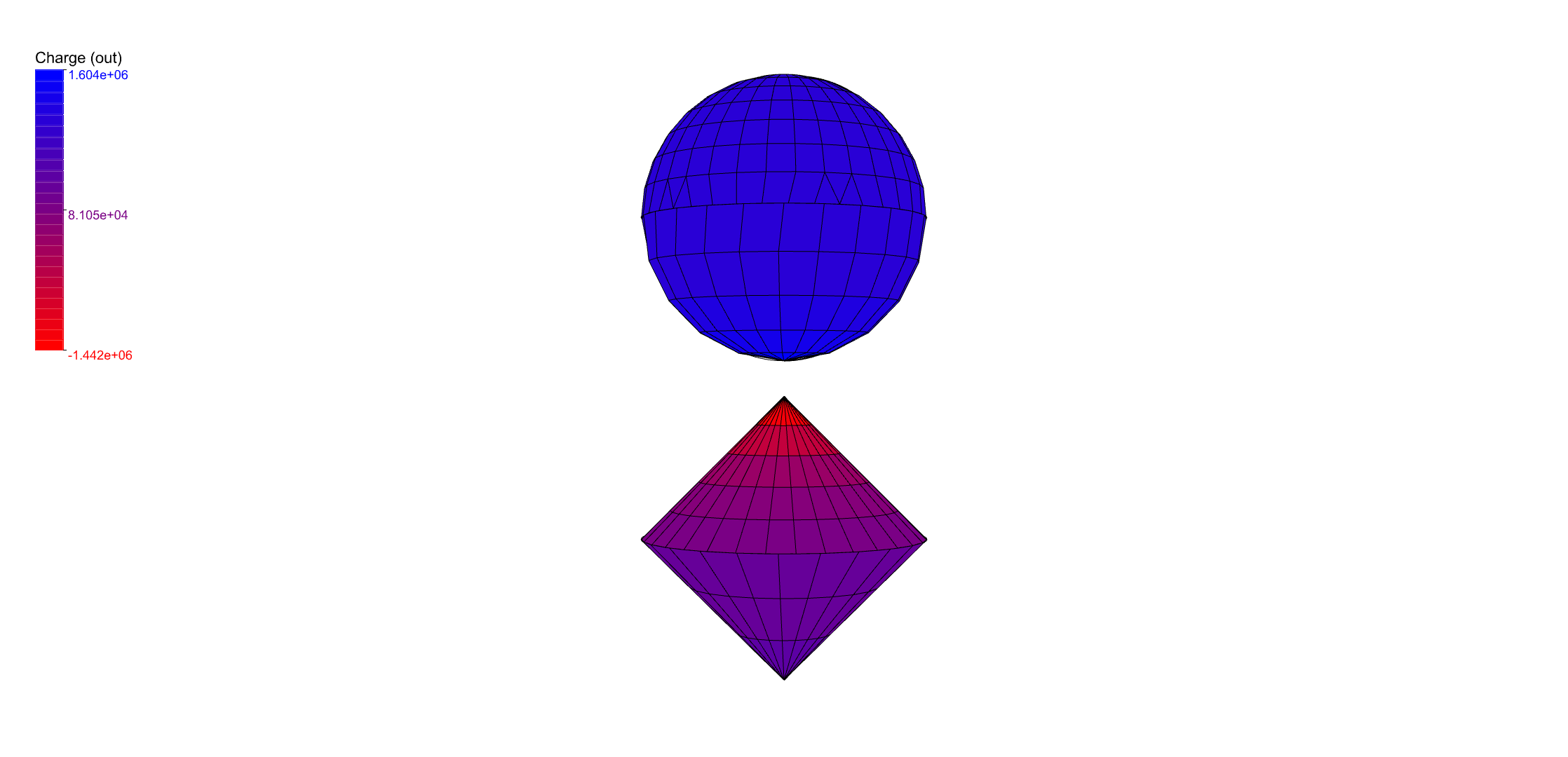}}

                \begin{tikzpicture}[every node/.style={\myBlack}]
            \myaddgraphic{FloatingStudy_Bicone_Quadratic_p2_n6_v0-6_2025-06-24_173255}{trim={1.4cm 15.2cm 56.6cm 2.5cm},clip,height=0.7\textwidth,angle=-90}{0}
            {
                \node[below] at (0.05,0){\tiny\num[tight-spacing = true]{-1.442e6}};
                \node[below] at (0.5,0){\tiny\num[tight-spacing = true]{8.105e4}};
                \node[below] at (0.95,0){\tiny\num[tight-spacing = true]{1.604e6}};
            }{}
        \end{tikzpicture}
        \subcaption{Sphere and bicone}
        \label{fig:FloatingStudyBicone}
    \end{subfigure}

    \caption{Floating potentials: computed surface charges for the three examples considered using quadratic isoparametric mesh ($\pu=\BEMFEOrder=2$) that are non-conforming. }   
    \label{fig:FloatingStudy}  
\end{figure}

For each test case, we compare isoparametric metric quadratic meshes ($\pu=\BEMFEOrder=2$) that are either (i) conforming or (ii) non-conforming.
\Cref{fig:FloatingStudy} illustrates the meshes employed and the surface charges obtained.
\Cref{tab:floatingStudyConformVsNonConform} summarizes the accuracy of the relative error of the computed potential $\alpha$ to the reference one $\alpha_{ref}$ as well as the degree of freedom (DoF) for the different test cases and meshes. 
\begin{table}[t]
\centering
\normalsize
\begin{tabular}{lcccccc} 
\toprule
 & \multicolumn{2}{c}{Two spheres far} 
 & \multicolumn{2}{c}{Two spheres near} 
 & \multicolumn{2}{c}{Sphere and bicone} \\ 
 & $\epsilon_{rel}$ & DoFs 
 & $\epsilon_{rel}$ & DoFs 
 & $\epsilon_{rel}$ & DoFs \\
\cmidrule(rl){2-3}
\cmidrule(rl){4-5}
\cmidrule(rl){6-7}
Conform   & \num{1.25199e-03} & 3900 & \num{8.28640e-04} & 3900 &\num{1.03715e-03} & 3054 \\
Non-conf. & \num{1.25238e-03} & 6456 & \num{8.31370e-04} & 6456 &\num{7.73848e-04} & 5124 \\
\bottomrule
\end{tabular}
\caption{Floating potentials by conforming (conform) and non-conforming (non-conf.) quadratic isoparametric meshes ($\pu=\BEMFEOrder=2$): Relative errors $\varepsilon_{rel}$ of the computed floating potential $\alpha$ and the related degrees of freedoms (DoFs) for each test case depicted in \cref{fig:FloatingStudy}.}
\label{tab:floatingStudyConformVsNonConform}
\end{table}
In all test cases, both the conforming and non-conforming meshes provide approximations that are close to the reference solutions.

Choosing quadratic meshes was motivated by the improved accuracy compared to linear meshes demonstrated in the previous example, cf.~\cref{sec:SphericalElectrodes}. At the same time, floating potentials are constant, and hence, it is reasonable to ask if increasing order $\BEMFEOrder$ of the shape functions for discretizing $\alpha$ is beneficial.
To assess the impact of $\BEMFEOrder$, we compute the floating potentials for all test case with a quadratic non-conforming mesh but with constant shape functions, i.e., $\pu=2$ and $\BEMFEOrder=0$. 
\Cref{tab:floatingStudyIsoparametric} reports the comparison of this discretization with its quadratic isoparametric counterpart regarding relative error, the DoFs, and the computational time.
While there are only slight deviations in accuracy, using $\BEMFEOrder=0$ yields a significant DoF reduction and, thus, a noticeable decrease in the computational time.
\begin{table}[t] 
    \normalsize
\centering
\begin{tabular}{lccccc} 
\toprule
 & \multicolumn{2}{c}{$\epsilon_{rel}$} 
 & \multicolumn{2}{c}{DoFs} 
  & rel.~time\\ 
  & $\BEMFEOrder=2$ & $\BEMFEOrder=0$  
  & $\BEMFEOrder=2$ & $\BEMFEOrder=0$  
 & $\BEMFEOrder=2/\BEMFEOrder=0$\\
\cmidrule(rl){2-3}
\cmidrule(rl){4-5}
\cmidrule(rl){6-6}
Two spheres far   & \num{1.25238e-03} & \num{1.23671e-03} & 6456 & 766  & \num{1.20} \\
Two spheres near & \num{8.31370e-04} & \num{7.24190e-04} & 6456 & 766  & \num{1.78} \\
Sphere and bicone& \num{7.73848e-04} & \num{8.05813e-04} & 5124 & 612  & \num{1.82} \\
\bottomrule
\end{tabular}
\caption{%
    Floating potentials by non-conforming meshes quadratic meshes ($\pu=2$) with different shape function order $\BEMFEOrder$: Relative errors $\varepsilon_{rel}$ of the computed floating potential $\alpha$, and the related degrees of freedoms (DoFs), and the relative computational times (rel,~time) for each test case depicted in \cref{fig:FloatingStudy}.}
\label{tab:floatingStudyIsoparametric}
\end{table}

\ifthenelse{\boolean{showSubfileBibliography}}{
    \bibliography{VEGA}%
}{}

%% file: sectionResultsParallelPlatesTrim.tex
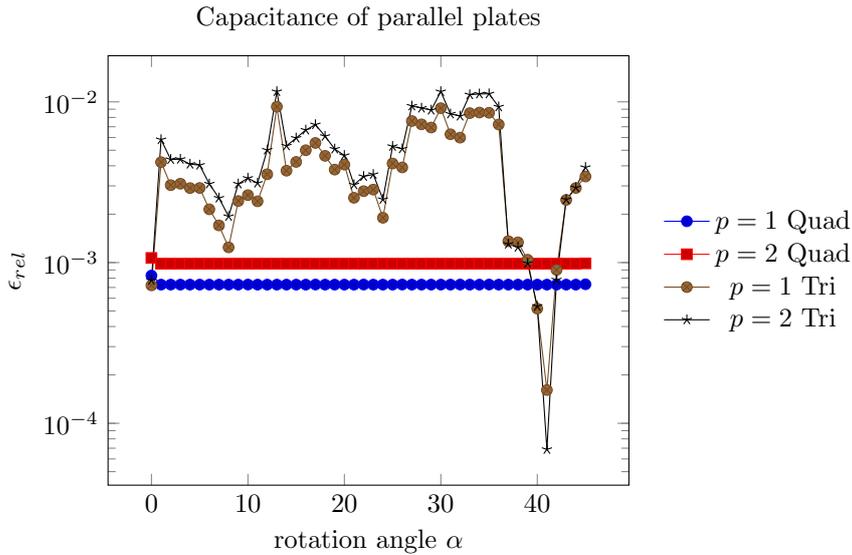
\begin{figure}[b!]
    \begin{tikzpicture}[every node/.style={black},]
    
        \begin{axis}[
            ymode=log, 
            xlabel={rotation angle $\alpha$}, ylabel={$\epsilon_{rel}$}, %
            legend style={draw=none, /tikz/every even column/.append style={column sep=8pt}},
            scale=1.0,
            title={Capacitance of parallel plates}, 
            legend style={at={(1.05,0.5)},anchor=west},
        ]

        \addplot table [ x index = 0, y index = 1 ]{\dirdata/ParallelPlatesTrimStudy_Quad_Linear_p1_2025-06-14_152012.log};
        \addplot table [ x index = 0, y index = 1 ]{\dirdata/ParallelPlatesTrimStudy_Quad_Quadratic_p2_2025-06-14_152012.log};
        \addplot table [ x index = 0, y index = 1 ]{\dirdata/ParallelPlatesTrimStudy_Tri_Linear_p1_2025-06-14_153726.log};
        \addplot table [ x index = 0, y index = 1 ]{\dirdata/ParallelPlatesTrimStudy_Tri_Quadratic_p2_2025-06-14_153726.log};

        \legend{
            $p=1$ Quad,
            $p=2$ Quad,
            $p=1$ Tri,
            $p=2$ Tri,
            };
    
    \end{axis}
    \end{tikzpicture}
    \caption{Capacitance of two parallel plates: relative error $\epsilon_{rel}$ related to the rotation angle $\alpha$ of triangular (Tri) and quadrilateral (Quad) meshes with different degrees $p$.}
    \label{fig:ParallelPlatesTrimStudy}
\end{figure}
\begin{figure}[th]
    \centering    
    \begin{subfigure}[t]{0.44\textwidth}    
        \myfbox{\includegraphics[trim={16.0cm 0.5cm 17.5cm 3cm},clip,width=0.92\textwidth]{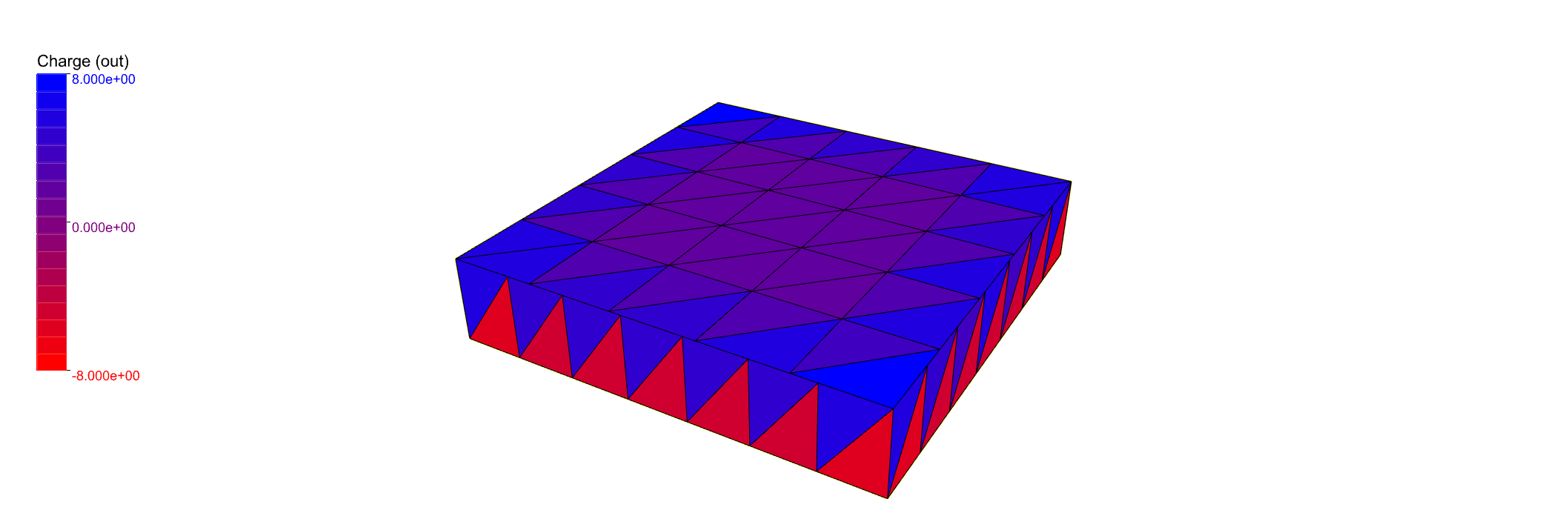}}
    \subcaption{Tri: $\alpha=0$}    
    \end{subfigure}
    \hfill
    \begin{subfigure}[t]{0.44\textwidth}  
        \myfbox{\includegraphics[trim={16.0cm 0.5cm 17.5cm 3cm},clip,width=0.92\textwidth]{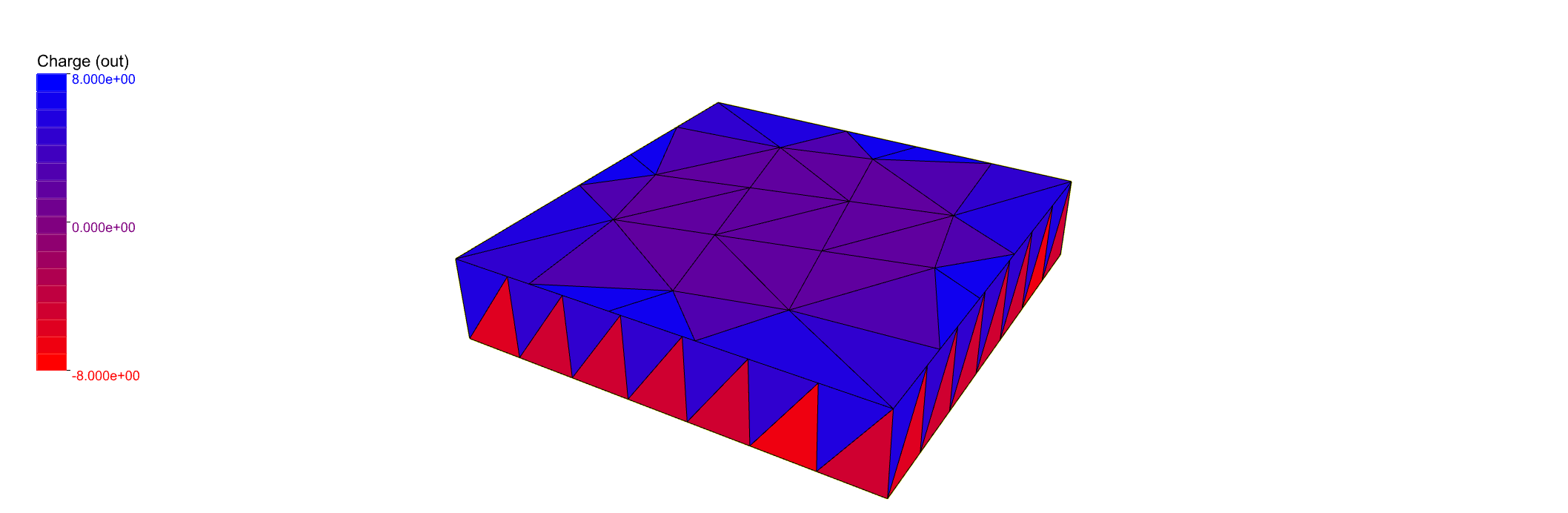}}
        \subcaption{Tri: $\alpha=13$}
    \end{subfigure}
    \begin{subfigure}[t]{0.44\textwidth}
        \myfbox{\includegraphics[trim={16.0cm 0.5cm 17.5cm 3cm},clip,width=0.92\textwidth]{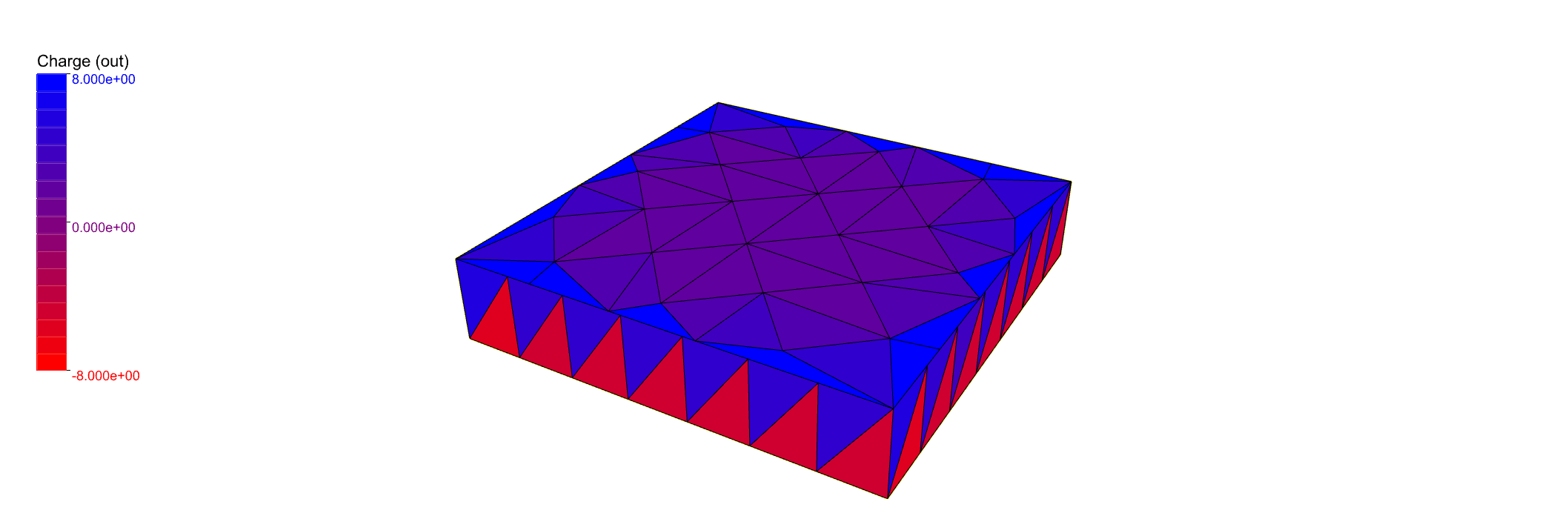}}
        \subcaption{Tri: $\alpha=41$}
    \end{subfigure}
    \hfill
    \begin{subfigure}[t]{0.44\textwidth}
        \myfbox{\includegraphics[trim={16.0cm 0.5cm 17.5cm 3cm},clip,width=0.92\textwidth]{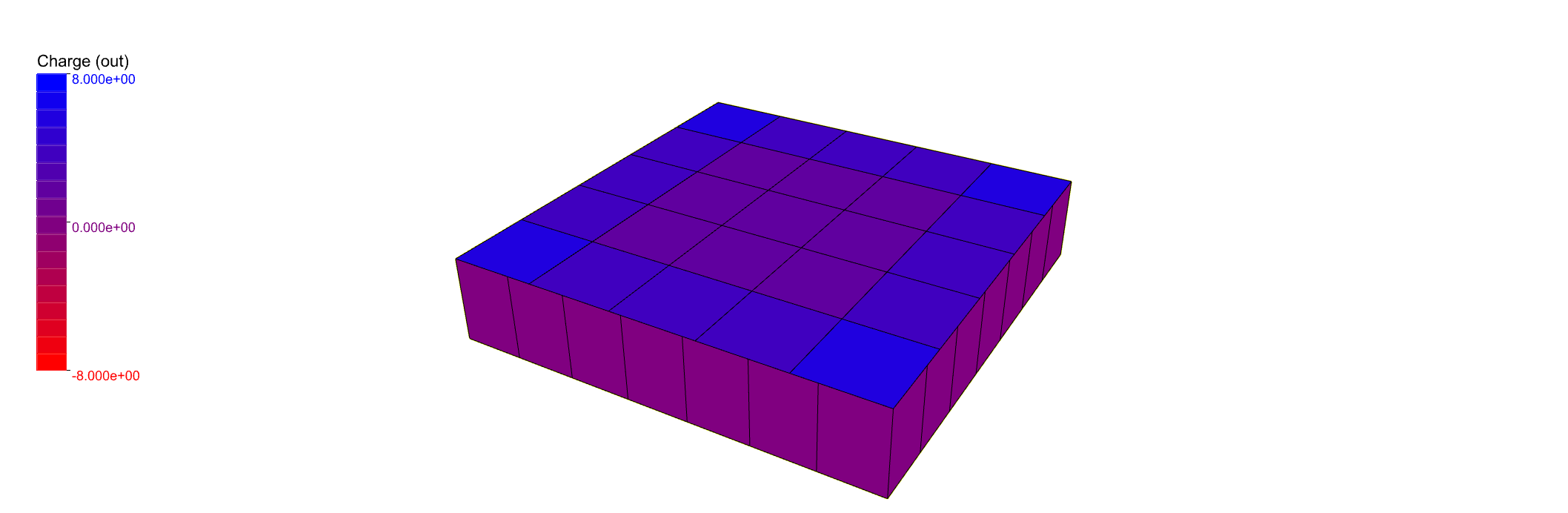}}
        \subcaption{Quad: $\alpha=41$}
    \end{subfigure}
    \begin{subfigure}[t]{1\textwidth}
        \centering
        \def\tkzscale{0.5}
        \tikzsetnextfilename{VEGAParallelPlatesScaleA}
        \includegraphics{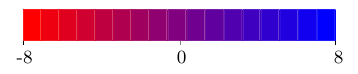}    
    \end{subfigure}
    \caption{Capacitance of two parallel plates: numerical results of the outgoing charge obtained by some linear meshes ($p=1$) from \cref{fig:ParallelPlatesTrimStudy}.}
    \label{fig:ParallelPlatesTrimCapture}
\end{figure}

This example examines the robustness of the proposed approach in translating the CAD model to the analysis mesh.
Here, we consider the capacitance of two parallel plates with the same measures. The capacitance is given as
\begin{align} 
    C = \frac{\varepsilon A}{d}
\end{align}
where $A$ is the area of the plates and $d$ is the distance between them.
For the example, we chose square plates with $A=1$ at $d=0.2$, and set the voltage of the top and bottom plates to $\pm 1$.
To assign the relative permeability $\varepsilon=10000$, the boundaries of the square plates are connected by perpendicular  side surfaces. 
This restriction of the domain introduces a modeling error because the reference solution of $C$ considers an unbounded setting. 
Anyway, we are interested in the change of accuracy due to different discretization.
In particular, the square plates are modeled as surfaces trimmed by the side planes.
\begin{figure}
    \centering
    
  \centering
  \def\tkzscale{0.3}
  \tikzsetnextfilename{VEGAParallelPlatesSetUp}
  \input{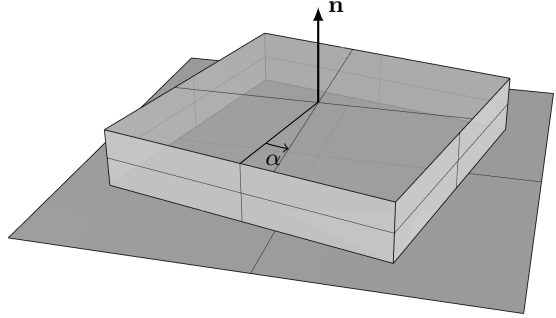}
  \caption{Capacitance of two parallel plates: two plates trimmed by four side surfaces. The top plate shows the trimmed surface, while the bottom one shows the initial  geometry. The trim curves vary with the rotation $\alpha$ around the normal vector $\mathbf{n}$.}
  \label{fig:ParallelPlatesTrimSetUp}

\end{figure}
As shown in \cref{fig:ParallelPlatesTrimSetUp}, different trimming situations are constructed by rotating the parallel plates by an angle $\alpha$ around the normal vector $\mathbf{n}$. 
For each trim situation, the same meshing settings are applied using either Rhino's  default triangulation or its QuadRemesh option.
\Cref{fig:ParallelPlatesTrimStudy} presents the relative error, $\epsilon_{rel}$, related to the angle $\alpha$ for the various mesh types.
Furthermore, \cref{fig:ParallelPlatesTrimCapture} illustrates some related meshes and the results of the computed charge.
%

Linear and quadratic meshes yield similar accuracy for the example considered since the model consists only of planar surfaces.
Considering the different mesh types, however, the results indicate that Rhino's default mesh generator is more dependent on the parametrization of the underlying geometric surface. 
The relatively new QuadRemesh option provided the same mesh for all $\alpha$, except for  $\alpha=0$, where a visually indistinguishable perturbation occurs.


%% file: sectionResultsBushing.tex
\begin{figure}
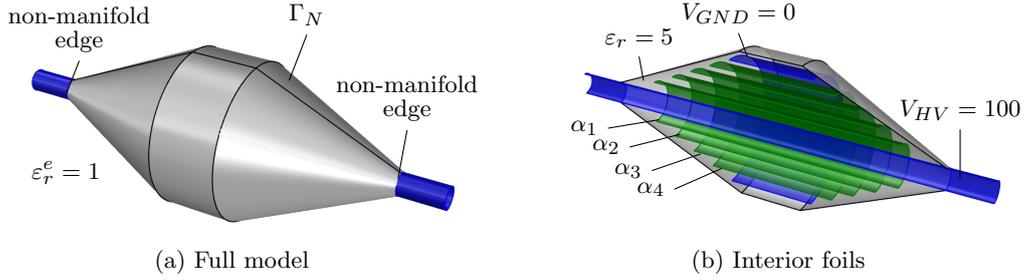

  \centering
  \begin{minipage}[b]{0.45\textwidth}    
    
    \def\tkzscale{0.27}
    \centering
    \tikzsetnextfilename{VEGABushingNonManifoldSetUpFull}
    \input{\TikzPath/VEGABushingNonManifoldSetUpFull}
    \subcaption{Full model}
    \label{fig:bushingBCFull}

  \end{minipage}
  \hfill
  \begin{minipage}[b]{0.45\textwidth} 
    
    \def\tkzscale{0.27}
    \centering
    \tikzsetnextfilename{VEGABushingNonManifoldSetUp}
    \input{\TikzPath/VEGABushingNonManifoldSetUp}
    \subcaption{Interior foils}
    \label{fig:bushingBCCut}

    \end{minipage}
  \caption{High voltage bushing CAD model: (a) cylindrical conductor shown in blue embedded in a solid dielectric with the boundary $\Gamma_N$ shown in gray, and (b) interior foils where the blue one is grounded, while the green ones are floating potentials with unknown $\alpha_i$.}
  \label{fig:bushingBC}
\end{figure}
This example considers the design of a high-voltage bushing, as shown in \cref{fig:bushingBC}.
At the center of the model is a cylindrical conductor with a radius of $r$ and a voltage of $V_{HV}=100$.
It is embedded in a solid dielectric material with the relative permittivity $\varepsilon_r=5$, while for the exterior domain $\varepsilon^e_r=1$.
The boundary $\Gamma_N$ of the solid dielectric is represented by two truncated cones (radii $r$ and $7r$, height $12r$) and a cylinder (radius $7r$, height $8r$).
Inside the solid material, five metallic cylindrical foils surround the conductor. The distance between them is $r$, and their height is chosen such that the distance to $\Gamma_N$ in the longitudinal direction is also $r$. 
The outermost foil is grounded; thus, the voltage is prescribed as $V_{GND}=0$.
The other foils are floating potentials $\Gamma_F$ that shall enforce a uniform potential distribution along the bushing's surface.
As highlighted in \cref{fig:bushingBCFull}, the model has two non-manifold edges where the cylindrical conductor leaves the dielectric.

\begin{figure}
  \centering
  \begin{subfigure}[t]{0.44\textwidth}  
      \myfbox{\includegraphics[trim={11cm 3.4cm 11cm 0.4cm},clip,angle=-100,width=\textwidth]{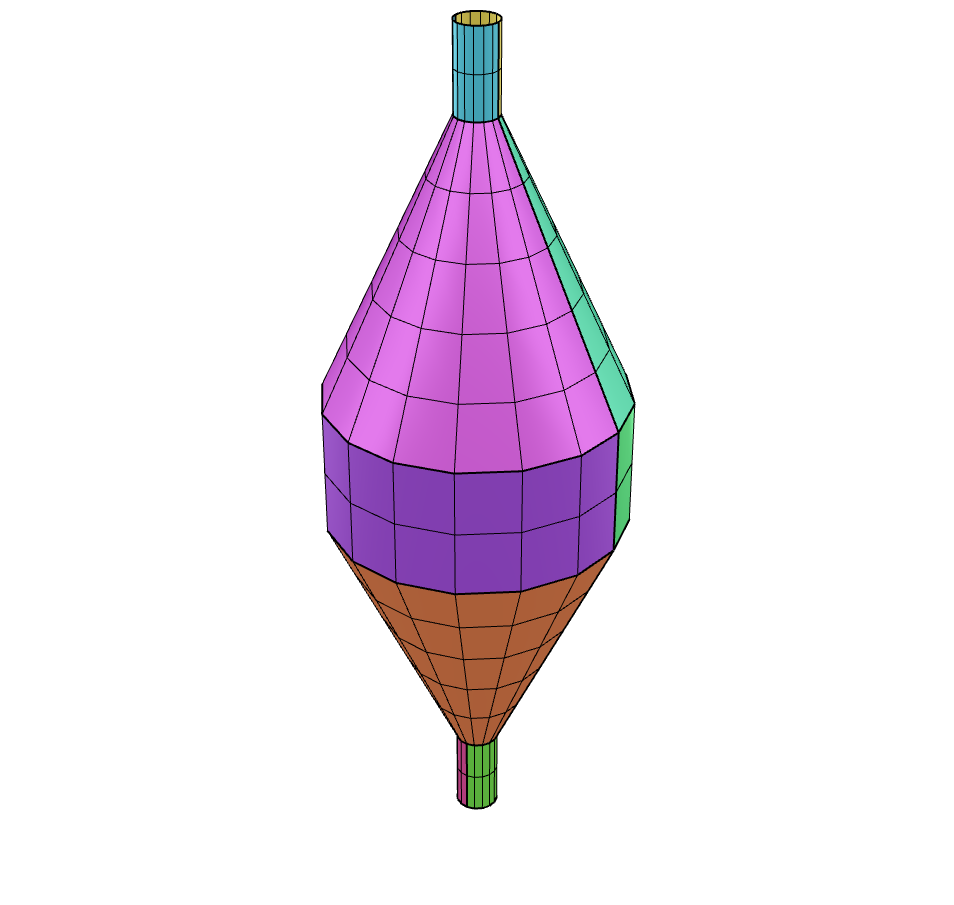}}
      \subcaption{Conforming mesh}
  \end{subfigure}
  \hfill
  \begin{subfigure}[t]{0.44\textwidth}
      \myfbox{\includegraphics[trim={11cm 3.4cm 11cm 0.4cm},clip,angle=-100,width=\textwidth]{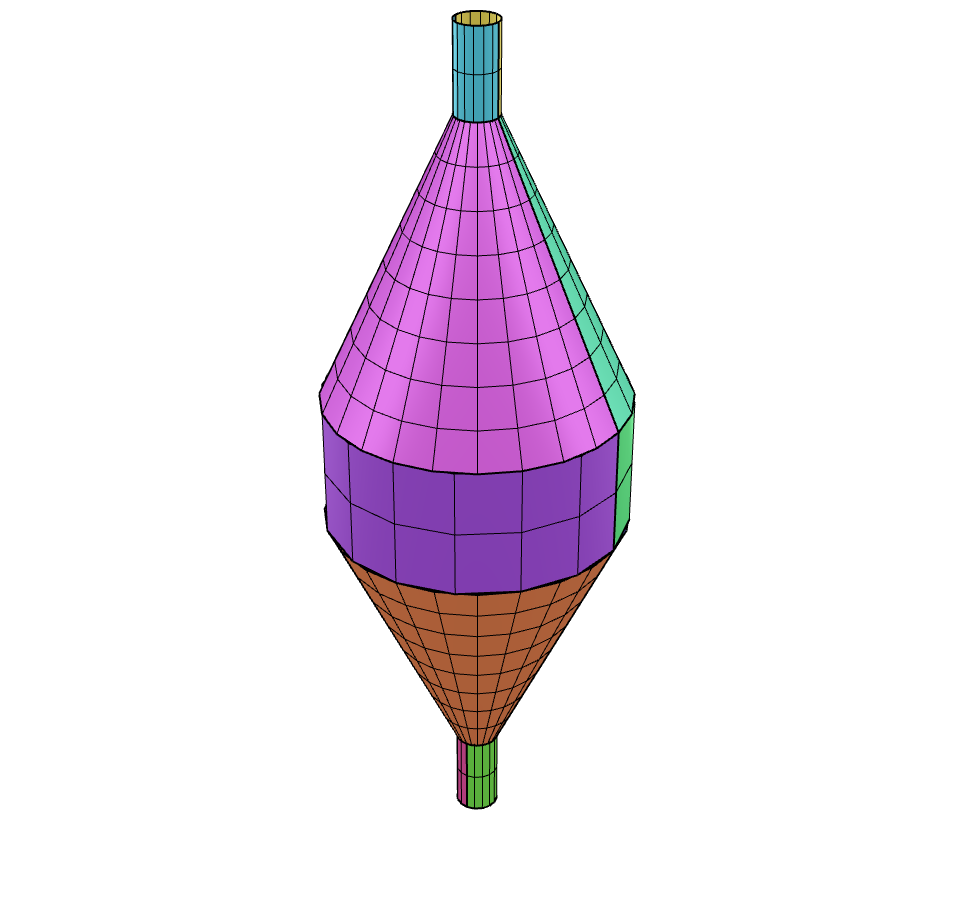}}
      \subcaption{Non-conforming mesh}
  \end{subfigure}
  \caption{High voltage bushing meshes}
  \label{fig:bushingMeshes}
\end{figure}

We employ the conforming and the non-conforming quadratic meshes shown in \cref{fig:bushingMeshes}.
Note that also the non-manifold edges are non-conforming in the latter mesh.
\Cref{tab:bushingStudy} compares the degrees of freedom of the meshes and the resulting approximate floating potentials $\alpha_i$, $i=1,\dots,4$.
\begin{table}[ht] 
    \normalsize
\centering
\begin{tabular}{lcccccc} 
\toprule
 & Nodes & DoFs & $\alpha_1$ & $\alpha_3$ & $\alpha_3$ & $\alpha_4$\\
\midrule
Conform   & 3366 & 6336 & \num{71.5158943409} & \num{52.2808145350} & \num{35.8145518293} & \num{19.3699561704} \\
Non-conf. & 4318 & 8352 & \num{71.4610196286} & \num{52.1993125419} & \num{35.7592048306} & \num{19.3530448068} \\
\bottomrule
\end{tabular}
\caption{High voltage busing: computed floating potentials $\alpha_i$ using the conformal and non-conformal quadratic isoparametric meshes shown in \cref{fig:bushingMeshes}.
 }
\label{tab:bushingStudy}
\end{table}

\ifthenelse{\boolean{showSubfileBibliography}}{
    \bibliography{VEGA}%
}{}